\documentclass[useAMS]{mn2e}
\usepackage{graphicx}
\usepackage{graphics}
\usepackage{amsmath}
\usepackage{amssymb}

\usepackage{times}

\title[Local radio galaxies at 20\,GHz]
      {The local radio-galaxy population at 20\,GHz}

\author[Sadler et al.]{
\parbox[t]{\textwidth}
{Elaine M.\ Sadler$^1$\thanks{E-mail: ems@physics.usyd.edu.au}, Ronald D.\ Ekers$^2$, 
Elizabeth K.\ Mahony$^3$, Tom Mauch$^{4,5}$, Tara Murphy$^{1,6}$
}
\vspace*{6pt} \\
$^{1}$Sydney Institute for Astronomy, School of Physics, The University of Sydney, NSW 2006, Australia \\
$^{2}$Australia Telescope National Facility, CSIRO, PO Box 76, Epping, NSW 1710, Australia \\ 
$^{3}$ASTRON, the Netherlands Institute for Radio Astronomy, Postbus 2, 7990 AA, Dwingeloo, The Netherlands \\
$^{4}$Oxford Astrophysics, Department of Physics, Keble Road, Oxford OX1 3RH  \\
$^{5}$SKA Africa, 3rd Floor, The Park, Park Road, Pinelands, 7405, South Africa \\ 
$^{6}$School of Information Technologies, The University of Sydney, NSW 2006, Australia \\
}

\date{Accepted 0000 December 08. Received 0000 December 08; in original form 0000 December 08}

\pagerange{\pageref{firstpage}--\pageref{lastpage}} 
\pubyear{2013}

\begin{document}

\maketitle

\label{firstpage}

\begin{abstract}
We have made the first detailed study of the high-frequency radio-source 
population in the local universe, using a sample of 202 radio sources from 
the Australia Telescope 20\,GHz (AT20G) survey identified with 
galaxies from the 6dF Galaxy Survey (6dFGS).  
The AT20G-6dFGS galaxies have a median redshift of z=0.058 and span a wide 
range in radio luminosity, allowing us to make the first measurement of 
the local radio luminosity function at 20\,GHz. 

Our sample includes some classical FR-1 and FR-2 radio galaxies, 
but most of the AT20G-6dFGS galaxies host compact (FR-0) 
radio AGN which appear lack extended radio emission even at lower 
frequencies. Most of these FR-0 sources show no evidence 
for relativistic beaming, and the FR-0 class appears to be a mixed 
population which includes young Compact Steep-Spectrum 
(CSS) and Gigahertz-Peaked Spectrum (GPS) radio galaxies. 

We see a strong dichotomy in the Wide-field Infrared Survey Explorer (WISE) 
mid-infrared colours of the host galaxies of FR-1 and FR-2 radio sources, 
with the FR-1 systems found almost exclusively in WISE `early-type' 
galaxies and the FR-2 radio sources in WISE `late-type' galaxies. 

The host galaxies of the flat- and steep-spectrum radio sources have 
a similar distribution in both K--band luminosity and WISE colours, 
though galaxies with flat-spectrum sources are more likely to show 
weak emission lines in their optical spectra. We conclude that these 
flat-spectrum and steep-spectrum radio sources mainly represent different 
stages in radio-galaxy evolution, rather than beamed and unbeamed radio-source 
populations. \\
\end{abstract}


\section{Introduction} 
Measurements of the radio-source population in the local universe provide an essential 
benchmark for studying the co-evolution of galaxies and their central black holes over 
cosmic time. 
The local radio-source population has now been mapped out in detail at 1.4\,GHz through the 
combination of large-area radio continuum and optical redshift surveys (Condon et al. 2002; 
Sadler et al.\ 2002; Best et al.\ 2005a; Mauch \& Sadler 2007; see also De Zotti et al.\ 2010 
for a recent review), and the radio luminosity functions of both star-forming galaxies and 
active galactic nuclei (AGN) have been accurately measured. 
Members of the two classes overlap in radio luminosity, but can usually be distinguished 
using optical spectra (Sadler et al.\ 1999; Best et al.\ 2005b).  

Much less is known about the local radio-source population at other frequencies, 
and the recent completion of a sensitive, large-area radio continuum survey at 20\,GHz, 
the AT20G survey (Murphy et al.\ 2010) provides a first opportunity to study the 
high-frequency radio properties of nearby galaxies in a systematic way.  

The radio emission from active galaxies at 20\,GHz arises mainly from the galaxy core, 
rather than from extended radio lobes (e.g Sadler et al.\ 2006, De Zotti et al.\ 2010, 
Massardi et al.\ 2011a, Mahony et al.\ 2011).  The AT20G survey therefore allows us to 
identify some of the youngest radio galaxies in the local universe, with radio spectra peaking 
above 5\,GHz, which can provide new insights into the earliest stages of radio-galaxy 
evolution (Snellen et al.\ 2000; Hancock et al.\ 2009).  

Our aim in this paper is to map out the overall properties of the local radio-source 
population at 20\,GHz, compare this with earlier studies of local radio sources selected 
at 1.4\,GHz (Best et al.\ 2005a; Mauch \& Sadler 2007; Best \& Heckman 2012) and use 
the radio spectral-index information available from the AT20G sample to test whether 
the host galaxies of flat-spectrum and steep-spectrum radio sources are drawn from the 
same population. 

Section 2 describes the construction of the first 20\,GHz-selected sample of nearby 
galaxies, and provides a data table for the 202 southern (dec $<$0$^\circ$) galaxies 
in this sample. The radio properties of the sample are discussed in \S3, and 
the local radio luminosity function at 20\,GHz derived in \S4. \S5 discusses 
the optical and infrared properties of our galaxy sample. \S6 compares the 
local radio-source population at 20\,GHz with that seen in earlier studies 
at 1.4\,GHz, and \S7 presents our conclusions and some suggestions for further 
work.  Some notes on individual sources are added in Appendix A. 

Throughout this paper, we assume H$_{\rm 0}$ = 71 km\,s$^{-1}$\,Mpc$^{-1}$, 
$\Omega_{\rm M}$ = 0.27 and $\Omega_{\lambda}$ = 0.73.

\section{The AT20G--6dFGS galaxy sample}

We assembled the galaxy sample studied in this paper by matching radio sources from the 
Australia Telescope 20\,GHz Survey catalogue (AT20G; Murphy et al.\ 2010) with nearby galaxies 
from the Third Data Release of the 6dF Galaxy Survey (6dFGS DR3; Jones et al.\ 2009).  The 
6dFGS was chosen because it is a large-area survey well-matched to the area covered by 
AT20G, and shallow enough in redshift that the effects of cosmic evolution within the sample 
volume can be neglected.  

In assembling the AT20G-6dFGS sample we used a similar methodology to that of Mauch \& Sadler 
(2007), who assembled and studied a sample of several thousand nearby radio sources selected 
at 1.4\,GHz from the 6dFGS DR2 (Jones et al.\ 2004) and NVSS (Condon et al.\ 1998) catalogues. 
By doing this, we can make direct comparisons between two galaxy samples selected from 
the same optical survey but at very different radio frequencies. 

The AT20G source catalogue covers the whole southern sky\footnote{As noted by Massardi et 
al.\ (2011a), the AT20G catalogue has low completeness in a strip of sky at 16--18 hours 
RA at declination north of $-15^\circ$ because of bad weather during the final observing 
run. The 6dFGS also has a small number of fields which could not be observed during the 
survey, mainly in the RA range 6--12 hours with declination south of $-40^\circ$.  
Diagrams showing the survey completeness as a function of position on the sky can be found 
in Figure\,1 of Massardi et al.\ (2011a) for AT20G and Figure 1(c) of Jones et al.\ (2009) 
for 6dFGS. } 
(declination $<0^\circ$ and Galactic latitude $|b|>1.5^\circ$)
and includes 5890 sources above a 20\,GHz flux density limit 
of 40\,mJy. The 6dFGS catalogue contains infrared JHK photometry and optical redshifts for 
a sample of about 125,000 southern (declination $<0^\circ$ and Galactic latitude $|b|>10^\circ$) 
galaxies brighter than K = 12.75 mag. The median redshift of the 6dFGS galaxies is $z=0.053$. 

\subsection{Source selection} 

We matched the AT20G catalogue (Murphy et al.\ 2010) 
against the 6dFGS DR3 spectroscopic catalogue for galaxies in the main K-band sample 
(progID=1 in the 6dFGS catalogue), taking into account the following points: 

\begin{enumerate}
\item 
Most AT20G sources are unresolved on scales of 10-15 arcsec, and are associated with the 
radio cores of galaxies and QSOs (Sadler et al.\ 2006).  For these objects, making an 
optical identification is generally straightforward.  
\item 
Around 5-6\% of AT20G sources show extended structure within the 2.4\,arcmin 
ATCA primary beam at 20\,GHz, and are flagged as extended in the catalogue 
(Murphy et al.\ 2010).  The AT20G catalogue position for these sources corresponds 
to the peak flux in the image.  This is usually the flat-spectrum core, and 
for these sources the optical identification will again be straightforward.  
In a small number of the strongest peak is a hotspot in the lobes or jet, rather 
than the core, and extra effort is needed to make the correct optical identification. 
\item
Previous work on the AT20G sources generally used a 2.5\,arcsec cutoff radius in making 
optical identifications (e.g. Sadler et al.\ 2006; Massardi et al.\ 2008).  
While this is appropriate for the AT20G sample as a whole (where distant QSOs are the dominant 
source population), a larger matching radius should be used for 
the 6dFGS galaxies because of their large angular size and the relatively low surface density 
of these bright galaxies. 
\end{enumerate} 

We began by setting a cutoff radius of 60\,arcsec for candidate AT20G/6dFGS matches.  
This produced a total of 425 candidates, 218 of which were galaxies in the main 
6dFGS sample (progID=1) with the remainder belonging to one of the ``additional target'' 
samples carried out in parallel with the 6dFGS (Jones et al.\ 2009).  These additional 
target objects (which include samples of QSOs, radio and infrared-selected AGN as 
well as other galaxies which are fainter than the K=12.75\,mag cutoff) are not discussed 
here, but are analysed in a separate paper (Mahony et al.\ 2011). 

We then visually inspected all the candidate matches, looking at the 20\,GHz AT20G 
images, optical overlay plots and (lower-resolution) low-frequency radio images from 
the 843\,MHz SUMSS and 1.4\,GHz NVSS images (Bock et al.\ 1999; Condon et al.\ 1998).  
We also cross-matched the full AT20G catalogue with the lower-frequency southern 2Jy 
(Morganti et al.\ 1993) and MS4 (Burgess \& Hunstead 2006) bright radio-source samples 
to check whether any AT20G sources were identified with hotspots of nearby radio galaxies 
with very large angular sizes (which would have been missed by our 1\,arcmin cutoff radius).  

\begin{table*}
\centering
\caption{AT20G-6dFGS radio galaxies with a large ($>$10\,arcsec) offset between the optical 
and 20\,GHz positions. The AT20G positions and flux densities of each component are 
listed, along with the offset $\Delta$ between the radio and optical positions, 
the fractional linear polarization at 20\,GHz (m20) and a classification of the 
radio structure. 
\label{tab_mult}}

\begin{tabular}{@{}lrlrrrrrrl@{}}
\hline
Source & comp & \multicolumn{1}{c}{AT20G name} & \multicolumn{2}{c}{AT20G position} & 
\multicolumn{1}{c}{$\Delta$} & \multicolumn{1}{c}{S$_{20}$} & $\pm$ & m20 &\multicolumn{1}{c}{Notes} \\
 &  & & \multicolumn{2}{c}{(J2000)} & \multicolumn{1}{c}{arcsec} & \multicolumn{2}{c}{(mJy)} & \% & \\
 \hline
\multicolumn{10}{l}{(a) Galaxies associated with two or more catalogued AT20G sources } \\
PKS\,0043-42 & 1 & J004613-420700 & 00 46 13.32 & -42 07 00.1 &  71 &  363 &  15 &    7.4 & Hotspot \\
             & 2 & J004622-420842 & 00 46 22.30 & -42 08 42.6 &  72 &  139 &  14 &   17.9 & Hotspot \\
Pictor A     & 1 & J051926-454554 & 05 19 26.34 & -45 45 54.5 & 245 & 1464 &  55 &   38.0 & Hotspot \\ 
             & 2 & J051949-454643 & 05 19 49.70 & -45 46 43.7 &   0 & 1107 &  54 & $<$1.0 & Core \\
             & 3 & J052006-454745 & 05 20 06.47 & -45 47 45.4 & 186 &  426 &  10 &    8.6 & Hotspot \\
PKS\,0634-20 & 1 & J063631-202857 & 06 36 31.24 & -20 28 57.6 & 356 &   55 &   3 &$<$14.7 & Hotspot \\
             & 2 & J063633-204239 & 06 36 33.00 & -20 42 39.3 & 466 &  183 &   6 &   12.1 & Hotspot \\
PKS\,1717-00 & 1 & J172019-005851 & 17 20 19.74 & -00 58 51.2 & 126 &  313 &   7 &$<$11.0 & Hotspot \\
             & 2 & J172034-005824 & 17 20 34.24 & -00 58 24.6 &  94 &  122 &   6 & $<$8.1 & Hotspot \\
PKS\,1733-56 & 1 & J173722-563630 & 17 37 22.24 & -56 36 30.0 & 185 &  208 &   8 & $<$9.2 & Hotspot \\ 
             & 2 & J173742-563246 & 17 37 42.95 & -56 32 46.4 &  97 &  488 &  19 &    5.9 & Hotspot \\
\hline
\multicolumn{10}{l}{(b) Galaxies associated with a single catalogued AT20G source which is offset by more than 10\,arcsec from the nucleus  } \\
PKS\,0000-550 &  & J000311-544516 & 00 03 11.04 & -54 45 16.8 &  19 &   95 &   3 &$<$17.0 & Resolved double \\
PKS\,0349-27  &  & J035145-274311 & 03 51 45.09 & -27 43 11.4 & 149 &  122 &   8 &   24.8 & Hotspot \\
PKS\,0625-53  &  & J062620-534151 & 06 26 20.58 & -53 41 51.4 &  16 &  253 &   4 & $<$7.8 & Resolved double \\
PKS\,0625-545 &  & J062648-543214 & 06 26 48.91 & -54 32 14.0 &  21 &  106 &   3 &$<$12.9 & Resolved triple \\     
PKS\,0651-60  &  & J065153-602158 & 06 51 53.67 & -60 21 58.4 &  21 &   44 &   2 &   23.0 & Resolved double \\
PKS\,0806-10  &  & J080852-102831 & 08 08 52.49 & -10 28 31.9 &  55 &  131 &   5 & $<$7.9 & Hotspot \\ 
Hydra A       &  & J091805-120532 & 09 18 05.82 & -12 05 32.5 &  12 & 1056 &  52 &   13.6 & Resolved double \\
MRC\,0938-118 &  & J094110-120450 & 09 41 10.74 & -12 04 50.6 &  47 &   44 &   2 &     $<$18.6 & Resolved triple \\
PKS\,1251-12  &  & J125438-123255 & 12 54 38.55 & -12 32 55.7 &  68 &   83 &   2 &        23.4 & Resolved double \\
PKS\,1801-702 &  & J180712-701234 & 18 07 12.55 & -70 12 34.5 &  12 &   62 &   3 &     $<$13.5 & Resolved double \\ 
PKS\,1954-55  &  & J195817-550923 & 19 58 17.08 & -55 09 23.3 &  14 &  581 &  12 &        43.5 & Resolved double \\  
PKS\,2053-20  &  & J205603-195646 & 20 56 03.52 & -19 56 46.8 &  16 &  159 &   3 &     $<$14.4 & Resolved double \\   
PKS\,2135-147 &  & J213741-143241 & 21 37 41.17 & -14 32 41.9 &  60 &  256 &   7 &         4.5 & Hotspot \\ 
PKS\,2158-380 &  & J220113-374654 & 22 01 13.79 & -37 46 54.3 &  50 &  174 &   5 &     $<$18.3 & Hotspot \\
\hline
\end{tabular}
\flushleft
\end{table*}

In most cases there was good agreement between the AT20G and 6dFGS positions. 
About 20\% of the candidate matches showed extended or double structure in the 
AT20G image, and in a few cases there was more than one catalogued AT20G source near 
the same 6dFGS galaxy, suggesting that the AT20G catalogue may be listing several 
discrete components of a single radio source (see \S2.2). 

Our next step was to accept as genuine IDs the 183 candidate matches with a position 
difference of less than 10\,arcsec.  Monte Carlo tests imply that all matches out to this 
radius are likely to be genuine, with less than one match expected to occur by chance.  
Of these 183 sources, 24 (13\%) were flagged as extended in the AT20G catalogue and 
the remainder were unresolved on scales of 10-15\,arcsec at 20\,GHz. 

\subsection{Sources with large radio-optical offsets} 

On the basis of our visual inspection (and cross-comparison with low-frequency radio images 
where necessary), we accepted a further 19 galaxies with AT20G-6dFGS position offsets 
$>10$\, arcsec as genuine IDs.  These are listed in Table\,\ref{tab_mult}. 

All but one of the 19 galaxies in Table\,\ref{tab_mult} are associated with sources flagged as 
extended in the AT20G catalogue (the exception is J031545-274311, which is an 
unresolved hotspot in one lobe of the radio galaxy PKS\,0349-27).  Five of the 6dFGS galaxies 
in Table\,\ref{tab_mult} are associated with two or more sources which are listed 
separately in the AT20G catalogue. 

Most of the objects in Table\,\ref{tab_mult} (11/19) are nearby FR-2 radio galaxies, 
where the catalogued AT20G source corresponds to one of the hotspots and so is 
offset from the optical position of the host galaxy.  The remaining sources 
have extended radio emission in which the brightest region at 20\,GHz is slightly 
offset from the galaxy nucleus. 
Adding the 19 galaxies from Table\,\ref{tab_mult} to the 183 galaxies identified 
the matching process described in \S2.1 gives a total of 202 galaxies 
in our final AT20G-6dFGS sample.

\subsection{Optical spectroscopy and spectral classification} 

Of the 202 galaxies in our final sample, 139 have a good-quality 6dFGS optical 
spectrum.  The remaining 62 galaxies were not observed in the 6dFGS survey 
because a published redshift was already available in the literature and so 
they were given lower priority when scheduling optical spectroscopy for the 
6dFGS. 

For each galaxy where a 6dFGS spectrum was available, we visually
classified the optical spectrum in the same way as Mauch \& Sadler (2007) 
to determine the dominant physical process responsible for the radio emission.   

Each object is first classified as either a star-forming galaxy (if the spectrum 
shows emission lines with ratios characteristic of star-formation regions) or 
an active galactic nucleus (AGN) if no evidence of star formation is seen. 
The AGN class is then further sub-classified into objects which have strong, 
weak or no emission lines in their spectra. Table\,\ref{tab_type} lists the 
different classifications used and the number of objects in each class. 

\begin{table} 
\caption{Spectral classes visually assigned to the 6dFGS-AT20G objects, 
as described in \S2.2.  
As discussed later in this paper (\S5.3), objects classified as AGN with strong emission 
lines (classes `Ae' and `AeB') will generally have an [OIII] emission-line equivalent 
width of at least 5\,\AA. }
\label{tab_spec}
\begin{tabular}{rlrr} 
\hline
Class & Type of spectrum & 6dF  & 6dF  \\
      &                  & only & +2Jy \\
\hline
Aa   & AGN, pure absorption-line spectrum &  67 & 72 \\ 
Aae  & AGN, weak narrow emission lines   &  40 & 51 \\ 
Ae   & AGN, strong narrow emission lines  & 25  & 27 \\  
AeB  & AGN, strong emission with broad Balmer lines & 7  & 9 \\ 
SF   & HII region-like emission spectrum & 0  & 1 \\ 
..   & No spectral class, redshift from literature &  63 &  42 \\ 
&& \\
Total  &   & 202 & 202\\
\hline
\end{tabular}
\label{tab_type}
\end{table} 

As noted by Mauch \& Sadler (2007), the 6dFGS spectra are obtained through 
6\,arcsec fibres which correspond to a projected diameter of about 6.8\,kpc 
at the median redshift of the survey ($z\sim0.05$).  As a result, the 6dF fibres 
include an increasing fraction of the total galaxy light for higher-redshift 
galaxies and so galaxies with weak emission lines in their nuclei may be 
harder to recognize at higher redshift.  

Less information is available for the 63 galaxies without 6dFGS spectra, 
but 21 of these galaxies are members of the 2\,Jy radio galaxy sample 
(Morganti et al.\ 1993) for which optical spectra at 3500--5500\,\AA\ have 
been published by Tadhunter et al.\ (1993; see their Figure 1) along 
with detailed notes on the spectral features. For these 21 objects, we 
made a spectral classification using the Tadhunter et al.\ (1993) spectra.  
These spectral classifications are marked with quality flag `J' in Table\,3.  
For objects with spectra from both 6dFGS and Tadhunter et al.\ (1993), 
the two classifications agree well. 
For the remaining objects where no 6dFGS spectrum was available, no spectral 
classification was generally made. In these cases, the catalogued 6dFGS redshift 
is included in Table\,3 (see below) but the quality flag and spectral class 
are left blank. 

\subsection{Origin of the radio emission}
As can be seen from Table\,\ref{tab_type}, 159 of the 160 20\,GHz-selected galaxies 
with good-quality optical spectra are classified as AGN and only one (NGC\,253) 
as a star-forming galaxy.  

This is quite different from the 1.4\,GHz-selected NVSS-6dFGS sample 
(Mauch \& Sadler 2007), which contains roughly 60\% star-forming galaxies 
and 40\% AGN. 
The difference is not simply due to the higher flux limit of the AT20G sample 
(40\, mJy, compared to 2.4\,mJy for NVSS), since at least 15\% of the Mauch \& Sadler 
(2007) sources stronger than 40\,mJy at 1.4\,GHz are star-forming galaxies, but  
also reflects differences in the radio spectral index distribution of AGN and 
star-forming galaxies over the frequency range 1--20\,GHz. 

Murphy et al.\ (2010) note that the AT20G survey is 
insensitive to extended 20\,GHz emission on angular scales larger 
than about 45\,arcsec, making it difficult to detect diffuse 
synchrotron emission from the disks of nearby spiral galaxies. 
In practice, however, the relatively low radio luminosity and 
steeply-falling radio spectrum of the disk emission from `normal' 
galaxies means that we would expect to detect very few such 
objects above the 40\,mJy AT20G survey limit even if the brightness 
sensitivity was not an issue. 

NGC\,253, the lowest-redshift galaxy in our sample, is the only galaxy in which 
the 20\,GHz radio emission appears to arises from a central starburst rather 
than an AGN.  The main supporting evidence for this is the lack of a parsec-scale 
central radio source with the high brightness temperature characteristic of AGN cores 
(Sadler et al.\ 1995; Lenc \& Tingay 2006).  NGC\,253 is also one of only two starburst 
galaxies so far detected as high-energy gamma-ray sources by the Fermi satellite 
(Abdo et al.\ 2010).

\subsection{The main data table}

Table\,3 lists radio and optical measurements for the 202 galaxies in the final AT20G-6dFGS 
sample.  Galaxies with unresolved 20\,GHz sources are listed first, followed by galaxies 
where the 20\,GHz radio source is flagged as extended in the AT20G catalogue and/or 
has multiple components. 

The column headings are as follows: 
\begin{itemize}
\item[(1)] Source name from the AT20G catalogue.  For galaxies which are identified with 
two or more AT20G sources (see Table\,\ref{tab_mult}), we list a commonly-used source name instead. 
\item[(2)] The 20\,GHz radio position (J2000) as catalogued by Murphy et al\, (2010). 
\item[(3,4)] The catalogued 20\,GHz flux density, and its error (Murphy et al.\ 2010). For sources 
with multiple AT20G components, we list the sum of the component flux densities. 
\item[(5,6)] Where available, the catalogued 8.4\,GHz flux density and its error (Murphy et al.\ 2010).  
\item[(7,8)] Where available, the catalogued 5\,GHz flux density and its error (Murphy et al.\ 2010). 
\item[(9)] For sources north of declination $-40^\circ$, the total 1.4\,GHz flux density measured 
from the NVSS catalogue (Condon et al.\ 1998).  For sources with more than one NVSS component 
the listed flux is the sum of the components, as described by Mauch \& Sadler (2007). 
\item[(10)] For sources south of declination $-30^\circ$, the total 843\,MHz flux density measured 
from the SUMSS catalogue (Mauch et al.\ 2003).  For sources with more than one SUMSS component 
the listed flux is the sum of the components. 
\item[(11)] 6dFGS name. 
\item[(12)] Offset between the AT20G and 6dFGS positions, in arcsec. 
\item[(13)] Total infrared K-band magnitude K$_{\rm tot}$ from the 2MASS extended source catalogue 
(Jarrett et al.\ 2000), as listed in the 6dFGS database. 
\item[(14)] Optical redshift, as listed in the 6dFGS catalogue (Jones et al.\ 2009). 
\item[(15)] 6dFGS redshift quality, q, where q=4 represents a reliable redshift and q=3 a probable redshift 
(Jones et al.\ 2004). 
\item[(16)] Spectral classification for galaxies with a good-quality 6dFGS spectrum. 
Aa = absorption-line spectrum, Aae = absorption lines plus weak emission lines, 
Ae = strong emission lines (see Sadler et al. 1999,\ 2002). 
\item[(17)] Notes on individual sources.  
\end{itemize}

The final AT20G-6dFGS sample contains 202 galaxies, 42 of which are flagged in the 
AT20G catalogue as extended radio sources at 20\,GHz. 


\begin{figure}
\begin{center}
%
%
\includegraphics[width=8.5cm]{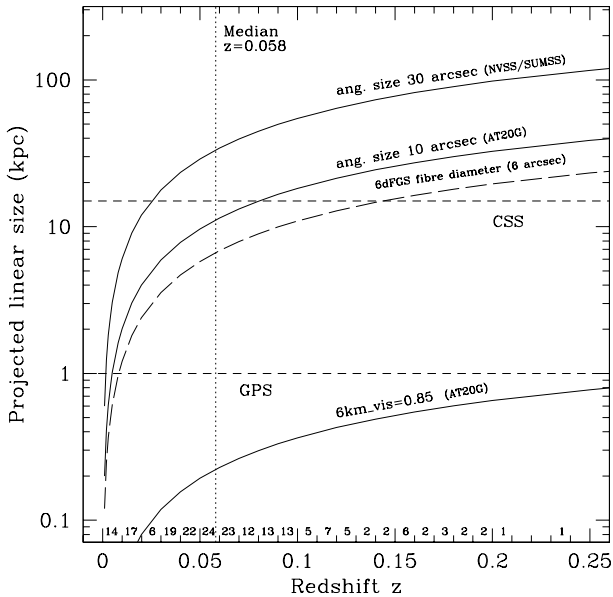}
\caption{Summary of the angular and linear scales probed by the radio 
data available for the AT20G-6dFGS galaxies. 
Solid lines show the approximate resolution limits of the low-frequency 
NVSS/SUMSS images, the 20\,GHz AT20G snapshot images and the ATCA 6\,km 
baseline at 20\,GHz (see Chhetri et al.\ 2012).  The angular size 
of the 6dFGS fibres used for optical spectroscopy is also shown 
for comparison. Horizontal lines show indicative sizes for the 
two main classes of `young' radio galaxies, the Compact Steep Spectrum 
(CSS) sources, and Gigahertz Peaked Spectrum (GPS) objects (O'Dea 1998). 
The vertical dotted line marks the median redshift of galaxies in our 
sample, and the number of galaxies in each individual $\Delta z=0.01$ 
redshift bin is shown at the bottom of the plot. 
\label{fig_ang1}}
\end{center}
\end{figure}

\section{Radio properties of the AT20G--6dFGS sample} 

We now consider the radio morphology and spectral-index distribution 
of the AT20G-6dFGS galaxies. 

The available radio data allow us to probe a range of size scales, 
as summarized in Figure\,\ref{fig_ang1}. 
At frequencies near 1\,GHz, the NVSS and SUMSS images can resolve 
extended radio structure on scales larger than about 30\,arcsec. 
In the redshift range covered by our sample, this corresponds to 
radio emission on scales of tens to hundreds of kpc, 
typical of classical FR-1 and FR-2 radio galaxies (Fanaroff \& Riley 1974). 
Around 25\% of the galaxies in Table\,3 have extended low-frequency 
radio emission in the NVSS/SUMSS images, as discussed below in \S3.1. 

At high frequency, the AT20G snapshot images have an angular resolution 
of 10--15\,arcsec (Murphy et al.\ 2010), corresponding to a projected 
linear size of 10--15\,kpc for galaxies at $z\sim0.06$ (the median redshift 
of galaxies in our sample).  In most cases, therefore, a source which is 
unresolved in the 20\,GHz images is confined within its host galaxy.  
This size scale is characteristic of Compact Steep Spectrum (CSS) 
radio sources, which are usually smaller than 15\,kpc in extent 
(Fanti et al.\ 1990).   

Measurements of the visibilities on the longest ($\sim$6\,km) ATCA baselines  
at 20\,GHz allow us to identify sources which are smaller than $\sim$0.2\,arcsec 
in size (Massardi et al.\ 2011a; Chhetri et al.\ 2012) and so confined to 
the central kiloparsec of their host galaxy. 
This size scale is characteristic of Gigahertz Peaked Spectrum (GPS) 
galaxies, which are generally smaller than a few hundred parsecs in size 
(Stanghellini et al.\ 1997). 
The high-frequency radio structure of the AT20G-6dFGS galaxies is discussed 
further in \S3.2. 

Finally, we note that the 6dFGS spectra are taken with 6-arcsec diameter 
optical fibres (Jones et al.\ 2009), and so sample a large fraction 
of the galaxy light for all but the closest AT20G-6dFGS objects. 
In particular, optical emission lines seen in the 6dFGS spectra 
could arise either from the nucleus or from ionized gas distributed 
more widely within the galaxy. 

\subsection{Radio morphology at frequencies near 1\,GHz} 
All the galaxies in AT20G-6dFGS sample have low-frequency radio 
images available from the NVSS (1.4\,GHz) or SUMSS (843\,MHz) surveys. 
NVSS and SUMSS are well-matched in sensitivity, and both surveys have 
similar angular resolution of around 45\,arcsec (i.e. about a factor 
of three lower than the AT20G 20\,GHz images). NVSS covers the sky north 
of declination $-40^\circ$, and SUMSS the region south of declination 
$-30^\circ$, so there is a 10-degree band at declination $-30<\delta<-40$ 
where images are available form both these surveys. 
The main motivation for studying these low-frequency images was to link 
the high-frequency core properties of nearby radio galaxies to the 
FR-1 and FR-2 classifications traditionally used in lower-frequency 
studies (Fanaroff \& Riley 1974).  

Of the 201 radio galaxies in our sample\footnote{Excluding the star-forming galaxy 
NGC\,253}, 65 (32\%) have extended radio emission (on scales larger than about 
30\,arcsec) in the 1.4\,GHz NVSS or 843\,MHz SUMSS images.  These 65 galaxies 
are listed in Table\,\ref{tab_fr}\ along with a classification as either FR-1 or FR-2. 
For 36 of the 65 objects we used existing FR classifications 
from the literature, mainly from the southern 2-Jy sample (Morganti, Killeen 
\& Tadhunter 1993) and the Molonglo Southern 4-Jy sample (Burgess \& Hunstead 2006). 
The other 29 extended sources had no published FR classification, and were classified by the 
authors as either FR-1 or FR-2 using the NVSS and SUMSS images.  The final 
column of Table\,\ref{tab_fr}\ indicates whether the galaxy is known to be a member of 
an Abell cluster, and in some cases gives additional information about the low-frequency 
radio structure. 

We classify the remaining 136 galaxies in our sample, for which the 1\,GHz 
emission is unresolved in the NVSS and SUMSS images as {\it `FR-0' radio galaxies}. 
The FR-0 classification was introduced by Ghisellini et al.\ (2011) as 
the name of a class of weak, compact radio sources described by Baldi 
\& Capetti (2009) but dating back to Slee et al.\ (1994) and possibly 
even earlier. 
In this paper, we adopt the FR-0 designation as a convenient 
way of linking the compact radio sources seen in nearby galaxies into 
the canonical Fanaroff-Riley classification scheme. 

Our final set of low-frequency classifications contains 49 FR-1, 16 FR-2 
and 136 FR-0 radio galaxies.  Thus only about one-third of the nearby 
radio AGN in our 20\,GHz-selected sample are classical FR-1/FR-2 radio 
galaxies at frequencies near 1\,GHz. 

\begin{table*}
\centering
\caption{Galaxies with extended (FR-1 or FR-2) radio emission at 1\,GHz.
The listed 408\,MHz flux density is from the Molonglo Reference Catalogue (MRC; Large et al.\ 1981).  
\label{tab_fr}} 
\resizebox{16.8cm}{!}{
\begin{tabular}{@{}llcllllrlll@{}}
\hline
Source & \multicolumn{1}{c}{AT20G} & \multicolumn{1}{c}{2\,Jy} & \multicolumn{1}{c}{MS4} & 
\multicolumn{1}{c}{$z$} & Spec.  & \multicolumn{1}{c}{S$_{20}$} & 
\multicolumn{1}{c}{S$_{0.4}$} & \multicolumn{1}{c}{FR} & Ref. & Notes\\
 & \multicolumn{1}{c}{name} & \multicolumn{1}{c}{name} & \multicolumn{1}{c}{name} &  
 & \multicolumn{1}{c}{class} & \multicolumn{1}{c}{(Jy)} & 
 \multicolumn{1}{c}{(Jy)} &  \multicolumn{1}{c}{class}  \\
\hline
PKS\,0001-531 & J000413-525458   & ...     & ...       & 0.0328 &  ..  & 0.065 &  1.25  & 1  & v &  \\ 
3C\,015       & J003704-010907   & 0034-01 & ...       & 0.0734 &  Aa  & 0.404 &  9.74  & 2  & a &  \\ 
PKS\,0043-42  & (Two sources)    & 0043-42 & B0043-424 & 0.1160 &  Aae & 0.438 & 21.0   & 2  & a, b &  \\ 
3C\,029       & J005734-012328   & 0055-01 & ...       & 0.0450 &  Aa  & 0.055 & 10.88  & 1  & a &  \\ 
NGC\,547      & J012600-012041   & 0123-01 & ...       & 0.0184 &  Aa  & 0.147 & 16.4   & 1  & a & In cluster Abell\,194  \\ 
NGC\,612      & J013357-362935   & 0131-36 & B0131-367 & 0.0305 &  Ae  & 0.440 & 17.1   & 2  & a,b &  \\ 
ESO\,198-G01  & J021645-474908   & ...     & B0214-480 & 0.0643 &  Aa  & 0.093 &  9.5   & 1  & b &  In cluster Abell\,239S\\ 
PKS\,0229-208 & J023137-204021   & ...     & ...       & 0.0898 &  Aa  & 0.160 &  1.87  & 1  & v &  \\  
PMN\,J0315-1906 & J031552-190644 & ...     & ...       & 0.0671 &  Ae? & 0.108 &  ..    & 1  & c &  In cluster Abell\,428 \\
PKS\,0344-34  & J034630-342246   & ...     & B0344-345 & 0.0535 &  ..  & 0.102 &  9.3   & 2  & b & Complex source, in cluster \\ 
PKS\,0349-27  & J035145-274311   & 0349-27 & ...       & 0.0657 &  Ae  &(0.122)&  8.75  & 2  & a & Lower limit at 20\,GHz \\ 
IC\,2082      & J042908-534940   & 0427-53 & B0427-539 & 0.0380 &  Aa  & 0.145 & 14.6   & 1  & a,b & In cluster Abell 463S \\ 
PKS\,0429-61  & J043022-613201   & ...     & B0429-616 & 0.0555 &  Aa  & 0.148 &  6.5   & 1  & b & In cluster Abell 3266 \\ 
Pictor\,A     & (Three sources)  & 0518-45 & B0518-458 & 0.0351 &  AeB & 6.320 & 166.0  & 2  & a, b &  \\ 
PKS\,0545-199 & J054754-195805   & ...     & ...       & 0.0551 &  Aa  & 0.042 &  1.81  & 1  & d &  \\ 
PKS\,0620-52  & J062143-524132   & 0620-52 & B0620-52  & 0.0511 &  Aae & 0.266 &  9.3   & 1  & a, b &  \\ 
ESO\,161-IG07 & J062620-534151   & 0625-53 & B0625-536 & 0.0551 &  Aa  & 0.253 & 26.0   & 1  & a, b &  In cluster Abell\,3391  \\ 
PKS\,0625-545 & J062648-543214   & ...     & B0625-545 & 0.0517 &  ..  & 0.106 &  7.86  & 1  & b &  In cluster Abell\,3395 \\ 
PKS\,0625-35  & J062706-352916   & 0625-35 & B0625-354 & 0.0549 &  Aa  & 0.688 &  9.23  & 1  & a, b & In cluster Abell\,3392 \\ 
PKS\,0634-20  & (Two sources)    & ..      & ..        & 0.0551 &  Ae  & 0.238 & 210.0  & 2  & e &  \\ 
PKS\,0651-60  & J065153-602158   & ...     & ...       & 0.1339 &  Aa  & 0.044 &  3.06  & 1  & v &  \\ 
PKS\,0652-417 & J065359-415144   & ...     & ...       & 0.0908 &  Aa  & 0.056 &  1.02  & 1  & v &  \\  
ESO\,207-G19  & J070459-490459   & ...     & ...       & 0.0419 &  ..  & 0.084 &  ..    & 1  & v & In cluster Abell\,3407 \\ 
PKS\,0707-35  & J070914-360121   & ...     & B0707-35  & 0.1108 &  ..  & 0.094 &  4.60  & 2  & b &  \\ 
PKS\,0803-00  & J080537-005819   & ...     & ...       & 0.0902 &  Aa  & 0.081 &  3.39  & 1  & v & Wide-angle tail, in cluster Abell\,623  \\ 
PKS\,0806-10  & J080852-102832   & 0806-10 & ...       & 0.1090 &  Ae  & 0.131 & 10.2   & 2  & a &  \\ 
ESO\,060-IG02 & J081611-703944   & ...     & ...       & 0.0332 &  Aa  & 0.062 &  2.56  & 1  & v & Complex source\\ 
PMN\,J0844-1001 & J084452-100059 & ...     & ...       & 0.0429 &  Aa  & 0.046 &  ..    & 1  & v &  \\
PMN\,J0908-1000 & J090802-095937 & ...     & ...       & 0.0535 &  Aa  & 0.060 &  ..    & 1  & v & Narrow-angle tail \\ 
PMN\,J0908-0933 & J090825-093332 & ...     & ...       & 0.1590 &  Aa  & 0.046 &  ..    & 1  & v & In cluster Abell 0754 \\ 
Hydra\,A      & J091805-120532   & 0915-11 & ...       & 0.0548 &  Aae & 1.056 & 132.0  & 1  & a &  In cluster Abell\,780 \\ 
PMN\,J0941-1205 & J094110-120450 & ...     & ...       & 0.1500 &  Aa  & 0.044 &  1.77  & 2  & v & Compact triple \\
NGC\,3557     & J110957-373220   & ...     & ...       & 0.0101 &  Aa  & 0.052 &  0.96  & 1  & f &  \\
PKS\,1118+000 & J112119-001316   & ...     & ...       & 0.0993 &  ..  & 0.108 &  1.50  & 1  & v & Wide-angle tail \\    
PKS\,1130-037 & J113305-040046   & ...     & ...       & 0.0520 &  Aa  & 0.108 &  1.18  & 1  & v & In cluster Abell\,1308 \\  
NGC\,4783     & J125438-123255   & 1251-12 & ...       & 0.0150 &  Aae & 0.083 & 14.7   & 1  & a &  \\ 
ESO\,443-G24  & J130100-322628   & ...     & ...       & 0.0170 &  ..  & 0.176 &  2.99  & 1  & g &  In cluster Abell\,3537 \\
PKS\,1308-441 & J131124-442240   & ...     & ...       & 0.0506 &  Aae & 0.044 &  1.07  & 1  & v & Giant radio galaxy \\ 
NGC\,5090     & J132112-434216   & 1318-43 & B1318-434 & 0.0112 &  Aae & 0.705 & 10.0   & 1  & a, b &  \\ 
NGC\,5128     & J132527-430104   & 1322-42 & B1322-427 & 0.0018 &  Aae & 28.35 & 2740.  & 1  & a, b &  \\ 
IC\,4296      & J133639-335756   & 1333-33 & B1333-33  & 0.0125 &  Aae & 0.323 & 30.8   & 1  & a, b & In cluster Abell\,3565 \\ 
ESO\,325-G16  & J134624-375816   & ...     & ...       & 0.0376 &  Aa  & 0.071 &  1.44  & 1  & v & Complex structure, in cluster Abell\,3570 \\ 
PKS\,1452-367 & J145509-365508   & ...     & ...       & 0.0946 &  Aae & 0.198 &  2.90  & 1  & v &  \\
PKS\,1637-77  & J164416-771548   & 1637-77 & B1637-771 & 0.0430 &  Aae & 0.399 & 13.5   & 2 & a, b &  \\ 
PKS\,1717-00  & (Two sources)    & 1717-00 & ...       & 0.0304 &  Aae & 0.435 & 61.28  & 2 & a &  \\ 
PKS\,1733-56  & (Two sources)    & 1733-56 & B1733-565 & 0.0985 &  Ae  & 0.696 & 20.3   & 2 & a, b &  \\ 
MRC\,1758-473 & J180207-471930   & ...     & ...       & 0.1227 &  Aa  & 0.114 &  1.06  & 1  & v &  \\ 
PMN\,J1818-5508 & J181857-550815 & ...     & ...       & 0.0726 &  Aa  & 0.075 &  ..    & 1  & v &   \\  
PKS\,1839-48  & J184314-483622   & 1839-48 & B1839-487 & 0.1108 &  Aa  & 0.305 &  9.08  & 1  & a, b &  \\ 
PMN\,J1914-2552 & J191457-255202 & ...     & ...       & 0.0631 &  Aa  & 0.147 &  1.06  & 1  & v & Head-tail source \\ 
MRC\,1925-296   & J192817-293145 & ...     & ...       & 0.0244 &  Aa  & 0.078 &  3.18  & 1  & v & Wide-angle tail, in cluster? \\ 
PKS 1954-55   & J195817-550923   & 1954-55 & B1954-552 & 0.0581 &  Aae & 0.581 & 14.8   & 1  & a, b &  \\ 
ESO\,234-G68  & J204552-510627   & ...     & ...       & 0.0485 &  Aa  & 0.054 &  0.96  & 1  & v &  \\ 
IC\,1335      & J205306-162007   & ...     & ...       & 0.0427 &  Aa  & 0.056 &  ..    & 1  & v &  \\
ESO\,106-IG15 & J205754-662919   & ...     & ...       & 0.0754 &  Aa  & 0.049 &  ..    & 1  & v & In cluster Abell\,2597? \\
PMN\,J2122-5600 & J212222-560014 & ...     & ...       & 0.0518 &  Aae & 0.058 &  ..    & 1  & v & Head-tail source?  \\ 
NGC\,7075     & J213133-383703   & ...     & ...       & 0.0182 &  Aa  & 0.046 &  1.79  & 1  & v &  \\
PKS\,2135-147 & J213741-143241   & 2135-14 & B2135-147 & 0.1999 &  AeB & 0.256 &  8.78  & 2 & a, b&  \\
PMN\,J2148-5714 & J214824-571351 & ...     & ...       & 0.0806 &  ..  & 0.048 &  1.67  & 1  & v & Head-tail source, in cluster? \\ 
PKS\,2148-555 & J215129-552013   & ...     & B2148-555 & 0.0388 &  ..  & 0.088 &  ..    & 1  & b & In cluster Abell\,3816 \\ 
ESO\,075-G041 & J215706-694123   & 2152-69 & B2152-699 & 0.0285 &  AeB & 3.400 & 61.6   & 2 & a, b &  \\ 
PKS\,2158-30  & J220113-374654   & ...     & B2158-380 & 0.0334 &  Ae  & 0.174 &  4.12  & 2 & b &  \\ 
PKS\,2316-423 & J231905-420648   & ...     & ...       & 0.0543 &  Aae & 0.150 &  ..    & 1  & v & Complex structure, in cluster Abell\,1111S \\
PKS\,2316-538 & J231915-533159   & ...     & ...       & 0.0958 &  Aa  & 0.075 &  ..    & 1  & v &  \\
PKS\,2339-164 & J234205-160840   & ...     & ...       & 0.0649 &  Aa  & 0.043 &  0.95  & 1  & v & Complex structure, in cluster? \\ 
\hline
\end{tabular}
}
\flushleft
References for FR\,1/2 classification: (a) 2Jy, Morganti et al.\ 1993; (b) MS4, Burgess \& Hunstead 2006; 
(c) Ledlow \& Owen 1996; (d) Zirbel \& Baum 1995; (e) Baum et al.\ 1988; (f) Birkinshaw \& Davies 1985; 
(g) Marshall et al.\ 2005; (v) visual classification by the authors, based on NVSS/SUMSS images. 
\end{table*}

\subsection{Radio morphology at 20\,GHz} 

The AT20G data provide a range of information on the radio morphology 
at 20\,GHz, as discussed by Murphy et al.\ (2010) and Massardi et al.\ (2011a). 
The angular resolution of the 20\,GHz AT20G snapshot images made with the 
ATCA  is typically $\sim10$\,arcsec (Murphy et al.\ 2010), corresponding  
to a linear size of $\sim10$\,kpc at the median redshift ($z=0.058$) of the 
AT20G-6dFGS galaxies. The 20\,GHz snapshot images therefore provide information 
on the high-frequency radio morphology of the AT20G-6dFGS galaxies on kiloparsec scales. 

We can recover some extra information on the smaller-scale 
structure of most AT20G sources by examining the 20\,GHz visibilities on 
baselines to the fixed ATCA `6\,km' antenna, which were not used 
for imaging (Murphy et al.\ 2010).  
These `6\,km visibility' measurements are available for 186  
of the AT20G-6dFGS galaxies (92\%), and provide information about 
the source compactness on sub-kpc scales as discussed in \S3.2.2. 

\subsubsection{20\,GHz morphology on kpc scales, from the AT20G snapshot images}

Murphy et al.\ (2010) flagged 337 of the 5890 sources in the AT20G catalogue 
(5.7\%) as extended (i.e. generally larger than about 10\,arcsec in size) 
at 20\,GHz.
If we remove the 82 objects flagged by Murphy et al.\ (2010) as Galactic 
or LMC sources, then the fraction of extended sources falls slightly, to 
308 out of 5806 or 5.3\%. 

The fraction of AT20G-6dFGS galaxies with extended 20\,GHz emission 
(42/202 or 20.8\%) is significantly higher than the 5\% fraction  
seen for all extragalactic sources in the AT20G survey (Massardi 
et al.\ 2011a).  This is not too surprising, since the AT20G-6dFGS 
galaxies are generally the lowest-redshift objects in the AT20G 
catalogue. 
The extended AT20G-6dFGS sources show a variety of radio morphologies, 
as summarized in Table\,\ref{tab_morph}.  The classifications in this 
table are based on visual inspection of the AT20G snapshot images.  
The radio galaxy NGC\,612 (J013357-010907), which has very extended 
20\,GHz radio emission imaged by Burke-Spolaor et al.\ (2009), 
is classifed here as a `wide triple'. 

Sources which are unresolved in the AT20G images will generally be smaller 
in extent than the galaxies which host them, but may still be several kpc 
in size.  Classical GPS and CSS radio sources, with typical sizes of $<1$\,kpc 
and $<15$\,kpc respectively (O'Dea 1998), are expected to be unresolved 
or marginally resolved sources in the AT20G images. 

For the great majority of AT20G-6dFGS galaxies (77\%), the high-frequency 
radio emission arises from an unresolved point source centred on the galaxy 
nucleus (i.e. a `radio core').  A further 11\% have an extended AT20G source 
centred on the galaxy nucleus. 
Only 12\% of the detected galaxies have 20\,GHz emission which is 
significantly offset from the galaxy nucleus or resolved into two 
or more components. 

\begin{table} 
\caption{20\,GHz radio morphology of the AT20G-6dFGS galaxy sample.  }
\label{tab_morph}
\begin{tabular}{rlr} 
\hline
\multicolumn{1}{c}{Description} & Class & Number \\
\hline
Single source, centred on galaxy nucleus &  Core         &  178 \\
\hspace*{0.5cm} (compact at 20\,GHz)     &               &  (159)   \\
\hspace*{0.5cm} (extended at 20\,GHz)    &               &  (19)   \\
Single source, offset from nucleus       &  Hotspot      &    4 \\
Two sources within a single ATCA beam    &  Compact double & 11 \\
Two separate AT20G detections            &  Wide double  &    4 \\
Three sources within a single ATCA beam    &  Compact triple &  3 \\
Three separate AT20G detections            &  Wide triple    &   2\\
Total &  & 202 \\
\hline
\end{tabular}
\end{table} 

\subsubsection{20\,GHz morphology on sub-kpc scales, from ATCA 6\,km visibilities}

Chhetri et al.\ (2012) introduced a second measure of compactness for the AT20G sources, 
based on a visibility determined by the ratio of the average scalar amplitude of the 
five long ($\sim4.5$\,km) ATCA baselines 
to the ten short (30--214\,m) baselines in the H214 configuration. Sources with a 
flux density ratio $>0.85$ were considered to be unresolved, and therefore 
have angular sizes smaller than about 0.2\,arcsec.  Measurements of the 
compactness parameter are available for 92\% of the AT20G-6dFGS galaxies 
(Chhetri et al., in preparation). 

\begin{figure}
%
%
\begin{center}
\includegraphics[width=8cm]{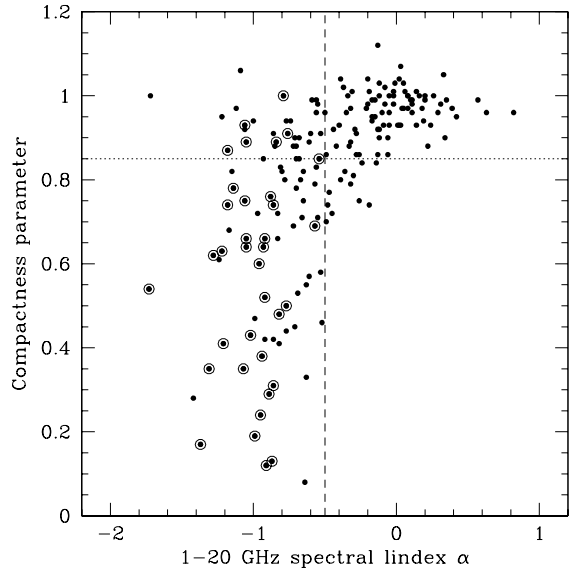}
\caption{Plots of the compactness parameter $R$ (defined as the ratio of the 20\,GHz 
flux densities measured on long and short baselines, as discussed in \S3.2.2) 
versus the 1-20\,GHz spectral index for AT20G-6dFGS galaxies. The vertical dashed 
line at $\alpha=-0.5$ shows the division between `steep-spectrum' and `flat-spectrum' 
radio sources, and sources above the horizontal dotted line at $R=0.85$ are expected 
to have angular sizes smaller than about 0.2\,arcsec (i.e. smaller than about 220\,pc 
at the median redshift of $z=0.058$ for this galaxy sample).  
Open circles show sources which are flagged as extended (on scales larger than 
10-15\,arcsec) in the AT20G catalogue. 
\label{fig_rajan} }
\end{center}
\end{figure}

Figure\,\ref{fig_rajan} shows the 20\,GHz compactness parameter versus the 
1-20\,GHz spectral index for the AT20G-6dFGS galaxies. The clean separation 
between steep-spectrum extended sources and flat-spectrum compact sources 
seen seen by Massardi et al. (2011a; their Figure\ 4) is also visible here, 
and the fraction of sub-kpc scale 6dFGS-AT20G sources (i.e. those with 
compactness parameter $>0.85$) is 87\% (81/93) for flat spectrum sources, 
but only 35\% (32/92) for steep-spectrum sources.  We find no significant 
correlation between the compactness parameter and either redshift or 
20\,GHz radio luminosity. 

\subsection{20\,GHz flux density measurements for extended sources} 
The AT20G survey images are insensitive to 
extended 20\,GHz emission on scales larger than about 45\,arcsec (Murphy et al.\ 2010). 
As can be seen from Figure\,\ref{fig_ang1}, this may affect the measured 20\,GHz 
flux densities of sources which are either at very low redshift ($z<0.01$) or extend well beyond 
their host galaxy. 

\begin{figure}
\begin{center}
%
%
\includegraphics[width=8cm]{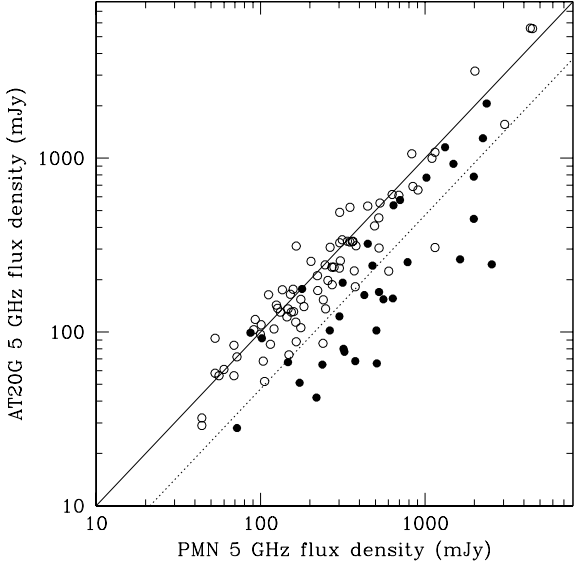}
\caption{Comparison of 5\,GHz flux density measurements from the AT20G 
catalogue (Murphy et al.\ 2010) and the PMN catalogue (Gregory et al.\ 1994). 
Filled circles show the FR-1 and FR-2 sources listed in Table\,\ref{tab_fr}, 
which have extended (angular size $>30$\,arcsec) low-frequency radio emission.  
Open circles represent galaxies which are unresolved in the low-frequency 
NVSS and SUMSS images. The solid diagonal line shows the relation 
S$_{\rm AT20G} =$ S$_{\rm PMN}$, and the dashed line S$_{\rm AT20G} = 
$ 0.47\,S$_{\rm PMN}$ as discussed in the text. 
\label{fig_pmn1}}
\end{center}
\end{figure}

Figure\,\ref{fig_pmn1} allows us to estimate the importance of this effect, 
by comparing the AT20G catalogue measurements at 5\,GHz with the Parkes-MIT-NRAO 
(PMN) catalogue (Gregory et al.\ 1994) which used the single-dish Parkes 
telescope (with a 4\,arcmin beam at 5\,GHz). The observations were taken 
more than a decade apart, so the flux density of individual sources may have 
varied, but the average value of the ratio S$_{\rm AT20G}$/S$_{\rm PMN}$ 
is 0.93$\pm$0.04 for the AT20G sources which are unresolved in both the 20\,GHz 
and NVSS/SUMSS images.  In this case, it seems plausible that the small departure 
from unity could be due to confusing sources within the Parkes beam, or small 
differences in the flux-density scale, rather than missing flux in the AT20G images. 

We therefore conclude that for the 70\% of AT20G-6dFGS sources which fall into 
the FR-0 class, the catalogued AT20G flux densities at 5, 8 and 20\,GHz 
represent an accurate measurement of the total radio flux density at these 
frequencies. 

For AT20G-6dFGS galaxies with extended low-frequency radio emission (i.e. the 
FR-1 and FR-2 radio galaxies listed in Table\,\ref{tab_fr}, which represent 
around 30\% of the AT20G--6dFGS sample), the ratio S$_{\rm AT20G}$/S$_{\rm PMN}$ 
is 0.47$\pm$0.05.  We therefore need to keep in mind that the listed AT20G flux 
densities for the FR-1 and FR-2 radio galaxies often reflect the high-frequency 
radio emission from the central core alone, rather than the core plus extended 
jets and lobes. 

Burke-Spolaor et al.\ (2009) made new 20\,GHz images of nine 
of the most extended sources in the AT20G sample, and their flux-density 
measurements were incorporated into the final AT20G catalogue (Murphy et al.\ 2010). 
Five of these sources (J013357-362935 (=NGC\,612), J133639-335756 (=IC\,4296). 
J215706-694123 (=ESO\,075-G41), Pictor A and Centaurus A) 
are also members of the AT20G-6dFGS sample. 
Additional radio observations, with better sensitivity to extended emission,  
would be valuable to measure the total high-frequency flux density accurately 
for the other AT20G-6dFGS sources which have extended 20\,GHz radio emission.  

\subsection{Radio spectral-index distribution}

The spectral-energy distribution of radio sources is commonly 
expressed in terms of a two--point spectral index $\alpha_a^b$ (where 
$S_\nu\propto\nu^\alpha$) between frequencies $\nu_a$ and $\nu_b$. 
Since many radio sources have curved rather than power-law continuum 
spectra, particularly at higher frequencies (Taylor et al.\ 2004; 
Sadler et al. 2006; Chhetri et al.\ 2012), it is important to keep in 
mind that the measured value of $\alpha$ may shift with observing 
frequency and/or redshift.  

Most radio-loud AGN have both a compact, flat-spectrum core and 
extended, steep-spectrum radio lobes, so the observed value of 
$\alpha$ reflects the relative dominance of compact (recent) versus 
extended (longer-term) radio emission. While most core-dominated 
radio sources have flat radio spectra, compact steep-spectrum 
(CSS) radio sources are also seen (Fanti et al. 1990; O'Dea 1998). 

At low frequencies, the radio AGN population is commonly divided into 
`steep-spectrum' ($\alpha\leq-0.5$) and `flat-spectrum' ($\alpha>-0.5$) 
sources (De Zotti et al.\ 2010; Chhetri et al.\ 2012).  
Steep-spectrum radio sources dominate in samples selected at frequencies 
near 1\,GHz.  For example, Mauch et al.\ (2003) measured a median 843--1400\,MHz 
spectral index of $-0.89$ for sources brighter than 50\,mJy at 843\,MHz. 
About 25\% of these sources were classified as flat-spectrum and 75\% as 
steep-spectrum.  

Samples selected at higher radio frequencies are known to contain more flat-spectrum 
sources (e.g De Zotti et al. 2010).  Sources brighter than 40\,mJy in the 20\,GHz--selected AT20G survey 
(Murphy et al.\ 2010) have a median 5--20\,GHz spectral index of $-0.28$, with 
69\% classified as flat-spectrum objects (Massardi et al.\ 2011a). 

\subsubsection{Near-simultaneous spectral indices at 5-20\,GHz} 
124 of the 202 galaxies in Table\,3 (61\%) have near-simultaneous radio flux 
density measurements at 5, 8 and 20\,GHz available from the AT20G catalogue 
(Murphy et al.\ 2010).  These multi-frequency data allow us to calculate 
high-frequency spectral indices between 5, 8 and 20\,GHz, $\alpha_5^{20}$, 
in the same way as Massardi et al.\ (2011a) have done for the full AT20G sample. 

As can be seen from Table\,\ref{alpha2}, the distribution of spectral behaviour 
for the AT20G-6dFGS galaxies is similar to that found by Massardi et al.\ (2011a) 
for the weaker (S$_{20}<100$\,mJy) 
sources in the full AT20G sample, with a roughly equal mix of flat and steep-spectrum 
objects and a much smaller fraction of peaked or upturning radio spectra. 

The main difference is that the local AT20G-6dFGS galaxies contain almost no 
`inverted-spectrum' sources with both $\alpha_5^8>0$ and $\alpha_8^{20}>0$.  
This is consistent with a picture in which the inverted-spectrum AT20G population 
(with a spectral peak above 20\,GHz) is dominated by flares from 
relativistically-beamed objects (blazars), as discussed by Bonaldi et al.\ (2013).  
We would not expect these beamed objects to be present in the AT20G-6dFGS sample 
of nearby, K-band selected galaxies. 

Bonaldi et al.\ (2013) estimate that the fraction of genuine `high-frequency peaker' 
(HFP) radio galaxies in the full AT20G sample is $<0.5$\ per cent, implying that we would 
expect to see no more than one such object in the 6dFGS-AT20G sample of 201 galaxies. 
In fact, our sample does contain one galaxy (AT20G\,J212222-560014) which appears 
to have a genuine HFP radio spectrum.  Hancock et al.\ (2010) note that J212222-560014 
has a radio spectrum peaking above 40\,GHz and shows no evidence for variability at 
20\,GHz, consistent with the behaviour expected for a very young GPS radio galaxy. 

The median spectral indices measured for the local AT20G-6dFGS galaxies, 
$\widetilde{\alpha}_5^8=-0.28\pm0.07$ and 
$\widetilde{\alpha}_8^{20}=-0.26\pm0.05$ are similar to those measured 
for the AT20G sample as a whole ($\widetilde{\alpha}_5^8=-0.16$ and 
$\widetilde{\alpha}_8^{20}=-0.28$; Massardi et al.\ 2011a), though it 
should be noted that (in contrast to the full AT20G sample) the median 
radio spectral index for the local AT20G-6dFGS galaxies is no steeper 
at 8-20\,GHz than at 5-8\,GHz. 

This lack of curvature in the median 5--20\,GHz radio spectrum for the 
AT20G-6dFGS galaxies is almost certainly due to their low redshift. 
Chhetri et al.\ (2012) show that the radio spectra of compact AT20G 
sources start to steepen above a rest frequency of about 30\,GHz.  
For the nearby galaxies in the AT20G-6dFGS sample, unlike the more distant 
sources in the AT20G catalogue as a whole, this high-frequency curvature 
has not been shifted into the 8-20\,GHz spectral range, so it is not 
surprising that the median 5--8 and 8--20\,GHz spectral indices are 
similar for the local AT20G-6dFGS sample. 

\begin{table*} 
\caption{High-frequency spectral-index classifications for the 
6dFGS-AT20G AGN, compared to the results for the full AT20G sample 
from Table\,2 of Massardi et al.\ (2011a). }
\begin{tabular}{llrrrrrr} 
\hline
Spectrum & Class &  \multicolumn{1}{c}{AT20G-6dFGS} & \multicolumn{1}{c}{Full AT20G}  & \multicolumn{1}{c}{Full AT20G, } \\
 &  &  \multicolumn{1}{c}{galaxies} & \multicolumn{1}{c}{sample} & \multicolumn{1}{c}{S$_{20}<100$\,mJy} \\
        &        & No. (per cent) & No. (per cent) & No. (per cent) \\
\hline
$-0.5<\alpha_5^8<+0.5$ {\bf and} & Flat (F$_{\rm high}$)   &  55 (44.7) & 1766 (53.0) & 694 (45.0) \\
$-0.5<\alpha_8^{20}<+0.5$ &  & \\
$\alpha_5^8<0$, $\alpha_8^{20}<0$ & Steep (S$_{\rm high}$) &  54 (43.9) & 1086 (32.6) & 619 (40.1) \\
$\alpha_5^8>0$, $\alpha_8^{20}>0$ & Inverted (I$_{\rm high}$) &  1 (0.8) & 195 (5.8)  & 66 (4.3) \\
$\alpha_5^8>0$, $\alpha_8^{20}<0$ & Peak (P$_{\rm high}$) &  6 (4.9) & 183 (5.5)  & 92 (5.9) \\
$\alpha_5^8<0$, $\alpha_8^{20}>0$ & Upturn (U$_{\rm high}$) &  7 (5.7) & 102 (3.1)  & 73 (4.7) \\
Any     &   & 123  (100)  & 3332  (100) & 1544 (100) \\
\hline
\end{tabular}
\label{alpha2}
\end{table*} 

\begin{figure}
%
%
\begin{center}
\hspace*{-0.5cm}
\includegraphics[width=8cm]{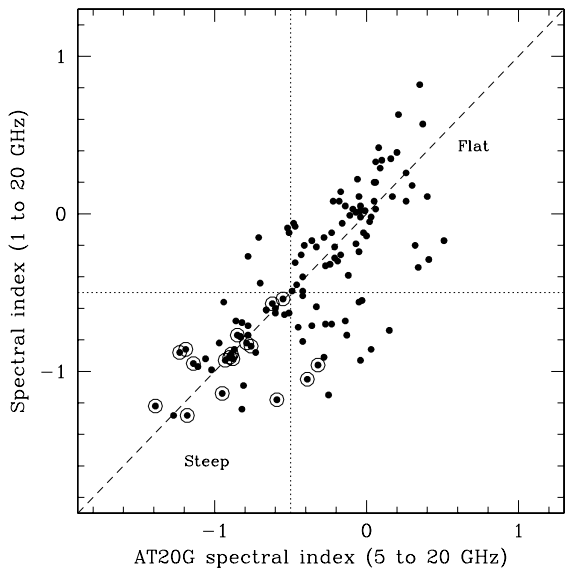}
\caption{Comparison of radio spectral indices measured 
at 5--20\,GHz from near-simultaneous AT20G data, and 
at 1--20\,GHz by cross-matching the AT20G and NVSS/SUMSS source catalogues. 
As in Figure\,2, the points plotted within larger open circles 
correspond to galaxies with extended 20\,GHz sources. 
\label{fig_alpha_comp}}
\end{center}
\end{figure}

\subsubsection{1-20\,GHz spectral indices} 
All 202 galaxies in the AT20G-6dFGS sample have low-frequency radio data available 
from the 1.4\,GHz NVSS (Condon et al.\ 1998) or 843\,MHz SUMSS (Mauch et al.\ 2003) 
surveys, allowing us to calculate a 1--20\,GHz spectral index $\alpha_1^{20}$ for 
each galaxy.  These spectral-index values need to be used with some caution since 
(i) the radio measurements were made several years apart and some sources may be 
variable, and (ii) the low- and high-frequency radio measurements have slightly 
different spatial resolution, which may affect the measured spectral index for 
extended sources.  

Figure\,\ref{fig_alpha_comp} compares $\alpha_1^{20}$ with the near-simultaneous 
$\alpha_5^{20}$ values for the 6dFGS galaxies which have multi-frequency 
AT20G data.   The median 1--20\,GHz spectral index for the 123 galaxies with 
5, 8 and 20\,GHz data, $-0.39\pm0.05$, is slightly steeper than the median 
5--20\,GHz spectral index of $-0.27\pm0.05$ for the same group of galaxies.  
This is mainly because (as we discuss later in \S 3.5), some of the flat-spectrum 
objects seen at 5--20\,GHz are embedded in more diffuse steep-spectrum lobes 
which contribute to the total flux density measured at frequencies near 1\,GHz. 
 
\begin{figure}
%
%
\begin{center}
\includegraphics[width=7cm]{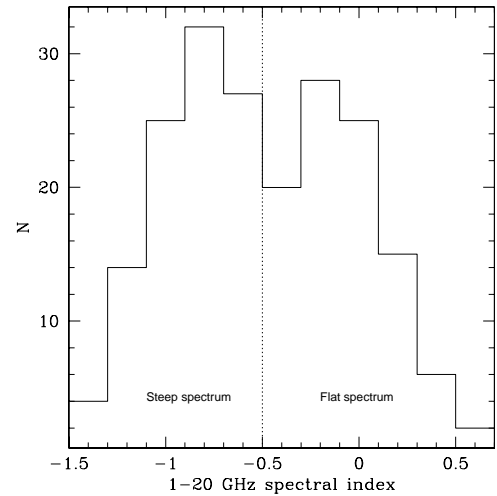}
\caption{Distribution of 1--20\,GHz radio spectral index 
for galaxies in the AT20G-6dFGS sample.  }  
\label{fig_hist}
\end{center}
\end{figure}

In the analysis which follows, we will use the values of $\alpha_1^{20}$ 
(which are available for the whole AT20G-6dFGS sample) as a guide to separating 
the `flat-spectrum' and `steep-spectrum' radio-source populations (though it is 
important to keep in mind that, as can be seen from Figure\,\ref{fig_alpha_comp}, 
roughly 15\% of the AT20G-6dFGS galaxies will have radio spectra which are `steep' at 
low frequencies and `flat' at higher frequencies).   

Figure\,\ref{fig_hist} shows the distribution of $\alpha_1^{20}$ for the 202 sources in 
the AT20G-6dFGS sample. The median 1--20\,GHz spectral index for the full AT20G-6dFGS 
sample is $-0.53\pm0.04$ and although this plot suggests that the distribution may 
be bimodal, the data are statistically consistent with a normal distribution with 
a mean value of $\alpha_1^{20}=-0.498$ and a standard deviation of 0.503. 

\subsection{Candidate GPS and CSS radio sources} 
In \S3.1, we divided the AT20G-6dFGS radio galaxies into three subclasses 
(FR-0, FR-1 and FR-2) based on their low-frequency radio morphology.  
We now use the radio morphology and spectral index information presented 
in \S3.2 snd \S3.4 to identify possible members of the class of Compact Steep Spectrum
(CSS) and Gigahertz-Peaked Spectrum (GPS) sources which are generally thought to 
represent the earliest stages of radio-galaxy evolution (O'Dea 1998). 

We have chosen to sub-divide the FR-0 class (i.e. the AT20G-6dFGS galaxies whose radio emission 
is unresolved in the 1\,GHz NVSS/SUMSS images) as follows: 
\begin{itemize} 
\item
{\bf Candidate GPS sources (FR-0g).}\ These are radio sources with a compactness parameter $R\geq0.85$ 
and 1-20\,GHz spectral index $\alpha_1^{20}>0$, i.e. likely to be less $<1$\,kpc in size 
and have a radio spectrum peaking above 1\,GHz (many of these sources peak at or above 5--10\,GHz). 
\item
{\bf Candidate CSS sources (FR-0c).}\ These are radio sources with a steep 1-20\,GHz spectral 
index $\alpha_1^{20}<-0.50$, i.e. less than 10-20\, kpc in size and likely to have a radio 
spectrum peaking below 1-5\,GHz. 
\item
{\bf Unclassified compact sources (FR-0u).}\ These are sources which can't be classsified as either
CSS or GPS using the currently available data, either because they have no measured compactness
parameter or because they have $-0.5<\alpha_1^{20}<0$ (making it difficult to locate a radio 
spectral peak in these objects, which generally have only a few data points available). 
\end{itemize}

These results confirm that the AT20G-6dFGS sample contains a high fraction of possible 
CSS and GPS galaxies.  
At least 83 objects (41\% of the total sample) are candidate CSS/GPS sources, and the 
fraction rises to 67\% if we include the unclassified FR-0u objects.  

Further analysis of the CSS/GPS candidate sources is difficult at this stage, 
though we discuss the possible effects of relativistic beaming in \S6.2. 
Higher-resolution (VLBI) radio images of these objects, together with 
improved multi-wavelength radio data to measure their spectral turnover 
frequency, are needed to estimate the source ages and establish how 
many of them are genuinely young radio galaxies.  

\section{The local radio luminosity function (RLF) at 20\,GHz} 

The local radio luminosity function (RLF) is the global average space 
density of radio sources at the present epoch (Auriemma et al.\ 1977; 
Condon et al.\ 1989), and provides an important benchmark for studying 
the cosmic evolution of radio-source populations (De Zotti et al.\ 2010). 
The local radio luminosity function of galaxies at 20\,GHz provides a 
particularly useful benchmark for the study of high-redshift radio galaxies 
(since, for example, 1.4 and 5\,GHz measurements of a galaxy at redshift 
$z\sim3$ correspond to 8 and 20\,GHz in the object's rest frame).  
High-frequency data also provide important constraints for the `simulated skies' 
(Wilman et al.\ 2008) which are increasingly used in the science planning for 
future large radio telescopes like the Square Kilometre Array. 

\subsection{Calculating the local RLF} 
To calculate the local radio luminosity function of galaxies at 20\,GHz, we used 
the same methodology as Mauch \& Sadler (2007).  

For the radio data, we set a flux--density limit of 50\,mJy at 20\,GHz and assumed 
a differential completeness of 78\% above 50\,mJy and 93\% above 100\,mJy 
(Massardi et al.\ 2011a). 

For the 6dFGS data, we used a magnitude limit of K=12.75\,mag. and assumed a 
spectroscopic completeness of 92.5\% (Jones et al.\ 2009). 
The sky area  covered by the AT20G-6dFGS sample was assumed to be the total 6dFGS 
area of 5.19\,sr (16,980 deg$^2$). 

We calculated the local radio luminosity function using the 1/$V_{\rm max}$ method of Schmidt (1968), 
where $V_{\rm max}$ is the maximum volume within which a galaxy can satisfy all the sample 
selection criteria (for the AT20G-6dFGS sample, $S_{\rm 20\,GHz}\geq50$\,mJy, $K\leq12.75$\,mag. 
and $0.003<z<0.200$).  

\begin{figure}
\begin{center}
%
%
\includegraphics[width=8.5cm]{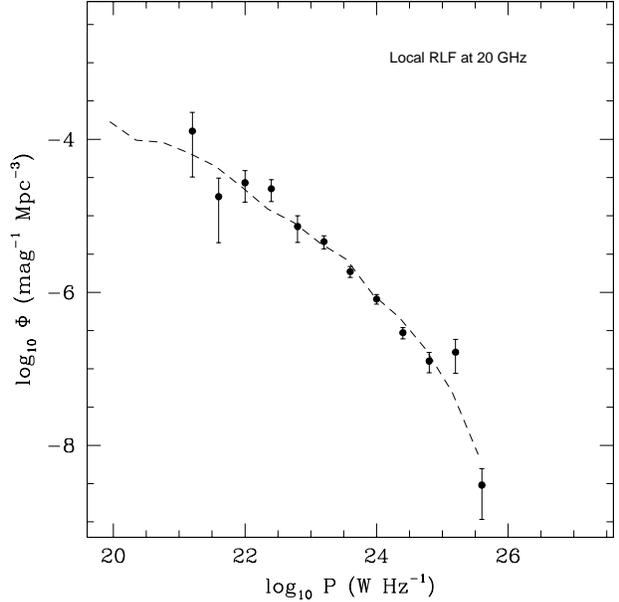}
\caption{The local radio luminosity function for galaxies at 20\,GHz  
(filled circles).  
The dashed line shows the 1.4\,GHz RLF for AGN from Mauch \& Sadler (2007), 
shifted in radio power by adopting a 1.4--20\,GHz spectral index of $-0.74$ 
as discussed in the text.  
\label{fig_rlf}}
\end{center}
\end{figure}

\begin{table}
\centering
\caption{Local radio luminosity function (RLF) at 20\,GHz for radio AGN. }
\label{tab_rlf}
\begin{tabular}{@{}crr@{}}
\hline
log$_{10}$\,P$_{20}$  & N & \multicolumn{1}{c}{log $\Phi$} \\
(W\,Hz$^{-1}$)        &   & (mag$^{-1}$\,Mpc$^{-3}$) \\
\hline
21.2 &    1   &  $-3.89^{+0.24}_{-0.60}$ \\ 
21.6 &    1   &  $-4.75^{+0.24}_{-0.60}$ \\ 
22.0 &    3   &  $-4.57^{+0.16}_{-0.25}$ \\ 
22.4 &    7   &  $-4.65^{+0.12}_{-0.17}$ \\ 
22.8 &    6   &  $-5.14^{+0.14}_{-0.21}$ \\
23.2 &   20   &  $-5.34^{+0.08}_{-0.09}$ \\ 
23.6 &   28   &  $-5.73^{+0.07}_{-0.08}$ \\ 
24.0 &   39   &  $-6.09^{+0.06}_{-0.07}$ \\ 
24.4 &   33   &  $-6.53^{+0.07}_{-0.08}$ \\ 
24.8 &   22   &  $-6.90^{+0.11}_{-0.15}$ \\ 
25.2 &   10   &  $-6.78^{+0.17}_{-0.27}$ \\
25.6 &    2   &  $-8.52^{+0.22}_{-0.45}$ \\ 
&& \\
\multicolumn{3}{l}{$<$V/V$_{\rm max}>$ = 0.45$\pm$0.02 } \\
\hline
\end{tabular}
\end{table}

\subsection{Results} 
The local 20\,GHz radio luminosity function measured from the AT20G-6dFGS galaxy 
sample is listed in Table\,\ref{tab_rlf} and plotted in Figure\,\ref{fig_rlf}. 

We derive a mean value of V/V$_{\rm max}$ (0.45$\pm$0.02) for these galaxies, 
which is slightly lower than the value of 0.50 expected for a complete sample.  
This is almost certainly because the total 20\,GHz flux density for some extended AT20G 
sources is underestimated, as discussed in \S3.3. If these sources have measured 
flux densities below 50\,mJy, they will be excluded from the RLF sample when they 
should have been included. 
If we separate the extended 20\,GHz sources from those which are 
unresolved, we find 
$<$V/V$_{\rm max}>$ = 0.47$\pm$0.02 for unresolved sources and 
$<$V/V$_{\rm max}>$ = 0.37$\pm$0.05 for the extended sources, 
confirming that the small incompleteness in our overall sample 
arises mainly from the extended 20\,GHz sources. 
Because of the relatively small size of the AT20G sample and the likely incompleteness 
in the extended-source population, we have not attempted to fit a functional form to 
the 20\,GHz RLF.  Instead, we used the parameterized form of the 1.4\,GHz local RLF 
from equation (6) of Mauch \& Sadler (2007), and fitted this to the 20\,GHz data by 
making a simple shift in radio power set by a single characteristic 1-20\,GHz spectral 
index $\alpha_0$\ for the RLF as a whole. The best-fitting value, $\alpha_0=-0.74$, 
is shown by the dashed line in Figure\,\ref{fig_rlf}. 

It is important to note that the shift of $\alpha_0=-0.74$ which provides the 
best match for the 1.4\,GHz and 20\,GHz radio luminosity functions is steeper than the 
median 1-20\,GHz spectral index of $\widetilde{\alpha}_1^20=-0.53\pm0.04$ which 
we found for the AT20G-6dFGS galaxies in \S3.4.2.  
The reason for this difference is not yet completely clear, but one plausible 
explanation is that the value of $-0.74$ represents a characteristic 1--20\,GHz 
spectral index for the local radio-galaxy population as a whole.  Since the individual 
objects which make up this population have a broad spread in radio spectral 
index, the flat-spectrum members of this population are more likely to be detected 
at 20\,GHz than the steep-spectrum objects and so this will flatten the observed 
spectral-index distribution for sources selected at 20\,GHz.

\section{Optical and infrared properties of the AT20G--6dFGS sample} 

\subsection{Redshift and infrared K-band luminosity } 
\begin{figure}
\begin{center}
%
%
\hspace*{-0.5cm}
\includegraphics[width=8cm]{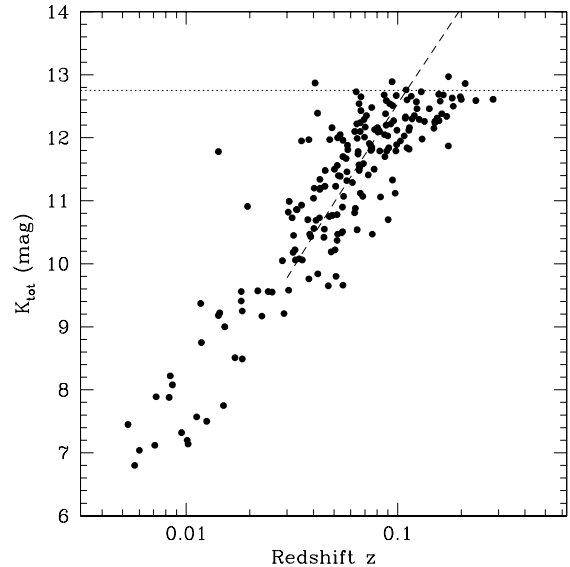}
\caption{Plot of K-band apparent magnitude versus redshift 
for galaxies in the 6dFGS-AT20G sample. 
The horizontal dotted line shows the K=12.75 magnitude limit 
of the 6dFGS, and the dashed line shows an extension of the 
radio-galaxy K-$z$ relation derived by Willott et al.\ (2003)  
over the redshift range $0.05<z<2$. 
\label{fig_kz}}
\end{center}
\end{figure}

Figure\,\ref{fig_kz} shows the distribution of the sample galaxies 
in K-band apparent magnitude and redshift.  Since the infrared 
K-band light in these nearby galaxies arises mainly from old giant 
stars, the K-band luminosity is closely related to the stellar mass 
of the galaxy. 

Most of the galaxies in Table\,3 lie close to the well-known K--$z$ relation 
for radio galaxies (Lilly \& Longair 1984; Willott et al.\ 2003; De Breuck 
et al. 2002). The K-$z$ relation derived by Willott et al.\ (2003) 
and plotted in Figure\,\ref{fig_kz}, 
$$ K = 17.37 + 4.53{\rm log}_{10}z -0.31({\rm log}_{10}z)^2$$ 
is very close to the expected K-magnitude evolution of a 
passively-evolving galaxy which formed at high redshift 
($z\sim10$) and has a present-day ($z\sim10$) luminosity of around 
3\,L$_*$. We therefore find that the AT20G-6dFGS radio sources are 
hosted by galaxies which have K-band luminosities matching those 
of powerful radio galaxies in the distant universe. 

\subsection{The radio/optical luminosity diagram} 
\begin{figure}
\begin{center}
%
%
\hspace*{-0.5cm}
\includegraphics[width=8.5cm]{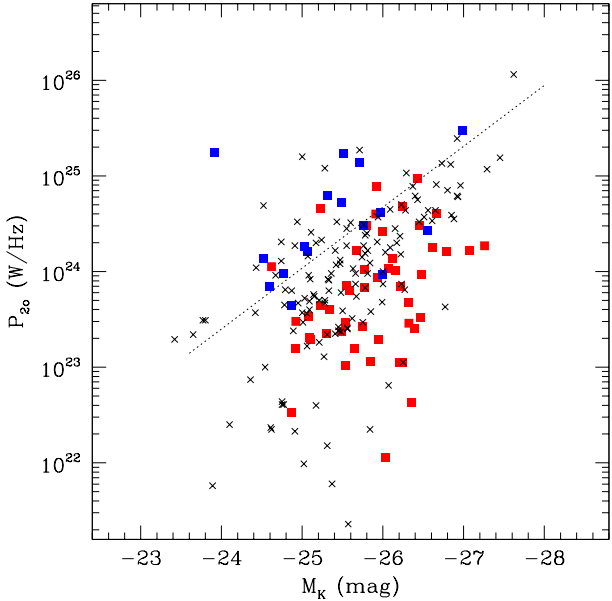}
\caption{Distribution of the AT20G-6dFGS galaxy sample in (infrared) K-band absolute 
magnitude and 20\,GHz radio power. FR-1 radio galaxies are shown as filled red squares, 
FR-2 radio galaxies as filled blue squares and compact (FR-0) sources as black crosses. 
The dashed line corresponds to the 1.4\,GHz FR-1/FR-2 dividing line from Ledlow \& Owen 
(1996), shifted to K-band and 20\,GHz by assuming (i) a typical galaxy colour of 
(R-K) = 3.0\,mag.\ and (ii) a characteristic radio spectral index $\alpha_1^{20}=-0.74$, 
as discussed in \S4.2 }
\label{fig_maglum}
\end{center}
\end{figure}

Figure\,\ref{fig_maglum} shows the distribution of the sample galaxies in 
radio and K-band luminosity.  A radio $k$-correction of the form 
$k_{\rm radio}(z)=(1+z)^\alpha$ has been applied, where $\alpha$ is 
the radio spectral index ($S_\nu\propto\nu^\alpha$). Different symbols 
show galaxies classified as FR-1, FR-2 or FR-0 (compact) on the basis 
of their 1\,GHz radio morphology, as discussed in \S3.   
As with the NVSS-6dFGS sample of Mauch et al.\ (2007), almost all the 
AT20G-6dFGS galaxies are optically luminous (M$_{\rm K}$ brighter than $-24$\,mag.), 
but there is no obvious correlation between radio and optical luminosity for 
galaxies above this K-band luminosity threshold. 

There is a clear tendency for FR-2 radio galaxies to have a higher radio luminosity 
than FR-1 radio galaxies of similar stellar mass, implying that the relation found by 
Ledlow \& Owen (1996; see their Figure 1), in which the FR-1/FR-2 division 
is a strong function of optical luminosity, also holds at 20\,GHz. Interestingly, 
the compact (FR-0) sources cover the full range in 20\,GHz radio power 
spanned by the FR-1 and FR-2 objects, and a few are also found in 
low-luminosity (M$_{\rm K}$ fainter than $-24$\,mag.) galaxies. 
The radio galaxy Pictor A (AT20G J051949-454643) is a notable outlier in this plot, 
having a total 20\,GHz radio power of $1.8\times10^{25}$\,W\,Hz$^{-1}$ despite 
being one of the least optically-luminous galaxies in the AT20G-6dFGS sample 
(with M$_{\rm K}=-23.9$\,mag). 

\subsection{High-excitation and low-excitation radio galaxies}
Recently, several authors (e.g. Hardcastle et al.\ 2007, Best \& Heckman 2012 and 
references therein) have proposed a fundamental dichotomy between `high-excitation radio galaxies' 
(HERGs), in which the AGN is fuelled in a radiatively efficient way by a classic accretion disk, 
and `low-excitation radio galaxies' (LERGs) in which the accretion rate is significantly lower 
and radiatively inefficient.  Best \& Heckman (2012) derive accretion rates of one and ten per cent 
of the Eddington rate for HERGs, in contrast to a typical accretion rate below one per cent Eddington 
for the LERGs in their sample. 

Observationally, HERGs are characterised by strong optical emission lines with line ratios characteristic 
of highly-excited gas (e.g. Kewley et al.\ 2006), while LERGs generally show weak or no optical emission 
lines.  Ideally we would use a well-determined quantity such as emission-line luminosity to 
classify the AT20G-6dFGS radio galaxies, but this is difficult since the 6dFGS spectra are 
not flux-calibrated. 
Instead, we make a qualitative separation by associating the `Ae' and `AeB' radio galaxies 
in our sample with the HERG class, and `Aa' and `Aae' objects with the LERG class. 

There are several reasons why this appears reasonable: 

\vspace*{-0.2cm}
\begin{enumerate}
\item
The 6dFGS spectra have a resolution of 5-6\,\AA\ in the blue and 10-12\,\AA\ in the red 
(Jones et al.\ 2004).  Since our Ae classification requires that a galaxy show optical 
emission lines which are strong relative to the stellar continuum, these objects are 
likely to have an [OIII] equivalent width well above the value of 5\,\AA\ which Best \& Heckman 
(2012) use as one of the distinguishing criteria for of their HERG class. 
\item 
At least 25\% of the Ae objects in our sample show broad Balmer emission characteristic of 
high-excitation Seyfert galaxies, while many of the narrow-line Ae objects also have a 
literature classification as Seyfert galaxies (see the notes in Appendix A). 
\item
We know that most early-type galaxies show weak optical emission lines (generally with low-excitation 
LINER-like spectra as described by Heckman 1980) if one looks carefully enough (e.g. Phillips et al.\ 1986). 
In some cases, these emission lines may be excited by hot post-AGB stars rather than an AGN (Bertelli et 
al.\ 1989; Cid Fernandes et al. 2011). 
For fibre spectroscopy these weak emission lines may be easier to recognize in lower-redshift galaxies, 
simply because the fibre contains less of the stellar light from the surrounding galaxy, as discussed 
by Mauch \& Sadler (2007).  As a result, we expect to see significant overlap between our Aa and Aae 
classes and so it seems plausible to associate all these objects with the LERG class. 
\end{enumerate}

If we make this separation, then 23\% of the AT20G-6dFGS radio sources are classified as 
high-excitation (HERG) systems and 77\% as low-excitation (LERG).  
While LERGs are the majority population, the HERG fraction is higher than that seen in 
comparable radio-source samples selected at 1.4\,GHz.  Only $\sim12\%$ of the radio AGN in 
the Mauch \& Sadler (2007; MS07) are classified as Ae galaxies, and the HERG fraction in the 
Best \& Heckman (2012; BH12) sample of radio AGN is even lower than this. 

The higher HERG fraction seen in the 20\,GHz sample is not simply an effect of 
comparing objects with different radio luminosities (the AT20G sources are typically 
more powerful than those selected from NVSS, because the AT20G catalogue has a 40\,mJy 
flux density limit, compared to 2.5\,mJy for NVSS).  To check this, we compared the HERG 
fraction in several bins of 1.4 GHz (not 20\,GHz) radio power for the AT20G-6dFGS, BH12 
and MS07 galaxy samples.  For radio powers in the range 10$^{23.8}$ to 
10$^{25.6}$\,W\,Hz$^{-1}$ the HERG fraction in the AT20G-6dFGS sample was at least three 
to five times higher than in the corresponding BH12 and MS07 samples. In the highest-power 
bin (10$^{25.6}$ to 10$^{26.5}$\,W\,Hz$^{-1}$), which contains only a relatively small number 
of objects, all three samples showed a high HERG fraction of around 40--50\%.

\begin{table*}
\centering
\caption{AT20G-6dFGS radio sources known to be associated with spiral galaxies. The galaxy 
classification in column 8 is taken from either the RC3 catalogue (de Vaucouleurs et al.\ 1991) 
or the ESO/Uppsala catalogue (Lauberts 1982), and the galaxy T-type in column 9 is from the RC3 
catalogue. } 
\label{tab_spiral}
\begin{tabular}{@{}llllllllllll@{}}
\hline
AT20G name & \multicolumn{1}{c}{Alt.} &  \multicolumn{1}{c}{Redshift}  & \multicolumn{1}{c}{Spectral}  & 
\multicolumn{1}{c}{M$_{\rm K}$} & \multicolumn{1}{c}{log\,P$_{20}$} & \multicolumn{1}{c}{$\alpha_1^{20}$} & 
Galaxy &  \multicolumn{1}{c}{T-type} & Notes \\
           &  \multicolumn{1}{c}{name}  & \multicolumn{1}{c}{$z$} & \multicolumn{1}{c}{class}  & 
	   \multicolumn{1}{c}{mag} &  \multicolumn{1}{c}{W\,Hz$^{-1}$} &  & class  & \\    
\hline
\multicolumn{10}{l}{(a) Nearby ($z\leq0.025$) galaxies classified as spiral in the RC3 and/or ESO/Uppsala catalogues} \\
  J004733-251717  & NGC\,253  & 0.0008 &  SF  & $-$22.74 & 20.93 &  $-$0.91  & Sc   & $+$5.0 & Nuclear starburst \\ 
  J024240-000046  & NGC\,1068 & 0.0038 &  Ae  & $-$25.31 & 22.18 &  $-$0.86  & Sb   & $+$3.0 & Seyfert 2 galaxy \\  
  J123959-113721  & NGC\,4594 & 0.0036 &  ..  & $-$25.57 & 21.36 &  $-$0.05  & Sa   & $+$1.0 & \\ 
  J130527-492804  & NGC\,4945 & 0.0019 &  ..  & $-$23.89 & 21.76 &  $-$0.64  & Scd  & $+$6.0 & AGN (Liner) \\  
  J130841-242259  & IRAS\,13059-2407& 0.0142 & Ae & $-$22.11 & 22.41 & $-$0.80 & Sc? & ... & Drake et al.\ (2003)  \\ 
  J131949-272437  & NGC\,5078 & 0.0071 &  Aae & $-$25.37 & 21.78 &  $-$0.59   & Sa   & $+$1.0 &  \\ 
  J133608-082952  & NGC\,5232 & 0.0228 &  Aae & $-$25.78 & 23.83 &  $+$0.16   & Sa   & $+$0.0 & \\ 
  J145924-164136  & NGC\,5793 & 0.0117 &   .. & $-$24.10 & 22.40 &  $-$1.00   & Sb   & $+$3.0 & Seyfert 2 galaxy \\ 
  J172341-650036  & NGC\,6328 & 0.0142 &   .. & $-$24.74 & 24.11 &  $-$0.08   & Sab  & $+$1.8 & AGN (Liner) \\
  J220916-471000  & NGC\,7213 & 0.0060 &   .. & $-$25.01 & 21.99 &  $+$0.01   & Sa   & $+$1.0 & Seyfert 1 galaxy \\
\multicolumn{10}{l}{(b) More distant ($z>0.025$) galaxies noted in the literature as spirals  } \\
  J001605-234352  & ESO\,473-G07    & 0.0640 &  Ae  & $-$24.79 & 23.81 &  $-$0.53  &  S... &  $+$5.0  &  \\ 
  J031552-190644  & PMN\,J0315-1906 & 0.0671 &  Ae? & $-$24.62 & 24.05 &  $+$0.03  &  S..  & ...      & Ledlow et al.\ (2001) \\ 
  J220113-374654  & AM 2158-380     & 0.0334 &  Ae  & $-$24.87 & 23.65 &  $-$0.82  &  S?   &  ..      & Drake et al.\ (2003) \\ 
\hline
\end{tabular}
\flushleft
\end{table*}

\subsection{Optical morphology of the host galaxies} 
Most of the powerful radio-loud AGN in the local universe are hosted by massive elliptical 
galaxies (e.g. Lilly \& Prestage 1987; Owen \& Laing 1989; Veron-Cetty \& Veron 2001), 
though some exceptions are known (e.g. Ledlow et al.\ 2001; Hota et al. 2011), and   
we also know that nearby spiral galaxies can host compact radio-loud 
AGN (e.g. Norris et al. 1988; Sadler et al.\ 1995). 

Some of the galaxies in the AT20G-6dFGS sample are at low enough redshift ($z<0.025$) that 
a reliable classification of their optical morphology is available from the RC3 
(de Vaucouleurs et al. 1991) and/or ESO-Uppsala (Lauberts 1982) galaxy catalogues.  
We have therefore used these classifications, where available, to look at the host galaxy 
properties of the AT20G-6dFGS sample. 

Of the 34 lowest-redshift ($z<0.025$) galaxies in Table\,3, twenty-four are classified 
as early-type (E or S0) galaxies and ten as late-type (spiral/disk) galaxies, implying 
that at the lowest 20\,GHz luminosities probed by our sample 
($10^{21}$ to $10^{23}$ W\,Hz$^{-1}$) around 30\% of the host galaxies are spirals.  

Although the morphologically-classified subsample (34 AT20G-6dFGS galaxies with 
$z\leq0.025$) is small, some general patterns can be seen. In particular, all nine 
FR-1 radio galaxies in this redshift range have E/S0 host galaxies but almost 
half the compact FR-0 sources are in spiral galaxies.  

Table\,\ref{tab_spiral} lists the thirteen galaxies in our sample which are known 
to have spiral or disk-like optical morphology (this table also includes some objects 
with $z>0.025$).  Six of these galaxies (NGC\,253, NGC\,1068, NGC\,4594. IRAS\,13059-2407, 
NGC 5078 and NGC\,5232) also belong to the sample of dusty `infrared-excess IRAS galaxies' 
identified by Drake et al.\ (2004), and PMN\,J0315-1906 has been identified by 
Ledlow et al.\ (2001) as a rare example of an FR-1 radio source hosted by a spiral galaxy. 

Since reliable morphological classifications are not available for 
the more distant galaxies in the 6dFGS-AT20G sample, the next section discusses the 
use of mid-infrared photometry from WISE as an alternative way of investigating 
host-galaxy properties. 

\subsection{Mid-infrared photometry from WISE}

Near- and mid-infrared photometry (at 3.4, 4.6, 12 and 22\,$\mu$m) is now available 
for all the AT20G-6dFGS galaxies from the WISE mission (Wright et al.\ 2010).  
The majority of galaxies in our sample 
(141/202) are flagged as extended 2MASS objects (ext\_flag=5), and for these we used 
the elliptical aperture magnitudes as recommended by Wright et al.\ (2010). For the remaining 
galaxies, which were flagged as unresolved in the WISE catalogue, we used the profile-fit 
magnitudes.  

We restricted our analysis to galaxies with WISE 3.6$\mu$m magnitude 
fainter than 6.0, since galaxies brighter than this may be too large 
on the sky to allow accurate photometry with WISE. We also excluded five  
galaxies which had anomalous WISE colours because of contamination from 
a companion galaxy or bright foreground star. This left a total of 193 
AT20G-6dFGS galaxies with good-quality WISE photometry. 

\subsection{The WISE two-colour plot} 
\begin{figure}
\begin{center}
%
%
\hspace*{-0.5cm}
\includegraphics[width=8.5cm]{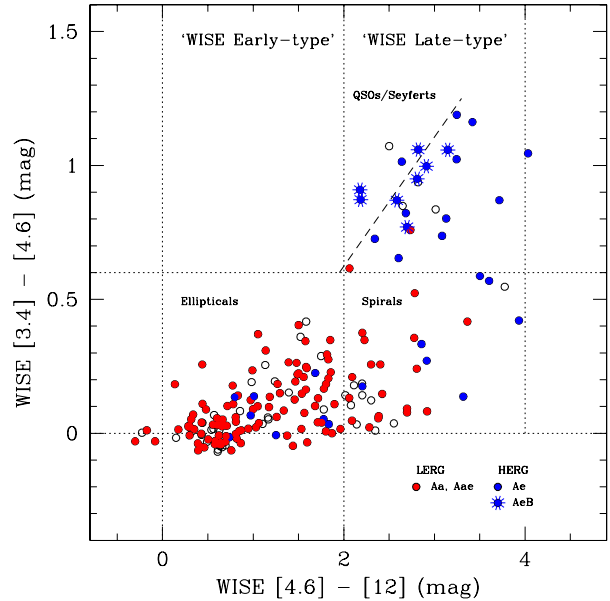}
\caption{WISE colour-colour plot for the AT20G galaxies, which are colour-coded according to 
their optical spectral classification as outlined in Table\,2 and \S2.3.  
Aa and Aae galaxies (LERGs) are shown by red filled circles and Ae galaxies (HERGs) 
by blue filled circles. Galaxies with broad emission lines (class AeB) are shown by 
blue stars, and galaxies with no spectral classification (see \S2.3) by black open circles.  
Dotted lines show the regions in which different galaxy populations are expected to lie, 
as discussed in \S5.6 of the text, and the diagonal dashed line shows the WISE blazar 
sequence from Massaro et al.\ (2012).  }
\label{fig_wisespec}
\end{center}
\end{figure}

Figure\,\ref{fig_wisespec} shows a WISE colour-colour plot for the full 
AT20G-6dFGS sample, based on Figure 12 of Wright et al.\ (2010). The errors on individual 
data points are typically $<$0.05\, mag in $[3.4] - [4.6]$ colour and $<$0.1\, mag in 
$[4.6] - [12]$ colour. 
The dotted horizontal and vertical lines are also based on the work of Wright et al.\ (2010), 
who divide elliptical and spiral galaxies at a WISE $[4.6] - [12]$ colour of +1.5 mag.\  
and note that the most powerful optical AGN lie above a $[3.4] - [4.6]$ colour of +0.6 mag. 
For nearby AT20G-6dFGS galaxies with a reliable optical classification (see \S4.3 above), 
there is excellent agreement between the optical and WISE galaxy classes. We also see 
that all the galaxies in which we detected strong, broad Balmer emission lines (class AeB) 
lie close to the line followed by blazars and radio-loud QSOs (Massaro et al.\ 2012). 

The WISE two-colour plot is becoming widely used for galaxy population studies, 
and its interpretation has recently been discussed by several authors.  
The [3.4] - [4.6] micron colour can be used to separate normal galaxies from AGN with 
a strongly-radiating accretion disk (Assef et al.\ 2010; Yan et al.\ 2013), while 
the [4.6] - [12] micron colour separates dusty star-forming galaxies (or galaxies in 
which the dust is heated by radiation from an AGN) from early-type galaxies with little 
or no warm dust. Donoso et al.\ (2012) have shown that the $[4.6] - [12]$ colour reflects 
a galaxy's specific star-formation rate (SSFR), i.e. the current star-formation rate 
divided by the total stellar mass. 

If we follow Wright et al.\ (2010) and set the dividing line between WISE `ellipticals' and `spirals' 
at a colour of $[4.6] - [12] = 1.5$\,mag., we find that almost half of our AT20G-6dFGS galaxies 
(93/193, or 48\%) fall into the `spiral' category.  If we set a more stringent cutoff for WISE 
`spirals' at $[4.6] - [12]\geq2.0$\,mag. (which roughly corresponds to the main locus of 
star-forming galaxies in Figure 8 of Donoso et al.\ 2012), we find a spiral fraction 
of 31\% (60/193) for the AT20G-6dFGS sample.  This is similar to the spiral fraction 
of $\sim30$\% derived from optical classification of the closest galaxies in our sample (see \S4.4), so 
we adopt $[4.6] - [12] = 2.0$\,mag. as a reasonable dividing line between host galaxies which lie in the 
`WISE elliptical' region and those which correspond to `WISE spirals'.  Since `elliptical' and `spiral' 
are morphological descriptions, while the WISE two-colour plot is based on the mid-infrared spectral-energy 
distribution, in the remainder of this paper we refer to the two classes of host galaxy as 
{\it `WISE early-type'} ($[4.6] - [12]<2.0$\,mag) and {\it `WISE late-type'} 
($[4.6] - [12]\geq2.0$\,mag) galaxies respectively. 

\subsubsection{WISE early-type galaxies} 
As discussed above, around 70\% of the AT20G-6dFGS galaxies have WISE colours which are typical 
of normal elliptical and S0 galaxies. Of the 25 `WISE early-type' galaxies with an 
RC3 or ESO morphological type, 22 are classified as E or S0 galaxies and three as early-type 
(Sa) spirals, showing that there is generally good agreement between the WISE classification 
and the optical morphology where both are available.

\subsubsection{WISE late-type galaxies} 
The remaining 30\% of AT20G-6dFGS galaxies fall into the `WISE late-type' class. 
Nine of them also have an RC3 or ESO morphological type -- seven are classified 
as spirals (ranging from Sa to Scd) and two as S0 galaxies, again implying 
that the WISE classification is generally consistent with the observed optical 
morphology. 

We note, however, that the `WISE late-type' galaxies in our sample may be 
a heterogeneous class, since they are selected because they contain significant 
quantities of (warm or hot) dust.  They include genuine spiral galaxies like 
those listed in Table\,\ref{tab_spiral}, as well as elliptical and S0 galaxies with dust lanes 
(e.g. NGC\,612 = AT20G\,J013357-362935; see Ekers et al.\ 1978) and composite objects like the 
`radio-excess IRAS galaxies' identified by Drake et al.\ (2004). 

\subsubsection{Host galaxies of HERGs and LERGs}
Table\,\ref{wise_tab2} relates the WISE classification (derived from Figure\,\ref{fig_wisespec}) 
to the HERG/LERG spectral class for the AT20G-6dFGS galaxies. 
We find that almost all the WISE `early-type' galaxies (92\%) have low-excitation (LERG) optical 
spectra, while the WISE `late-type' galaxies have a mix of HERG and LERG spectra. 

This result is in broad agreement with earlier studies of radio galaxies selected at 1.4\,GHz. 
Hardcastle et al.\ (2013) found that HERGs are typically about four times as luminous as LERGS 
in the far-infrared 250\,$\mu$m band, and interpret this as showing that HERGs are more likely 
to be located in star-forming galaxies.  Best \& Heckman (2012) found that HERGs in their 
local sample typically had lower stellar masses and younger stellar populations than LERGS, 
again consistent with a picture in which HERGs are commonly found in star-forming galaxies.

\begin{table}
\centering
\caption{Distribution of spectral classifications for the AT20G-6dFGS galaxies, split by WISE colours. }
\label{wise_tab2}
\begin{tabular}{@{}llrrrr@{}}
\hline
\multicolumn{2}{c}{WISE [4.6] - [12]\,$\mu$m}   & \multicolumn{2}{c}{Spectral class} & Median  & HERG  \\
\multicolumn{1}{c}{colour} & \multicolumn{1}{c}{class} & HERG & LERG & M$_{\rm K}$\,(mag) & fraction \\
\hline
  $<$2.0\,mag    & `Early-type'    &    8   &  100  &  $-25.67\pm0.08$ &  7\,\% \\ 
  $\geq$2.0\,mag & `Late-type'     &   26   &   20  &  $-25.41\pm0.16$ & 57\,\% \\ 
\hline
\end{tabular}
\end{table}

\begin{table*}
\centering
\caption{General radio properties of the AT20G-6dFGS sample, split by (low-frequency) radio 
morphology and 1-20\,GHz spectral index, and excluding the nearby starburst galaxy NGC\,253. 
The compact (FR-0) objects are also divided into three subclasses: FR-0g (candidate GPS), FR-0c 
(candidate CSS) and FR0-u (unclassified compact sources), as discussed in \S3.5 of the text.}
\label{tab_frclass}
\begin{tabular}{@{}llcrrrrcccc@{}}
\hline
\multicolumn{1}{l}{Class} &\multicolumn{1}{c}{1\,GHz radio} & \multicolumn{1}{c}{20\,GHz radio} & 
\multicolumn{1}{c}{N} & \multicolumn{1}{c}{Median} & \multicolumn{1}{c}{Median} & \multicolumn{1}{c}{Median log\,P$_{20}$} & \multicolumn{2}{c}{Spectral class} &  \multicolumn{2}{c}{WISE class}\\ 
   & \multicolumn{1}{c}{ morphology} & \multicolumn{1}{c}{ morphology} & & \multicolumn{1}{c}{redshift} & \multicolumn{1}{c}{M$_{\rm K}$ (mag)} & \multicolumn{1}{c}{ (W\,Hz$^{-1}$)} & 
   \multicolumn{1}{c}{LERG} & \multicolumn{1}{c}{HERG}  & \multicolumn{1}{c}{Early} & \multicolumn{1}{c}{Late} \\
& &\multicolumn{1}{c}{/spectral index} &&&&&  {\footnotesize (Aa/Aae)} & \multicolumn{1}{l}{\footnotesize (Ae/AeB)} \\
 \hline
FR-2                                 & Resolved,           &  Any &   16   & 0.0604 & $-25.39\pm0.25$ & $24.45\pm0.17$ & 36\% & 64\%  &  7\% & 93\% \\  
                                     & edge-brightened     &      &        &        &                 &                &      &       & & \\
FR-1                                 & Resolved,           &  Any &   49   & 0.0535 & $-25.92\pm0.11$ & $23.68\pm0.11$ & 98\% &  2\% &  93\% &  7\% \\ 
                                     & not edge-brightened &      &        &        &                 &                &      &       & & \\
FR-0 (all)                           & Unresolved          &  Any &  136   & 0.0653 & $-25.49\pm0.09$ & $23.97\pm0.09$ & 75\% & 25\%  & 67\%  & 33\% \\
                                     & (LAS $\leq$ 30\,arcsec)    &      &        &        &                 &                &      &  & &      \\
 &&&&& \\
FR-0g                             & Unresolved  &  Compactness param.            &  34 & 0.0662 & $-25.39\pm0.19$ & $24.18\pm0.15$ & 75\% & 25\%  & 64\% & 36\% \\
                                  &             & $>$0.85,{\bf and}\ $\alpha_1^{20}\geq0$ &  & & \\
FR-0c                             & Unresolved  &  $\alpha_1^{20}\leq-0.5$       & 49  & 0.0684 & $-25.48\pm0.16$ & $24.07\pm0.15$ & 67\% & 33\%  & 67\% & 33\% \\
FR-0u                             & Unresolved  &  (not in FR-0g                 & 53  & 0.0517 & $-25.55\pm0.13$ & $23.76\pm0.15$ & 84\% & 16\%  & 69\% & 31\% \\
                                  &             &   or 0c class)    &     &        &                 &                &      &      & &  \\
\hline
\multicolumn{3}{l}{All flat-spectrum sources ($\alpha_1^{20}>-0.5$)}    &          97  & 0.0576 & $-25.49\pm0.10$ & $23.91\pm0.10$ & 81\% & 19\%  & 68\%  & 32\%  \\
\multicolumn{3}{l}{All steep-spectrum sources ($\alpha_1^{20}\le-0.5$)} &         104  & 0.0596 & $-25.73\pm0.10$ & $24.02\pm0.10$ & 74\% & 26\%  & 69\%  & 31\%  \\
\hline
\multicolumn{3}{l}{All AT20G-6dFGS AGN }                                &         201  & 0.0581 & $-25.58\pm0.07$ & $23.97\pm0.07$ & 77\% & 23\%  & 68\% & 32\% \\
\hline
\end{tabular}
\flushleft
\end{table*}

\subsection{Comparison with the 1.4\,GHz-selected NVSS-6dFGS radio galaxy sample}

Finally, we compare the spectroscopic properties of the current (20\,GHz-selected) 
6dFGS-AT20G galaxy sample with the 1.4\,GHz-selected 6dFGS-NVSS sample complied by Mauch 
\& Sadler (2007).  This is a useful comparison, since both samples were selected from 
the 6dFGS galaxy catalogue and so differ only in their radio flux limit and the 
frequency at which they were selected. 

Table\,\ref{tab_comp}\ compares the spectroscopic properties of the two samples. 
The AT20G survey has a significantly brighter flux-density limit (40\,mJy at 20\,GHz) 
than the NVSS (2.5\,mJy at 1.4\,GHz), so a comparison with a brighter NVSS sub-sample 
(S$\geq100$\,mJy at 1.4\,GHz) is also included. 

One striking difference between the 6dFGS-NVSS and 6dFGS samples is the fraction 
of galaxies which show optical emission lines (class Aae, Ae or AeB), which is 
20--25\% for radio AGN selected at 1.4\,GHz, but more than twice as high (50\%) 
for galaxies selected at 20\,GHz.

\begin{table*}
\centering
\caption{A comparison between the 20\,GHz-selected AT20G-6dFGS sample discussed in 
this paper and the  1.4\,GHz-selected NVSS-6dFGS sample of radio AGN studied by 
Mauch \& Sadler (2007). }
\begin{tabular}{@{}lccc@{}}
\hline
   & \multicolumn{2}{c}{6dFGS-NVSS (1.4\,GHz)} & 6dFGS-AT20G (20\,GHz) \\ 
   & \multicolumn{2}{c}{Mauch \& Sadler (2007)} & This paper \\ 
   & all radio AGN & S$_{1.4}>100$\,mJy only &  all AGN\\
 \hline
No. of galaxies & 2784 & 271   & 201 \\ 
Magnitude limit & K$_{\rm tot}<12.75$ & K$_{\rm tot}<12.75$ &  K$_{\rm tot}<12.75$ \\
S$_{\rm lim}$ (mJy) & 2.5 & 100.0 & 40.0 \\ 
Area (sr) & 2.16 & 2.16 & 5.19  \\
Median $z$ & 0.073 & 0.0798 & 0.0596 \\ 
Median absolute magnitude  M$_{\rm K}$ & $-25.50$ & $-25.69$ & $-25.59\pm0.07$ \\
HERG fraction (Ae, AeB) (\%)              & 13$\pm1$ & 10$\pm2$  & 23$\pm3$ \\
Emission-line fraction (Aae, Ae, AeB)(\%) & 26$\pm1$ & 19$\pm3$  & 50$\pm6$ \\
\hline
\end{tabular}
\label{tab_comp}
\flushleft
\end{table*}
 
This is not a redshift-based selection effect, since the AT20G--6dFGS galaxies consistently 
show a higher emission-line fraction than NVSS--6dFGS galaxies at the same redshift.  
Instead, it arises from two effects: (i) a significantly higher HERG fraction in the 
AT20G-6dFGS sample as a whole (23\% HERGs, compared to 10-13\% for the NVSS-6dFGS sample, 
see Table\,\ref{tab_comp}), and (ii) a correlation between the 1--20\,GHz radio spectral 
index of a galaxy and the probability that it will show weak optical emission lines. 
We find that a radio-loud AGN selected at 1.4\,GHz from the 6dFGS sample is typically
three times more likely to be detected in the AT20G survey if it has an emission-line 
(Aae or Ae) spectrum.

\begin{figure}
%
%
\begin{center}
\includegraphics[width=8cm]{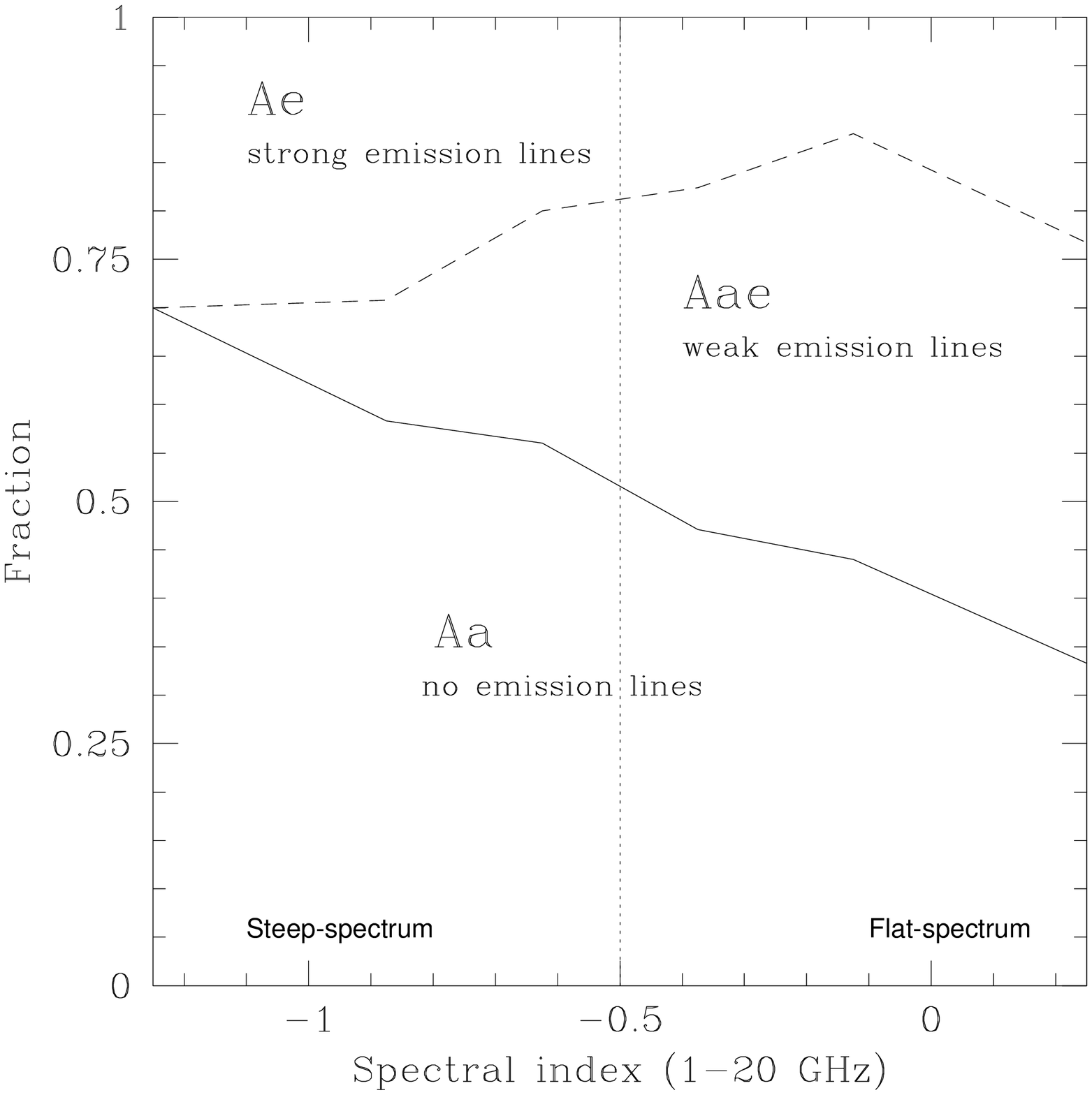}
\caption{Fraction of AT20G-6dFGS galaxies showing optical emission lines 
in the 6dFGS spectrum, plotted against the 1-20\,GHz radio spectral index. 
\label{alpha_frac}}
\end{center}
\end{figure}

Figure\,\ref{alpha_frac} shows that while galaxies with strong emission lines 
(class Ae) are seen across the full range of radio spectral index, the  
fraction of radio galaxies with weak optical emission lines (class Aae) 
increases rapidly as we move from steep-spectrum to flat-spectrum sources. 
Further work is needed to understand what produces this correlation.  
There are several possibilities. The sources may be small enough to be  
affected by free-free absorption in the innermost regions of the nucleus, 
giving rise to an observed radio spectrum which peaks at frequencies above 
5\,GHz, as appears to be the case in the nearby galaxy NGC\,1052 
(Kameno et al.\ 2001; Vermeulen et al.\ 2003). Alternatively, young radio sources 
may be preferentially triggered in galaxies where sufficient gas is present in the 
nucleus to fuel them.  

Whatever the physics involved, it appears that the AT20G sample contains a distinct class of 
compact, flat-spectrum radio sources which lie in galaxies with (generally weak) 
emission-line AGN. Because of their flat or peaked radio spectra, such objects are more 
likely to be seen in samples selected at higher radio frequencies and so they have not 
been studied in a systematic way in the past. 

\section{Discussion} 

\subsection{Radio and optical properties of the AT20G-6dFGS sample}
We have measured three main radio parameters for galaxies in our sample: the radio 
{\it morphology}, radio {\it luminosity}\ and {\it spectral index}.  
We also have three key optical/infrared measurements: 
the K-band (2.2\,$\mu$m) {\it absolute magnitude}, which is a proxy for galaxy 
stellar mass, the WISE {\it mid-infrared colours}, which can reveal the presence of 
dust heated by star formation, or a radiatively-efficient AGN, and the 
{\it spectral class}\ (Aa, Aae or Ae), which is thought to reflect the 
efficiency of gas accretion onto the central black hole.  

Four of these parameters (radio luminosity, radio morphology, galaxy stellar mass 
and spectral class) have been examined in previous large studies of radio-source 
populations, using galaxy samples selected at frequencies of 1.4\,GHz or below. 
In terms of the interplay of these four properties, our results (which we discuss 
briefly below) are in broad agreement with earlier work. 

The other parameters (radio spectral index and WISE mid-infrared colours) are 
less well-studied.  The WISE data (Wright et al.\ 2010) are relatively new, but 
are rapidly coming into wide use for studies of galaxy evolution. 
Earlier studies of local radio-source populations have generally not had any 
radio spectral-index information which allowed them to distinguish between 
steep-spectrum and flat-spectrum sources, though a flux-limited multi-frequency 
catalogue of northern radio sources extending up to 5\,GHz has been published 
by Kimball \& Ivezic (2008). 

Table\,\ref{tab_frclass} summarizes the overall properties of the AT20G-6dFGS galaxies 
as a function of Fanaroff-Riley class (FR\,0, 1 and 2) and radio spectral index. 
We can draw several conclusions from this: 

\subsubsection{Radio morphology:}\ 
Only one-third of the AT20G-6dFGS sources (32\%) are associated with classical FR-1 and FR-2 
radio galaxies -- the other two-thirds appear to be relatively compact sources even at 
low radio frequencies (0.8-1\,GHz). 

The host galaxies of FR-2 radio sources are typically less massive than the hosts of FR-1 
galaxies (as measured by their infrared K-band luminosity), and the FR-2 galaxies also 
have a much higher fraction of HERG (Ae/AeB) spectra, in broad agreement with the results 
of Best \& Heckman (2012) and earlier studies (e.g. Hardcastle et al.\ 2004). 

The compact (FR-0) radio galaxies have typical properties which are intermediate between 
the FR-1 and FR-2 systems in both stellar mass and spectral class. The FR-0 class 
contains a mix of HERG and LERG systems, and seems likely to be a composite class which 
includes the early stages of both FR-1 and FR-2 radio galaxies, as well as some 
sources in which the core is brioghtened by relativistic beaming. 
We discuss this question  further in \S6.2 and \S6.3. 

Baum et al.\ (1998) found that GPS radio sources tend to have lower emission-line 
luminosities than CSS sources. The fraction of galaxies with strong optical emission 
lines is slightly higher for the candidate CSS (FR-0c) sources ($33\pm8$\%) than 
in the GPS (FR-0g) sources ($25\pm8$\%), but the difference is not statistically 
significant in this relatively small sample. 

\subsubsection{Radio luminosity}
By selecting our sample from the AT20G catalogue, which has a 
cutoff flux density of 40 mJy, we are by definition selecting the more luminous 
radio galaxies in the local universe.  The median 20\,GHz radio luminosity of the 
AT20G-6dFGS galaxies is just below 10$^{24}$\,W\,Hz$^{-1}$, i.e. close to the 
dividing line between classical FR-1 and FR-2 radio galaxies.  

The FR-2 radio galaxies have a significantly higher median radio luminosity at 20\,GHz 
(by almost an order of magnitude) than the FR-1 objects, even though the FR-1 sources 
generally lie in more massive galaxies.  
Our data support the finding of Ledlow and Owen (1996) that FR-2 radio sources 
are more powerful than FR-1 sources in galaxies of similar stellar mass.  
It is interesting that the `Ledlow-Owen' line derived for the total radio 
power at 1.4\,GHz also these objects reasonably well at 20\,GHz, with the 
majority of FR-2 sources (69\%) lying above the line and the majority of 
FR-1 sources (92\%) below.

\begin{figure}
\begin{center}
%
%
\hspace*{-0.5cm}
\includegraphics[width=8.5cm]{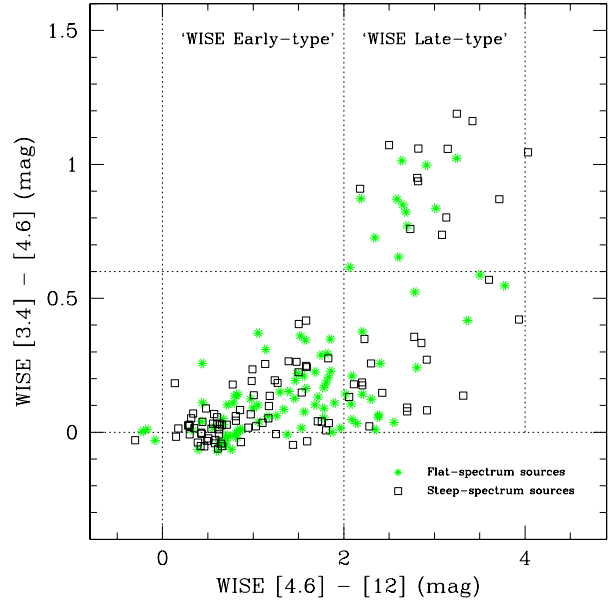}
\caption{WISE colour-colour plot for the AT20G-6dFGS galaxies with flat (green stars) 
and steep (black open squares) 1--20\,GHz radio spectral index.  }
\label{fig_wise_alpha}
\end{center}
\end{figure}

\subsubsection{Radio spectral index:} 

As discussed in \S3.4, roughly half of the AT20G-6dFGS galaxies 
have flat-spectrum radio sources and (as shown in Figure\, \ref{fig_rajan}) 
there is a close relationship between between radio morphology and spectral index, 
with flat-spectrum sources being more compact than steep-spectrum ones. 

Figure\,\ref{fig_wise_alpha}\ shows a WISE two-colour plot similar to Figure 9, 
but with now with separate symbols for steep-spectrum ($\alpha_1^{20}\leq-0.5$) 
and flat-spectrum ($\alpha_1^{20}>-0.5$) radio sources. 
We see a roughly equal mix of flat-spectrum and steep-spectrum radio sources  
in both the `WISE early-type' and `WISE late-type' galaxies, showing that 
the distribution of radio spectral index does not depend on the 
host-galaxy type.  
    
Table\,\ref{tab_frclass} also shows that the flat--spectrum and steep--spectrum radio 
sources in our sample are found in galaxies of similar median stellar mass. 
Both findings are consistent with a picture in which the flat-spectrum and steep-spectrum 
sources in the AT20G-6dFGS sample correspond to different stages of radio--galaxy evolution, 
rather than physically-distinct radio-source populations.  

\subsection{Relativistic beaming} 

De Zotti et al.\ (2010) note that radio surveys at frequencies of 5\,GHz and above 
are dominated by flat-spectrum sources (at least at higher flux densities), and 
that most of these flat-spectrum sources are likely to be brightened by relativistic beaming 
because they are observed along a line of sight which is very close to the jet axis. 
These beamed sources generally show rapid variability at radio wavelengths, and fall 
into two classes: Flat-spectrum radio quasars (FSRQs), which show strong optical emission 
lines with broad Balmer wings, and BL Lac objects (BLLs), in which the optical emission lines 
are weak or absent.  In unified models for AGN (Urry \& Padovani 1995; Jackson \& Wall 
1999), BLLs and FSRQs are assumed to be the beamed counterparts of FR-1 and FR-2 radio galaxies 
respectively. 

Most tests of the unified model have been carried out for populations of 
very luminous (radio power $>10^{25}$\,W\,Hz$^{-1}$) radio sources 
(e.g. Padovani \& Urry 1992; Cara \& Lister 2008), and the extent to which the much 
larger population of low-luminosity, flat-spectrum radio AGN is beamed remains 
unclear.  
The radio sources in the AT20G-6dFGS sample have typical radio 
luminosities of 10$^{23}$ to 10$^{25}$\,W\,Hz$^{-1}$ at 20\,GHz. 

Marcha at al.\ (2005) compared the emission-line luminosities of a 
sample of core-dominated sources (with core powers in the range $10^{21}$ to 
10$^{25}$\,W\,Hz$^{-1}$ at 5\,GHz) and a matched sample of FR-1 radio galaxies 
with similar redshift and 1.4\,GHz flux density.  They found that the core-dominated 
objects had significantly stronger optical emission lines than the matched FR-1 
sample, contradicting the predictions of a simple unified model.  These authors 
concluded that while a more sophisticated version of the unified model could 
fit their data, their results were equally consistent with a model in which 
relativistic beaming has little or no effect on the observed flux density of 
most low-luminosity flat-spectrum sources. 
Ekers et al.\ (1983) carried out a six-year program to monitor the 5\,GHz 
flux densities of a sample of low-luminosity compact radio sources, and also 
concluded that their results did not support simple relativistic beaming models. 

The AT20G-6dFGS galaxies in this paper are selected from within a 
sample of normal, nearby galaxies.  If all the compact flat-spectrum sources 
in this sample are brightened at 20\,GHz by relativistic beaming, then the effects 
at optical and mid-IR wavelengths must generally be quite subtle since only 
about 10\% of these sources lie near the WISE `blazar line' (Massaro et al.\ 2012) 
in Figure\,\ref{fig_wisespec} and most have mid-IR colours characteristic of 
normal galaxies rather than beamed AGN. 
  
A detailed test of beaming models for the AT20G-6dFGS sample would require 
high-quality radio imaging of these sources at VLBI resolution, and so is outside 
the scope of the present paper. 
We can however make some simple tests to estimate the fraction of AT20G-6dFGS sources 
which may have their observed 20\,GHz radio luminosity affected by relativistic 
beaming.  In the absence of detailed VLBI images of the galaxies in our sample, 
we consider (i) objects which are already identified in the literature as blazars, 
(ii) tests for radio variability, as used by Bonaldi et al.\ (2013) to distinguish 
between genuine High-Frequency Peakers and candidate blazars, and (iii) the evidence 
for a non-thermal contribution to the optical spectrum, as used by 
Marcha \& Caccianiga (2013) to identify low-luminosity BL Lacs.  

\subsubsection{Individual AT20G-6dFGS galaxies identified as beamed radio sources (blazars) } 
Four of the galaxies in our sample  (AT20G\,J024554-445939; J052223-072513; J130715-760245 
and J180957-455241) have compact (FR-0), flat-spectrum radio emission and show broad emission 
lines in their optical spectra.  All four can be regarded as flat-spectrum radio 
quasars (FSRQs), in which the high-frequency radio emission is probably beamed. 

A further three galaxies (AT20G\,J090802-095937; J151741-242220 and J231905-420648) are classified 
as radio-selected BL Lacs in the literature.  One of these, J151741-242220 (AP Lib; Carini et al.\ 1991) 
is a compact flat-spectrum source and a well-studied blazar which shows strong variability at both 
optical and radio wavelengths. 
The other two objects (J090802-095937, which is in the BL Lac sample studied by Nieppola et al.\ 2006, 
and J231905-420648, identified as a radio-selected BL Lac object by Roberts et al.\ 1998) are both 
associated with FR-1 radio galaxies in clusters, rather than flat-spectrum FR-0 sources.  
Whether these last two objects are beamed is therefore somewhat unclear. 

If we assume that all seven of these objects are blazars, we can estimate the {\it minimum}\ fraction 
of beamed radio sources in the local AT20G-6dFGS sample -- this is roughly 3.5\% (7/201) for the sample 
as a whole, and 6\% (5/83) for the sub-sample of compact flat-spectrum (FR-0g and FR-0u) sources. 

Since the classification of low-luminosity BL Lac objects in the literature is inhomogeneous 
and likely to be incomplete (particularly in the southern hemisphere; Massaro et al.\ 2009), 
we address the question of beaming by using two indicators which can be applied in a fairly 
uniform way to at least a subset of our sample.  These tests are described in the next two subsections, 
after which we derive the likely {\it maximum}\ beaming fraction in the AT20G-6dFGS sample.

\subsubsection{Radio-frequency variability of the compact, flat-spectrum (FR-0g and FR-0u) AT20G-6dFGS sources} 
Relativistically-beamed radio sources are expected to be strongly variable at GHz radio frequencies 
on timescales of months to years (e.g. Hovatta et al.\ 2007; Massardi et al.\ 2011b; Chen et al.\ 2013). 
In contrast, genuinely young GPS and CSS are not expected to be beamed, and typically vary in flux density 
by less than 10\% on timescales of a year or so (O'Dea 1998). 

Bonaldi et al.\ (2013) used measurements of radio variability to discriminate 
between candidate High Frequency Peakers (HFPs; i.e. very young radio galaxies) and candidate blazars. 
Over a 2--4 year timescale, they found a median variability index of 13.1\% at 5\,GHz and 15.0\% at 
9\,GHz for their sample of flat-spectrum sources, 95\% of which were blazars.  

We are not able to measure the variability of the AT20G-6dFGS sources at 20\,GHz, 
since at this frequency we only have data at a single epoch.   
We can, however, use 5 and 8\,GHz flux measurements from the ATPMN survey (McConnell et al.\ 2012) 
to estimate the variability at these frequencies.  Since the ATPMN survey was carried out with a different 
ATCA configuration which had significantly higher spatial resolution than the 5 and 8\,GHz 
observations carried out for the AT20G survey, we can only make this test for compact, flat spectrum 
sources which are unresolved in both surveys. 

The ATPMN observations were taken between 1992 November and 1994 March, and the AT20G observations between 
2004 November and 2007 May.  Thus the time interval between each pair of observations is between 10 and 15 years. 
We defined our 5 and 8 GHz variability sample as follows: 
\begin{enumerate}
\item 
We only included galaxies in the area of sky covered by the ATPMN survey, i.e. south of 
declination $-37^\circ$. 
\item
We restricted our analysis to compact (FR-0) sources with flat radio spectra from 1--20 GHz 
(i.e. $\alpha_1^{20}>-0.5$), to avoid introducing false `variability' in sources where 
extended 5/8 GHz structures included in the AT20G flux density measurement were partly 
resolved out by the smaller ATPMN beam. 
\end{enumerate}

This left a final sample of 26 flat-spectrum FR-0 galaxies, and for each of these objects we 
calculated a debiased variability index (as described by Sadler et al.\ 2006 and 
Massardi et al.\ 2011b) at both 5 and 8\,GHz.  

For the compact, flat-spectrum AT20G-6dFGS galaxies which were also observed in the ATPMN survey, 
we find a median variability index of $<6.4$\% at 5\,GHz and 6.5\% at 8\,GHz, over a time interval 
of 10-15 years.  Only 35\% of sources (9/26) varied by more than 10\% at 5\,GHz over this time interval, 
and 38\% (10/26) varied by more than 10\% at 8\,GHz.  Only three sources (J024326-561242, J121044-435437
and J194524-552049) varied by more than 20\% at either 5 or 8\,GHz over this 10-15 year interval. 

In practice, the fraction of sources with observed flux density variations of $>$10\% is likely to be 
an upper limit to the blazar fraction in our sample.  There are two reasons for this.  Firstly, the 
time interval over which we measure variability (10--15 years) is much longer than the 1--2 year 
interval for which O'Dea (1998) characterises the variability of young GPS and CSS and CSS 
radio galaxies, so it is plausible than even non-beamed compact sources could vary by up to 
15--20\% over this longer interval. Secondly, the mean flux ratio $<{\rm S}_{\rm AT20G}/{\rm S}_{\rm
ATPMN}>$ is $1.15\pm.05$ at 5\,GHz and $1.21\pm.05$ at 8\,GHz, implying that even though 
we have restricted our analysis to compact flat-spectrum sources, the AT20G flux measurements 
include some extended emission which is not seen in the higher-resolution ATPMN images. 
If so, the presence of this extended emission at one of our two epochs will produce a spurious rise 
in the measured variability level. 

Since we find that around two-thirds of the flat-spectrum FR-0 sources in our sample show a long-term 
variability in flux density of less than 10\% at 5--8 GHz, consistent with the behaviour expected for 
young radio galaxies rather than beamed radio sources, we estimate that the fraction of  
beamed sources in the combined FR-0g and FR-0u subsample is no higher than 35--40\%. 

A more detailed variability study of the AT20G-6dFGS sources, using a smaller time interval and 
better-matched array configuration, would allow more stringent limits to be placed on the likely 
beamed fraction. 

\subsubsection{Tests for a non-thermal contribution to the optical spectrum } 

Since the variability sample discussed above is relatively small, we make a second test using the 6dFGS optical spectra. 
Marcha \& Caccianiga (2013) note that it can be difficult to recognise the presence of low-luminosity BL Lac nuclei 
concealed within optically bright galaxies, and used the contrast across the Ca H$+$K break at 4000\,\AA\ to identify 
these weak beamed AGN.  Their selection criteria required that enough non-thermal flux from the AGN was present, 
in addition to the starlight of the host galaxy, to reduce the contrast across the 4000\,\AA\ break from 
the value of $\sim50\%$ seen in a normal early-type galaxy to $\leq40\%$. 

We applied the same test to the 6dFGS spectra of the FR-0 and FR-1 galaxies in our sample, including both 
flat-spectrum (FR-0g, FR-0u) and steep-spectrum (FR-0c, FR-1) sources.  The FR-2 sources were excluded from 
this test, since only a few of them have a suitable 6dFGS spectrum available.  

\begin{table}
\centering
\caption{Measurements of the Ca H+K break for the AT20G-6dFGS galaxies with weak or no optical 
emission lines (class Aa or Aae).  Galaxies with strong narrow (Ae) or broad (AeB) emission lines 
are also included for comparison.  }
\label{tab_opt}
\begin{tabular}{@{}lcccrr@{}}
\hline
               & \multicolumn{1}{c}{Flat-spectrum} & \multicolumn{2}{c}{Steep-spectrum}  \\
 Spectral type & \multicolumn{1}{c}{ FR-0u, 0g} & \multicolumn{1}{c}{ FR-0c} & \multicolumn{1}{c}{ FR-1} \\
\hline
  Aa/Aae, contrast $>40\%$    & 31    &   18  &  25 \\ 
  (normal galaxy)   & \\
  Aa/Aae, contrast $\leq40\%$ & 14    &    1  &   3 \\ 
  (BLL candidate)  & \\
  Ae                          &  6    &    7  &  -- \\
  AeB                         &  5    &    2  &  -- \\
  Total                       & 56    &   32  &  28 \\
  \hline
  {\bf Maximum beamed fraction:} & \\
  Galaxies with Aa/Aae spectra & 31\% (14/45) &  5\% (1/19)  &             \\ 
  All galaxies                 & 32\% (18/56) &  9\% (3/32)  & 11\% (3/28) \\
\hline
\end{tabular}
\end{table}

\begin{figure}
\begin{center}
%
%
\hspace*{-0.5cm}
\includegraphics[width=8.5cm]{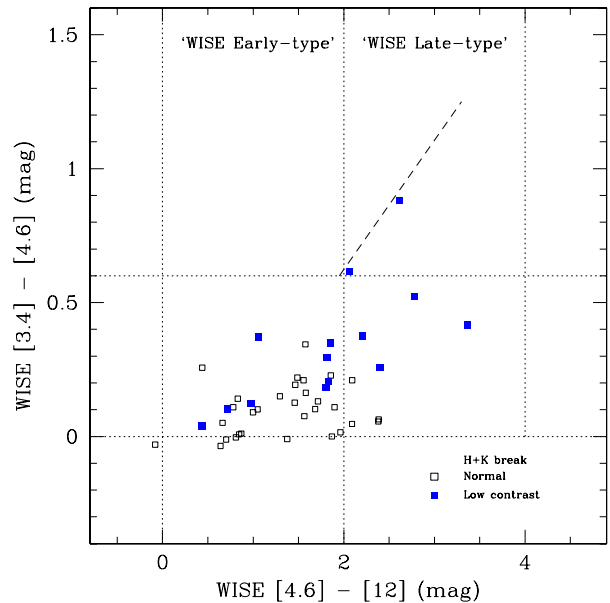}
\caption{WISE colour-colour plot for flat-spectrum FR-0 sources with 
weak or no optical emission lines (type Aa or Aae spectra). Open squares 
show galaxies with a `normal' 4000\,\AA\ break, and filled squares 
galaxies with a low-contrast break which may indicate the presence of 
a low-luminosity BL Lac nucleus. The diagonal dashed line show shows 
the location of the `WISE Blazar Strip' defined by Massaro et al.\ (2012). }
\label{fig_wise_split}
\end{center}
\end{figure}

Table\,\ref{tab_opt} summarizes the results of these measurements.  If we assume that all the galaxies with 
a low-contrast ($\leq40\%$) H$+$K break are low-luminosity BL Lacs (Marcha \& Caccianiga 2013), 
and that all the galaxies which show broad optical emission lines also contain beamed sources 
(Hardcastle et al.\ 1999), then the fraction of AT20G-6dFGS sources which are beamed is around 35\% 
for flat-spectrum sources and 11\% for steep-spectrum sources. 

In practice, it is likely that not all the galaxies with a `low-contrast' break are BL Lacs, since 
the presence of a young or intermediate-age stellar population will also reduce the size of the break 
and we have already shown (\S5.6.2) that around 30\% of the AT20G-6dFGS radio sources lie in galaxies 
with ongoing star formation.  As with the variability tests described in the \S6.2.2, the H$+$K break 
measurements therefore provide an upper limit to the fraction of AT20G-6dFGS galaxies in which the 
observed radio emission is strongly affected by beaming. 

Figure\,\ref{fig_wise_split} shows a WISE two-colour plot for the compact flat-spectrum sources for which 
we were able to measure the contrast across the H$+$K break. Filled squares show the galaxies with a 
low-contrast break, which could potentially host a low-luminosity BL Lac nucleus. 

As can be seen from Figure\,\ref{fig_wise_split} the WISE $[3.4]-[4.6]\,\mu$m colours of  the 
`low-contrast' galaxies are significantly offset from those of the galaxies with a `normal' H$+$K break. 
Two of the `low-contrast' galaxies, J091300-210320 and J151741-242220, have $[3.4]-[4.6]>0.6$ and lie 
close to the WISE blazar line of Massaro et al.\ (2012).  The remaining 12 `low-contrast' galaxies 
have a mean $[3.4]-[4.6]$ colour of 0.27$\pm$0.04\,mag., compared to a mean of 0.10$\pm$0.02\,mag 
for the 31 galaxies with a normal H$+$K break.  This shift in WISE $[3.4]-[4.6]$ colour is consistent 
with what might be expected if the `low-contrast' galaxies host a weak (beamed) BL Lac nucleus. 

\subsubsection{How many of the flat-spectrum AT20G-6dFGS sources are affected by relativistic beaming?}

The variability analysis in \S6.2.2 and the optical H$+$K break measurements presented in \S6.2.3 
both imply that no more than 30--35\% of the flat-spectrum sources in the AT20G-6dFGS sample are 
affected by relativistic beaming.  We showed in \S6.2.1 that the minimum beamed fraction is around 
6\%, so the overall fraction of relativistically-beamed sources within the flat-spectrum population 
is within the range 6--35\%. More detailed VLBI and variability studies are needed to refine this 
further.  Based on the evidence available so far, however, we conclude that at least two-thirds of 
the flat-spectrum sources in the AT20G-6dFGS sample are likely to be genuinely compact radio galaxies 
rather than low-power BL Lac objects.

\subsection{WISE colours and radio morphology}
\begin{figure}
\begin{center}
%
%
\hspace*{-0.5cm}
\includegraphics[width=8.5cm]{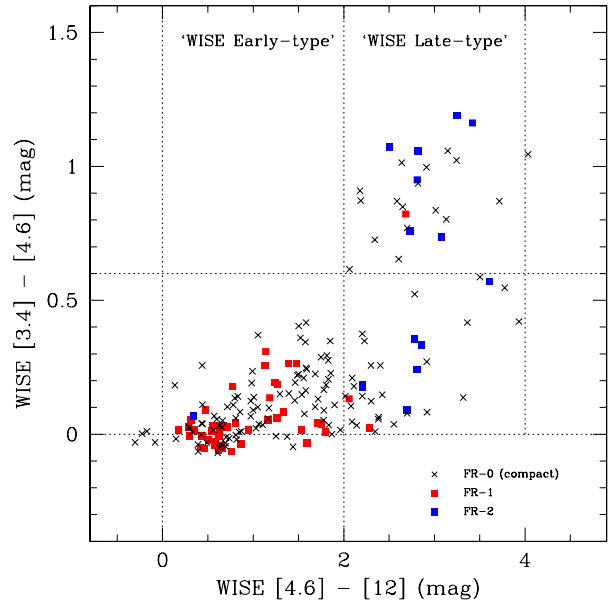}
\caption{WISE colour-colour plot for FR-1 (red squares), FR-2 (blue squares) and 
compact (FR-0, black crosses) radio galaxies in the 20 GHz AT20G-6dFGS sample.   }
\label{fig_wise_fr}
\end{center}
\end{figure}

Figure\,\ref{fig_wise_fr} shows a WISE colour-colour plot similar to that in 
Figures\,\ref{fig_wisespec} and \ref{fig_wise_alpha}, 
but now split by radio morphology into the FR-0, FR-1 and FR-2 classes defined 
in Table\,\ref{tab_frclass}. 

This plot shows a remarkable split in the mid-infrared colours of FR-1 and FR2 hosts, 
at $[4.6]-[12])\sim2.0$\,mag. 
For the 59 FR-1 and FR-2 galaxies with reliable WISE photometry available: 

\begin{itemize}
\item
93\% (41/44) of the FR-1 hosts have $[4.6]-[12]<2.0$\,mag, \\
\item
93\% (14/15) of the FR-2 hosts have $[4.6]-[12]\geq2.0$\,mag).  \\
\end{itemize}

\vspace*{-0.2cm}
In other words, there is a near-complete dichotomy between the host galaxies 
of our FR-1 and FR-2 radio sources, with FR-1 sources being found almost exclusively 
in WISE `early-type' galaxies and FR-2 sources in WISE `late-type' galaxies. 

This provides strong evidence that the host galaxies of FR-1 and FR-2 radio 
sources are drawn from different galaxy populations.  This is further supported 
by our earlier finding (\S6.1.1 and Table\,\ref{tab_frclass}) that the host galaxies 
of FR-1 radio sources are typically more massive than the FR-2 hosts. 
 
\subsection{An overall picture of the local radio-source population }

We now attempt to interpret the results presented so far in terms of an overall 
picture of the local radio-source population at 20\,GHz. 

We first assume that the distinction between high-excitation radio galaxies 
(HERGs), which have a radiatively-efficient accretion disk surrounding the black hole, 
and low-excitation galaxies (LERGs), in which accretion is radiatively inefficient, 
is an important one. 
 
\begin{figure}
%
%
\begin{center}
\includegraphics[width=8.5cm]{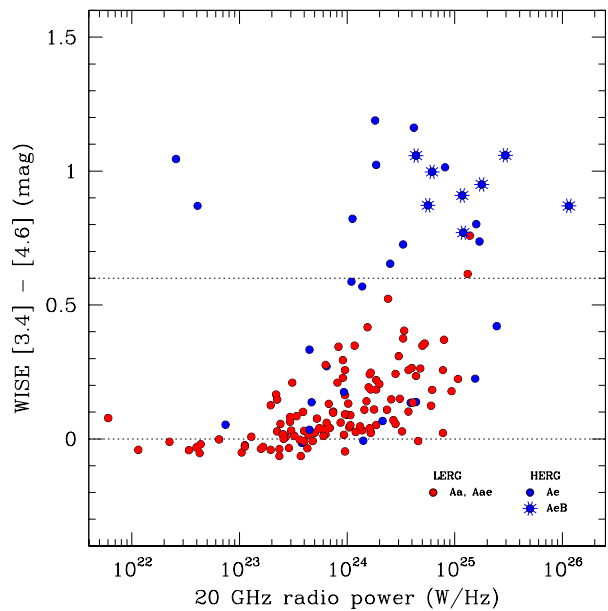}
\caption{WISE $[3.4]-[4.6]\,\mu$m colour, plotted against 20\,GHz radio power, 
for HERG (blue points) and LERG (red points) objects in the AT20G-6dFGS sample. 
The horizontal line at $[3.4] -[4.6]=0.6$\,mag.\ is the canonical dividing line between 
normal galaxies and radiatively-efficient AGN (Wright et al.\ 2010). } 
\label{wise_radio}
\end{center}
\end{figure}

Figure\,\ref{wise_radio}\ provides some support for this.  
As noted by Stern et al.\ (2012), the WISE $[3.4]-[4.6]\,\mu$m colour 
allows us to distinguish the power-law spectrum of a radiatively-efficient AGN 
from the blackbody stellar spectrum of a normal galaxy (which peaks near $\sim1.6\,\mu$m) 
in a way which is largely immune from the effects of dust extinction. 
Almost all the LERGs in our sample lie below the line at 
$[3.4] -[4.6]=0.6$\,mag.\ which separates normal galaxies from radiatively-efficient 
AGN (Wright et al.\ 2010), while most of the HERGs lie above this line.  
The clear separation of the blue and red points in Figure\,\ref{wise_radio} 
implies that the spectroscopic classification discussed in \S2.3 has allowed 
us to separate the two classes in a reliable way. 

Figure\,\ref{wise_radio} also shows that for low-excitation objects with 20\,GHz radio powers above about 
$10^{23}$\,W\,Hz$^{-1}$, the median WISE $[3.4]-[4.6]\,\mu$m colour moves to higher values 
as the 20\,GHz radio power increases.  Since there is no correlation between M$_{\rm K}$\ and 
$[3.4]-[4.6]$ colour for the AT20G-6dFGS galaxies, this implies that we may be seeing 
the weak signature of a radiative AGN in the more powerful LERGs in our sample. 

We can now split our sample into low-excitation and high-excitation populations, 
as summarized in Table\,\ref{tab_comp2}. 

\begin{table}
\centering
\caption{Summary of the radio and optical properties of low-excitation (LERG) and high-excitation 
(HERG) radio sources in the AT20G-6dFGS galaxy sample}
\begin{tabular}{@{}llll@{}}
\hline
\multicolumn{1}{c}{Class}   & \multicolumn{1}{l}{N} & \multicolumn{1}{c}{Host galaxies} & \multicolumn{1}{c}{Radio spectral} \\
                            &                       &                                   & \multicolumn{1}{c}{index $\alpha_1^{20}$}  \\ 
 \hline
\multicolumn{3}{l}{\bf Low-excitation populations} \\
FR-0, LERG  & 78 & WISE early-type and late-type   & 67\% flat, 33\% steep \\ 
FR-1, LERG  & 40 & Mainly WISE early-type        & Steep  \\
FR-2, LERG  &  5 & Mainly WISE late-type         & Steep  \\
&&& \\
\multicolumn{3}{l}{\bf High-excitation populations} \\
FR-0, HERG  & 26 & Mainly WISE late-type  & 50\% flat, 50\% steep  \\
FR-2, HERG  &  9 & WISE late-type     & Steep \\
\hline
\end{tabular}
\label{tab_comp2}
\flushleft
\end{table}

The {\it low-excitation systems} (LERGs) in Table\,\ref{tab_comp2} span the full range 
of radio morphologies (FR-0, FR-1 and FR-2).  The compact (FR-0) sources 
are seen across a wide range of host galaxy morphology, spanning both WISE 
`early-type' 
and WISE `late-type' galaxies. However, the radio sources which show extended 
radio emission at 1\,GHz (which we assume are the 
longer-lived counterparts of some of the FR-0 objects) are generally seen as FR-1 
radio galaxies if their host is a `WISE early-type' galaxy and (low-excitation)  
FR-2 radio galaxies if their host is a `WISE late-type' galaxy.  

This strongly suggests that some factor related to the host-galaxy morphology or 
large-scale environment, rather than the accretion rate alone, determines whether a 
young radio source undergoing radiatively-inefficient accretion evolves into an 
extended FR-1 radio galaxy or an FR-2 system. 

The {\it high-excitation systems} (HERGs) in Table\,\ref{tab_comp2}\ are a minority 
population in the AT20G-6dFGS 
sample, and their interpretation appears more straightforward.  Here we see only 
compact FR-0 and extended FR-2 systems, suggesting that compact radio sources with 
radiatively-efficient accretion disks are likely to evolve into FR-2 radio galaxies 
rather than FR-1 systems.

\subsection{Two radio-galaxy populations? }
The concept of a dual radio-source population in which the two populations undergo 
different cosmic evolution is a long-standing one, and is essential to explain the 
observed radio source counts (Longair 1966).  Several alternative models have 
been proposed for these two populations, as discussed in a recent review by De Zotti 
et al.\ (2010). 

Dunlop \& Peacock (1990) developed a model in which radio luminosity was the key parameter, 
with luminous radio sources undergoing rapid cosmic evolution while sources below some 
critical radio luminosity evolve more slowly or not at all.  
Wall (1980) identified the two populations with FR-1 and FR-2 radio galaxies, a model 
which was later extended by Jackson \& Wall (1999) to include 
flat-spectrum sources as the beamed counterparts of the steep-spectrum FR-1 and FR-2 sources. 
Willott et al.\ (2001) used emission-line strength rather than the FR-1/2 classification 
to define the two populations.  All these models were developed at a time when the local 
radio-source population was still poorly studied and the local radio luminosity function 
poorly determined, making it difficult to carry out a detailed comparison with the 
data. De Zotti et al.\ (2010) note that the possible dichotomies between evolutionary 
properties of low- versus high-luminosity and of flat- versus steep- spectrum AGN-powered 
radio sources remain an unresolved question. 

More recently, and using a much larger data set, Best \& Heckman (2012) have proposed 
a picture in which HERGs and LERGs constitute the two radio-source populations, 
with the observed redshift evolution of the radio luminosity function being driven 
at least partly by changes in the relative contribution of these two populations. 
In this picture, the key discriminant is the accretion rate onto the central black hole, 
with HERGs typically having accretion rates of 1--10 per cent of the Eddington rate 
and LERGs generally accreting at a rate below one per cent Eddington.  
These authors find that HERGs and LERGs show different rates of cosmic evolution at a 
fixed radio luminosity, with HERGs evolving strongly with redshift while LERGs show weak 
or no evolution.  They also identify LERGs with galaxies undergoing `radio-mode feedback', 
which acts to suppress further star formation (Croton et al.\ 2006), and HERGs with 
`quasar-mode' systems fuelled by the infall of cold gas, which may still have ongoing 
star formation.

Our results are in broad agreement with the Best \& Heckman (2012) picture. 
We find a clear distinction between the host galaxies of FR-1 and FR-2 galaxies in 
our sample, with the FR-1 objects lying almost exclusively in `WISE early-type' galaxies 
and the FR-2 objects in `WISE late-type' galaxies.  This implies that the host galaxy 
and its surrounding environment play an important role in the determining the overall 
properties of an individual radio galaxy, as has already been recognised by others 
(Heckman et al.\ 1986; Baum et al. 1995; Saripalli 2012; Ramos Almeida et al.\ 2012; 
Best \& Heckman 2012). 
We also see a fairly clear distinction between the optical spectra of the two classes, 
with 98\% of the FR-1 radio galaxies in our sample having LERG spectra while the 
majority (64\%) of our FR-2 radio galaxies are HERGs.  

As in earlier studies, however, we see a substantial population of FR-2 radio galaxies 
with LERG optical spectra. 
As Table\,\ref{tab_comp2} shows, the radio galaxies in our sample fall into 
{\it five}\ (rather than two or three) distinct sub-populations once we take into account 
the host-galaxy type, radio morphology and observed optical spectrum/accretion mode. 
If we also take into account that (as discussed in \S6.2) the flat-spectrum FR-0 
sources include a small subset of beamed objects, then the number of potential  
sub-populations is further increased. 

If the Best \& Heckman (2012) model is correct, then it should be possible to fit 
all these sub-populations into a dual-population model with one rapidly-evolving 
and one slowly-evolving (or non-evolving) radio-source population.  
To do this, we would need to know how each of the sub-populations in Table\,\ref{tab_comp2} 
evolves with redshift.  Such a test is clearly beyond the scope of the current paper, 
but would be interesting to carry out with a larger sample spanning a wider 
range in redshift. 

\section{Summary and future work} 
We have carried out the first detailed study of the high-frequency radio-source 
population in the local universe.  By selecting our sample at 20\,GHz, rather than 
at 1.4\,GHz as most other recent studies have done, we select galaxies based on the  
recent radio emission from their central regions rather than the longer-lived extended 
emission from their jets and lobes (which reflects the activity of the central black hole 
on longer timescales of up to 10$^7-10^8$ years). 

We now summarize the main results from this study, and highlight a number of areas 
where follow-up work would be particularly useful. Some follow-up work is already in 
progress, in particular a study of associated 21\,cm HI absorption in compact sources from 
the AT20G-6dFGS sample (Allison et al. 2012). \\

\noindent
(i) {\it Compact `FR-0' radio galaxies are the dominant source population in the AT20G-6dFGS 
galaxy sample}. 
These compact FR-0 sources are a heterogeneous population in terms of both host-galaxy 
morphology (75\% in early-type galaxies, 25\% in star-forming galaxies) and optical 
spectra (75\% LERG; 25\% HERG).  

Their observed properties are consistent with them being a mixture of several 
kinds of objects.  Some of the weaker flat-spectrum sources in our sample, like 
NGC\,1052 and NGC\,4594 (which have 20\,GHz radio power below 10$^{23}$\,W\,Hz$^{-1}$), 
are known to be less than a few parsecs in size (e.g.\ Slee et al.\ 1994; 
Sadler et al.\ 1995) and strongly affected by absorption. 
These low-luminosity sources could plausibly be maintained over much of the life of 
the host galaxy, and need not necessarily be young.  The FR-0 class is 
also likely to contain young Compact Steep-Spectrum (CSS) and Gigahertz-Peaked 
Spectrum (GPS) radio sources, as well as a minority population of beamed radio sources.  

Since the FR-0 objects make up around 70-75\% of the AT20G-6dFGS sources at all radio powers 
between about 10$^{22}$ and 10$^{26}$\,W\,Hz$^{-1}$ it seems unlikely that all of them 
can grow into long-lived FR-1 or FR-2 radio galaxies. 
Instead, it seems likely that many of the FR-0 sources have short duty cycles, 
so that they switch on and off regularly but are never `on' for long enough to 
drive large-scale jets and lobes (e.g. Conway 2002). 

A more detailed study of these objects would be very interesting, since they 
are likely to contain the largest and most complete sample of candidate CSS 
and GPS radio sources in the local universe.  

Such a study would ideally include detailed measurements of the radio spectrum  
over at least the range 1--20\,GHz, to determine the spectral shape and peak frequency; 
detailed 1.4\,GHz imaging to search for low surface-brightness jets and lobes 
which may lie below the limits of current large-area surveys; and high-resolution 
VLBI imaging to examine the parsec-scale source structure and look for evidence of 
long-term expansion or motion. 
While it appears that no more than 30--35\% of the compact flat-spectrum sources 
in our sample are brightened by relativistic beaming, it would clearly be valuable 
to quantify this more accurately. 
\\

\noindent
(ii) {\it We find a roughly-equal mixture of flat-spectrum and steep-spectrum 
radio galaxies in our sample}.  
The classical FR-1 and FR-2 radio galaxies in our sample have steep ($\alpha\leq-0.5$) 
radio spectral indices due to the extended low-frequency emission from 
their jets and lobes.  The compact (FR-0) 
sources include both flat-spectrum ($\alpha>-0.5$) and steep-spectrum sources, 
and we see no correlation between radio spectral index and radio luminosity or 
host-galaxy morphology.  We therefore conclude that flat-spectrum and steep-spectrum 
sources in our sample are not drawn from different parent-galaxy populations, 
but instead are likely to represent different evolutionary stages of the overall radio-galaxy 
population. 
\\

\noindent
(iii) {\it The FR-1 and FR-2 radio sources in our sample lie in different kinds of host galaxies.} 
Mid-infrared photometry from WISE (Wright et al. 2010; Donoso et al.\ 2012) 
shows that the host galaxies of the FR-1 and FR-2 galaxies in our sample are very different, 
with the FR-1 sources found almost exclusively in early-type galaxies and the FR-2 sources in late-type 
galaxies with dust and/or ongoing star formation. 
At 20\,GHz, we also find that FR-2 radio sources are more powerful (typically by 
almost an order of magnitude in 20\,GHz radio luminosity) than the FR-1 radio sources 
in galaxies of similar stellar mass.  The dividing line seen by Ledlow \& Owen 
(1996) at 1.4\,GHz also divides the FR-1 and FR-2 populations reasonably well 
at 20\,GHz if shifted appropriately in radio spectral index and (R-K) colour. 
\\

\noindent
(iv) {\it Galaxies with optical emission lines are more common in our 20\,GHz-selected 
sample than in a similar radio-galaxy sample selected at 1.4\,GHz. } 
HERG sources with strong optical emission lines make up $23\%$ of the 
AT20G-6dFGS sample, a significantly higher fraction than the 10-13\%  
seen in the lower-frequency 6dFGS-NVSS (Mauch \& Sadler 2007) sample. 
The HERG fraction is similar for flat-spectrum and steep-spectrum 
sources in the AT20G-6dFGS sample, 
but within the LERG population, weak optical emission-lines 
(class Aae) are much more common in galaxies which host flat-spectrum radio sources.  
The reason for the observed correlation between emission-line fraction and radio 
spectral index is not yet clear, and a more detailed study of the interstellar 
medium in these sources is needed to investigate whether the difference is 
mainly due to shock ionization of circumnuclear gas (Dopita et al.\ 1997), 
free-free absorption in the most compact sources (Kameno et al.\ 2005) 
or some other mechanism. 
\\

\noindent
(v) {\it Around 30\% of the AT20G-6dFGS radio galaxies are late-type, dusty or star-forming galaxies.  }
Although most of the radio-loud AGN in the AT20G-6dFGS sample are are hosted by massive early-type 
galaxies, both the catalogued optical morphology (where available) and the WISE infrared colours 
imply that $\sim30\%$ of the host galaxies are spirals or 
other dusty, late-type galaxies with some ongoing star-formation.  This is very similar to the 
$\sim35\%$ fraction of powerful 3CRR and 2Jy radio galaxies which were found by Dicken et al.\ (2012) 
to show ongoing star formation, based on the detection of infrared polycyclic aromatic hydrocarbon (PAH) 
spectral features. 
\\

Finally, we note that the increased continuum sensitivity now provided routinely by new wide-band correlators 
on large radio interferometers, including the Compact Array Broadband Backend correlator (CABB; Wilson et al.\ 2011) 
on the Australia Telescope Compact Array and the Wideband Interferometric Digital ARchitecture (WIDAR) 
correlator on the Jansky Very Large Array will make it far easier to carry out high-frequency and 
multi-frequency radio studies of large galaxy samples in the future. 
One goal of this paper is to give some idea of what these studies might discover, and how they 
might help to improve our understanding of the complex physical processes which shape the formation and 
evolution of radio galaxies. 

\section{Acknowledgements} 
We thank our colleagues in the AT20G and 6dFGS survey teams, who carried 
out the surveys which underpin much of the work presented here. 
We are also grateful to Tom Jarrett for several useful discussions on the WISE infrared data. 
We acknowledge financial support from the Australian Research Council through 
the award of an ARC Australian Professorial Fellowship to EMS.  
This research has made use of the NASA/IPAC Extragalactic Database (NED) which 
is operated by the Jet Propulsion Laboratory, California Institute of Technology, 
under contract with the National Aeronautics and Space Administration. 
\\

\appendix

\section{Notes on individual sources}

\begin{figure}
\begin{center}
\hspace*{-1.5cm}
\includegraphics[width=8.5cm,angle=270]{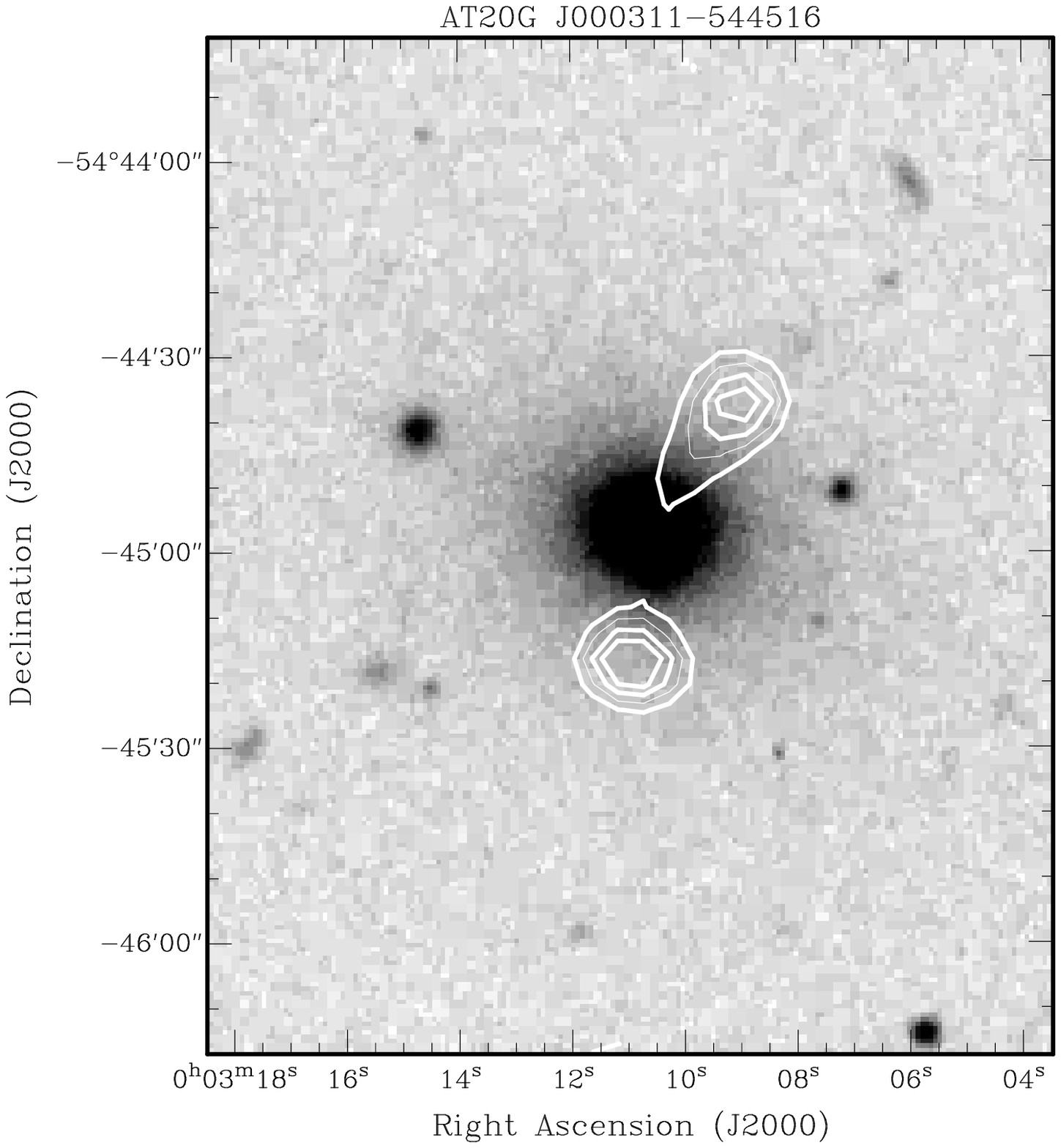}
\caption{The resolved AT20G source J000311-544516 
(PKS\,0000-550).  The greyscale is an optical B(J) image from 
Supercosmos and the white contours (at levels of 10, 15, 20 and 
25\,mJy/beam) show 20\,GHz emission from the AT20G followup image.  
The rms noise in the 20\,GHz AT20G snapshot image was 3.6\,mJy/beam.  
\label{fig_J0003}}
\end{center}
\end{figure}

\noindent
{\bf J000311-544516} (PKS\,0000-550) \\
The radio source is resolved into a 1 arcmin double at 20\,GHz (see Figure \ref{fig_J0003}), 
and the catalogued AT20G position corresponds to the southern hotspot. 
No core is seen in the 20\,GHz image, though a faint core is visible at 
5 and 8\,GHz. The source is only slightly resolved in the lower-frequency SUMSS 
image, with a largest angular size of 50\,arcsec. 

\noindent
{\bf J000413-525458} (PKS\,0001-531) \\
This resolved 20\,GHz source is the core of a wide ($\sim$6 arcmin) triple 
source at low frequency, and the 843\,MHz flux density listed in Table\,3 
is the sum of three SUMSS components. The AT20G images show a single source 
with an extension to the north-west in the direction of the northern low--frequency radio lobe. 
We tentatively classify this as an FR-1 radio galaxy based on the SUMSS image. 

\noindent
{\bf J001605-234352} (PKS\,0013-240, ESO\,473-G07) \\
This AT20G source is associated with the nucleus of a spiral galaxy.  
Slee et al.\ (1994) detected a parsec--scale radio core with an 8.4\,GHz 
flux density of 60\,mJy, suggesting that most of the observed 20\,GHz 
emission comes from a central AGN.  Allison et al.\ (2012) detected 
the 21cm HI line in absorption against the central continuum source. 

\noindent
{\bf J002736-540907} \\
The optical spectrum shows only weak absorption lines, suggesting this is 
a possible BL Lac object. 

\noindent
{\bf J002901-011341} (PKS\,0026-014) \\
The 2dFGRS spectrum of this galaxy (TGS\,819Z353) shows weak optical emission lines 
on a strong stellar continuum (spectral type Aae). 

\noindent
{\bf J003704-010907} (3C\,15, PKS\,0034-01) \\
The AT20G source is the core of a well-studied FR-2 radio galaxy in the 
2\,Jy sample of Morganti et al.\ (1993). 
The fractional linear polarization of 3.7\% measured at 20\,GHz is close 
to the NVSS value of 4.1\% at 1.4\,GHz. 

\noindent
{\bf J004613-420700} and {\bf J004622-420842} (PKS\,0043-42) \\
A well-studied FR-2 radio galaxy roughly $\sim$3\,arcmin in angular size. 
The hotspots are detected as two separate AT20G sources at 20\,GHz  
and are strongly polarized, with fractional linear polarizations of 7.4\% and 17.9\% 
for the northern and southern hotspots respectively. The galaxy nucleus  
is undetected at 20\,GHz, implying a core flux density below 10-15\,mJy. 
There is no 6dFGS spectrum, but Morganti et al. (1999) 
note that an optical spectrum shows only weak, low-ionization emission lines and
a stellar continuum typical of early-type galaxies. They remark that this
is an example of a powerful FR-2 radio galaxy with significantly
weaker emission lines that expected from the radio power--emission line
luminosity correlation.  We adopt the redshift of $z=0.116$ published by 
di Serego Alighieri et al.\ (1994).  

\noindent
{\bf J004733-251717} (NGC\,253) \\
A well-studied nearby spiral galaxy in the Sculptor group. The galaxy hosts a 
nuclear starburst as well as more extended star formation (Ulvestad \& Antonucci 1997), 
and has been detected as a gamma-ray source by the Fermi satellite (Abdo et al.\ 2010).
The radio source is extended in the AT20G image, and the listed AT20G flux density is 
a lower limit to the total value.  The flux density measured by WMAP at 23\,GHz is 1.3\,Jy. 
NGC\,253 is the closest galaxy (and lowest-luminosity 20\,GHz source) detected in the AT20G survey. 
Since no VLBI core component is detected (Sadler et al.\ 1995; Tingay 2004), 
the observed 20\,GHz emission probably arises from a compact nuclear starburst 
rather than an AGN.  

\noindent
{\bf J005734-012328} (PKS\,0055-01) \\
The AT20G source corresponds to the core of an FR-1 radio galaxy which 
belongs to the 2\,Jy sample of Morganti et al. (1993). 

\noindent
{\bf J012600-012041} (NGC\,547) \\
The AT20G point source is associated with NGC\,547, which lies in the cluster 
Abell 194 and has a close companion galaxy NGC\,545.  
The low-frequency emission is $\sim$10\,arcmin in extent 
and is resolved into several components in NVSS. The 1.4\,GHz 
flux density listed in Table\,3 is from Condon, Cotton \& Broderick (2002).    
Although this source is listed as a FR-2 radio galaxy in the 2\,Jy sample of 
Morganti et al. (1993), based on the original classification by O'Dea \& Owen (1985), 
more recent radio images suggest it should be reclassified as an FR-1 (Jackson 
et al.\ 2002). 

\noindent
{\bf J013357-362935} (NGC\,612) \\
This is a well-studied FR-2 radio galaxy in the 2\,Jy sample of Morganti et al. 
(1993). Ekers et al.\ (1978) noted that the host galaxy has a well-defined 
disk, and this was the first-known example of a powerful radio galaxy in disk 
galaxy. The 20\,GHz radio emission extends over at least 6\,arcmin, 
as shown in the mosaic image made by Burke--Spolaor et al.\ (2009). 
The 20\,GHz flux density listed in Table\,3 is also from Burke--Spolaor 
et al.\ (2009), and is a lower limit to the total value since the source is 
larger than the field imaged by these authors. 

\noindent 
{\bf J021645-474908} (ESO\,198-G01, PKS\.0214-48) \\
This is a resolved triple source at 20\,GHz, with a total extent of $\sim$1.5\,arcmin. 
The catalogued AT20G position corresponds to the core. This object is classified as an FR-1 
radio galaxy in the MS4 sample of Burgess \& Hunstead (2006). 

\noindent
{\bf J023137-204021} (PKS\,0229-208) \\
This AT20G source is the core of the radio galaxy PKS\,0229-208, which 
is an extended source (around 3 arcmin in extent) at low frequencies.  
The NVSS flux density listed in Table\,3 is for the central component 
(as also listed by Mauch \& Sadler 2007), and the extended emission at 
1.4\,GHz is split into at least six overlapping components in the NVSS 
catalogue. We tentatively classify this as an FR-1 radio galaxy on the 
basis of the NVSS image. 

\noindent
{\bf J024104-081520} (NGC\,1052) \\
This is a nearby and well-studied elliptical galaxy with a double-sided VLBI radio jet 
(Kameno et al.\ 2001).  Tingay et al.\ (2003) identify NGC\,1052 as a GPS radio source. 

\noindent
{\bf J024240-000046} (NGC\,1068, PKS\,0240-00) \\
NGC\,1068 is a nearby spiral galaxy with a well-studied active nucleus.  
The central radio source is resolved at 20\,GHz. 

\noindent
{\bf J031552-190644} (PMN\,J0315-1906)  \\
This galaxy has been studied in detail by Ledlow et al.\ (1998; 2001), 
who identify it as a rare example of an FR-1 radio source in a spiral host 
galaxy. Their 1.4\,GHz VLA image shows a faint jet extending south-west from 
the nucleus, and they also detect HI in absorption against the bright radio 
core. Only the core component is seen in the AT20G image. Keel et al.\ (2006) 
present HST and Chandra images of the host galaxy, noting that it has a 
very luminous bulge. The WISE infrared colours ($[3.4]-[4.6]=0.82$\,mag; 
$[4.6]-[12]=2.68$\,mag) imply that this is a late-type galaxy which 
probably hosts a radiatively-efficient accretion disk. 

\noindent
{\bf J034630-342246} (PKS\,0344-34) 
This AT20G source is the core of a radio galaxy in the MS4 catalogue of 
Burgess \& Hunstead (2006). The source has extended low-frequency 
emission, and is a 5\,arcmin double in the 843\,MHz SUMSS image.  
We classify this as an FR-2 radio galaxy on the basis of the 
low-frequency NVSS and SUMSS images. Raimann et al.\ (2005) note that the 
optical spectrum of this galaxy shows strong, narrow emission lines (class Ae). 

\noindent
{\bf J035145-274311} (PKS\,0349-27) \\
The listed AT20G source is the northern hotspot of an FR-2 radio 
galaxy (Morganti et al.\ 1993), and has a fractional polarization 
of 24.8\% at 20\,GHz.  The core and southern hotspot were not detected 
in the AT20G survey. The galaxy has an extended optical emission-line 
region with a disturbed velocity structure which may result from a 
recent collision or merger (Danziger et al.\ 1984, Grimberg et al.\ 1999). 

\noindent
{\bf J035257-683117} (PKS\,0352-686) \\ 
This source is core-dominated, and only slightly resolved in the 20\,GHz image.  
It is unresolved in the lower-frequency SUMSS image. 

\noindent
{\bf J042908-534940} (IC\,2082, PKS\,0427-53) \\ 
The AT20G source is the core of a head-tail or wide-angle tail FR-1 radio galaxy 
(Ekers 1969; McAdam et al.\ 1988) 
associated with a dumbbell galaxy (Carter et al.\ 1981) at the centre of a cluster.  
This galaxy is also in the 2\,Jy (Morganti et al.\ 1993) and MS4 (Burgess et al.\ 2006) 
radio samples, and Raimann et al.\ (2005) note that its optical spectrum shows weak 
optical emission lines (class Aae). 

\noindent
{\bf J043022-613201} (PKS\,0429-61) \\ 
The AT20G source is centred on the 6dFGS galaxy, but is significantly offset from 
the extended lower-frequency emission seen in the SUMSS image. 
The galaxy is classified as FR-1 by Burgess \& Hunstead (2006). 

\noindent
{\bf J045523-203409} (NGC\,1692) 
This is a compact ($\sim$30\,arcsec) double at 20\,GHz, with the emission 
probably arising from a pair of hotspots rather than a core. 
Tadhunter et al.\ (1993) note that no optical emission lines are detected in this
galaxy and the continuum appears typical of early-type galaxies.

\noindent
{\bf J050453-101451} (Arp\,187, PKS\,0502-10) \\ 
The radio source is associated with a disturbed or interacting galaxy listed in 
the Arp (1966) Atlas of Peculiar Galaxies. Although the 20\,GHz emission is flagged 
as extended in the AT20G catalogue, it is unresolved by NVSS at 1.4\,GHz. 

\noindent
{\bf J051949-454643, J051926-454554} and {\bf J052006-454745} (Pictor A, PKS\,0518-45) \\ 
A well-studied FR-2 radio galaxy, imaged in detail at the VLA by Perley, Roeser \& Meisenheimer 
(1997).  Three separately-catalogued AT20G sources correspond to the core and two hotspots. 
The 20\,GHz flux density listed in Tab;e 3 is from Burke-Spolaor et al.\ (2009).  
The optical spectrum (Eracleous \& Halpern 1994) shows strong, broad emission lines.  

\noindent
{\bf J054754-195805} (PKS\,0545-199) \\
The AT20G source is resolved at 20\,GHz, and the 1.4\,GHz emission is also 
extended. Zirbel \& Baum (1995) identify this an FR-1 radio galaxy.  

\noindent
{\bf J055049-314428} (ESO\,424-G27, PKS\,0548-317) \\
This is a compact ($\sim$30\arcsec separation) double in the AT20G image.  The catalogued 
AT20G position corresponds to one of two hotspots, rather than the core, and both hotspots 
lie just beyond the optical galaxy.  The lower-frequency NVSS source is only slightly 
extended, with an angular size of about 20\,arcsec in the NVSS catalogue. 

\noindent
{\bf J062143-524132} (PKS\,0620-52) \\
This source is core-dominated but slightly resolved at 20\,GHz.  The low-frequency 
emission seen in the SUMSS image is characteristic of a wide-angle tail (WAT) 
source, and the galaxy is classified as an FR-1 by Morganti et al.\ (1993). 

\noindent
{\bf J062620-534151} (ESO\,161-IG07, PKS\,0625-53) \\ 
This AT20G source is a resolved double at 20\,GHz, and is classified as an FR-1  
radio galaxy in the 2\,Jy (Morganti et al.\ 1993) and MS4 (Burgess et al.\ 2006) 
samples.  The optical ID is a dumbbell galaxy 
at the centre of a cluster (Lilly \& Prestage 1987; Gregorini et al.\ 1994).  
The catalogued AT20G position is offset by about 15\,arcsec from the optical 
galaxy, and probably corresponds to the southern hotspot of a compact double, rather than the core. 

\noindent
{\bf J062648-543214} (PKS\,0625-545) \\
This is a resolved triple source at 20\,GHz, about 1.5\,arcmin in extent.  
The AT20G catalogue position corresponds to the northern hotspot, 
but the core is also detected at 20\,GHz. 

\noindent
{\bf J062706-352916} (PKS\,0625-35) \\
This is the central galaxy of the cluster Abell\,3392, and is 
classified as an FR-1 radio galaxy by Morganti et al. (1993).

\noindent
{\bf J063631-202857} and {\bf J063633-204239} (PKS\,0634-20) \\
These two AT20G sources correspond to hotspots of a very extended ($\sim$15\,arcmin in 
angular size) FR-2 radio galaxy.  No core component is detected in the AT20G survey,  
The host galaxy is well-studied at both optical and radio wavelengths, and a detailed 
1.4\,GHz image was made at the VLA by Baum et al.\ (1988). This object is classified 
by Ishwara-Chandra \& Saikia (1999) as a Giant Radio Galaxy, with the overall projected 
size of the radio emission exceeding 1\,Mpc.  

\noindent
{\bf J065153-602158} (PKS\,0651-60) \\
The catalogued AT20G position appears to correspond to the northern hotspot of a $\sim$1\,arcmin 
double at 20\,GHz.  A second, fainter 20\,GHz source is seen at the position of the optical galaxy. 
The lower-frequency SUMSS emission extends over several arcmin, and we tentatively classify 
this as an FR-1 radio galaxy based on the low-frequency morphology. 

\noindent
{\bf J065359-415144} (PKS\,0652-417) \\
The 6dFGS DR3 catalogue lists the redshift of this galaxy as $z$=0.00, based on (foreground) 
Galactic nebular emission lines.  We have remeasured the correct redshift as $z$=0.0908, based 
on the position of the Ca H and K absorption lines in the 6dFGS spectrum. 
We tentatively classify this as an FR-1 radio galaxy on the basis of the 
843\,MHz SUMSS image, which shows extended jet-like emission. 

\noindent
{\bf J070240-284149} (NGC\,2325) \\
This is a nearby and well-studied elliptical galaxy with a dust lane (Veron-Cetty \& Veron 1988). 

\noindent
{\bf J070459-490459} (ESO\,207-G19) \\
The SUMSS image shows extended low-frequency emission, from which we classify this as 
an FR-1 radio galaxy. 

\noindent
{\bf J070914-360121} (PKS\,0707-35) \\
The AT20G source is the core of a radio galaxy which has an extended `double-double' structure 
at low frequencies.  The low-frequency emission has an FR-2 morphology extending over  
about 8\,arcmin on the sky.  The source has been studied in detail by Subrahmanyan et al.\ (1996), 
and is in the Ishwara-Chandra \& Saikia\ (1999) sample of giant radio sources with  
a projected size larger than 1\,Mpc. 

\noindent
{\bf J080537-005819} (PKS\,0803-00, 3C\,193) \\
The 20\,GHz emission extends over about 20-30\,arcsec.  The 1.4\,GHz VLA image 
made by Owen \& Ledlow (1997) shows that this is a wide-angle tail (WAT) galaxy 
in the cluster Abell\,623.  We classify this as an FR-1 radio galaxy based on the 
1.4\,GHz image. 

\noindent
{\bf J080852-102832} (PKS\,0806-10, 3C195) \\ 
The catalogued AT20G source is the southern hotspot of an FR-2 radio galaxy 
about 2.5\,arcmin in extent (Morganti et al.\ 1993).  The core and northern 
hotspot were not detected in the AT20G survey.  This galaxy is in the ``radio-excess 
IRAS galaxy'' sample of Buchanan et al.\ (2006), who note that the optical spectrum shows 
strong, narrow emission lines. 

\begin{figure}
\begin{center}
\hspace*{-1.5cm}
\includegraphics[width=8.5cm,angle=270]{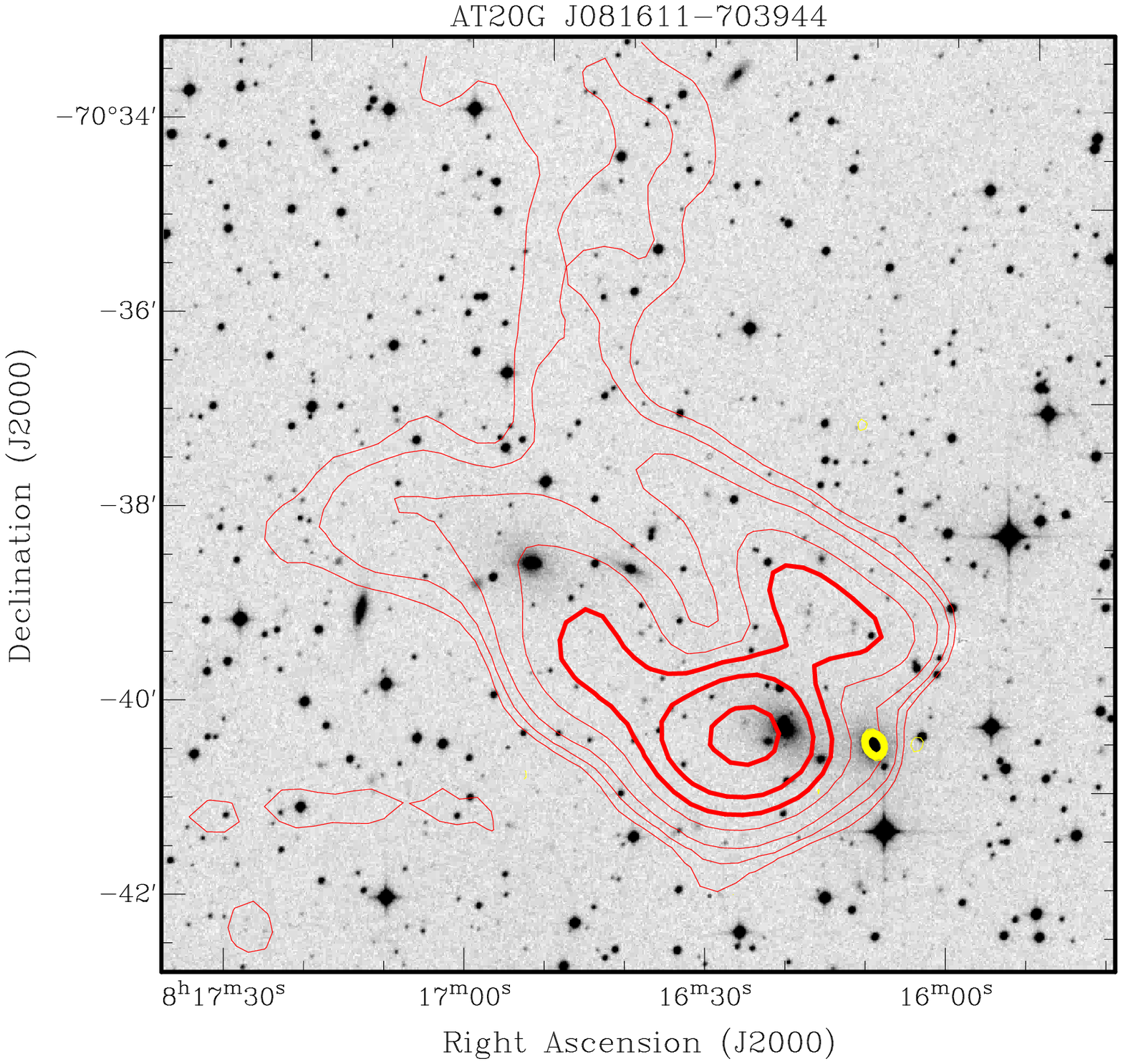}
\caption{The AT20G radio source J081611-703944, with yellow contours marking 
20\,GHz emission from the AT20G image and red contours the low-frequency 
emission from the 843\,MHz SUMSS image.  
\label{fig_j0816}}
\end{center}
\end{figure}

\noindent
{\bf  J081611-703944} (ESO\,060-IG02, PKS\,0816-705) \\
The AT20G source is centred on the 6dFGS galaxy g0816118-703945 at z=0.0333, 
but is significantly offset from the centroid of the low-frequency 
emission (see Figure\ref{fig_j0816}). 

\noindent
{\bf J084452-100059} (PMN\,J0844-1001) 
This galaxy has extended low-frequency emission in the 1.4\,GHz NVSS image, 
and we tentatively classify it as an FR-1 radio galaxy. 

\noindent
{\bf J090802-095937} (PMN\,J0908-1000 ) \\ 
This galaxy is listed as a BL Lac object by Nieppola et al.\ (2006), 
and was also detected as a ROSAT X-ray source (Bauer et al.\ 2000). 
The galaxy lies in an Abell cluster, and the NVSS image shows extended, 
complex low-frequency emission.  O'Dea \& Owen (1985) classify this as a 
narrow-angle tail radio source based on a higher-resolution 1.4\, GHz 
VLA image. We have classified it as an FR-1 radio galaxy based on 
the extended low-frequency emission seen in the NVSS image. 

\noindent
{\bf J090825-093332} (PMN\,J0908-0933) \\
The 6dFGS spectrum looks almost featureless and is classified as poor-quality. 
The listed redshift is taken from Christlein \& Zabludoff (2003). 
We tentatively classify this as an FR-1 radio galaxy based on the 
morphology of the low-frequency emission seen in the NVSS image. 

\noindent
{\bf J091300-210320} (MRC\,0910-208) \\ 
The 6dFGS spectrum shows weak absorption lines, and the galaxy is classified 
as a possible BL Lac object by Massaro et al.\ (2009). 

\noindent
{\bf J091805-120532} (PKS\,0915-11, 3C\,218, Hydra A) \\
This is a well-studied FR-1 radio galaxy in the Hydra cluster, 
and belongs to the 2-Jy sample of Morganti et al.\ (1993). 
The AT20G source is a compact double, probably corresponding 
to two inner hotspots rather than a core and jet. 

\noindent
{\bf J094110-120451} (MRC\,0938-118) \\
The AT20G image shows a resolved triple source, about 2 arcmin in extent. 
The catalogued AT20G position corresponds to the northern hotspot, but the 
core is also detected.  This source is a resolved double in the 1.4\,GHz NVSS image, 
and the 1.4\,GHz flux density listed in Table\,3 is the sum of the two NVSS components. 
We tentatively classify this as an FR-2 radio galaxy on the basis of the NVSS morphology. 

\noindent
{\bf J094409-015116} (PMN\,J0944-0151) \\
Buchanan et al.\ (2006) include this object in their sample of ``radio--excess IRAS galaxies''. 
Their optical spectrum shows strong, narrow emission lines superimposed on a stellar continuum 
with Balmer absorption features typical of post-starburst systems. 

\noindent
{\bf J105533-283134} (PKS\,1053-282) \\
This source appears to have a complex, diffuse structure at 20\,GHz, which is poorly imaged by 
the AT20G snapshot observation.  The source is less than 30\,arcsec in extent in the 
1.4\,GHz NVSS image. Slee et al.\ (1994) detected a parsec-scale radio core with an 
inverted radio spectrum at 2--8\,GHz. 

\noindent
{\bf J110957-373220} (NGC\,3557) \\
This is a well-studied nearby elliptical galaxy which hosts a FR-1 radio source 
(Birkinshaw \& Davies 1985). 

\noindent
{\bf J112119-001316} (PKS\,1118+000) \\  
The low-frequency NVSS emission is extended and offset from the AT20G emission, 
and the higher-resolution FIRST image shows that this is a wide-angle tail 
(WAT) radio galaxy with a bright core and complex extended structure at 1.4\,GHz. 

\noindent
{\bf J113305-040046} (PKS\,1130-037) \\
This AT20G source is the core of a radio galaxy which is extended at low frequencies 
and spans almost 10\,arcmin on the sky in the NVSS image. We tentatively classify this 
as an FR-1 radio galaxy based on the 1.4\,GHz NVSS image. 

\noindent
{\bf J122343-423532} (PKS\,1221-42) \\
The redshift of z=0.0266 listed in the 6dFGS catalogue is incorrect and the correct redshift 
is z=0.1706 (Simpson et al.\ 1993). The host galaxy and its stellar population have 
been studied in detail by Johnston et al.\ (2005), who note that this is a young compact 
steep-spectrum (CSS) radio source with double lobes located well within the optical galaxy. 

\noindent
{\bf J123959-113721} (NGC\,4594) \\
This is a well-studied nearby spiral (the `Sombrero Galaxy').  Sadler et al.\ (1995) 
found that at 8.4 GHz, the central radio source was smaller than 0.03\,arcsec 
(i.e. $<3$\,parsec) in angular size. 

\noindent
{\bf J124849-411840} (NGC\,4696 = PKS\,1245-41) \\
This is the central galaxy of the Centaurus cluster, recently studied in detail by 
Taylor et al.\ (2006).  The radio emission is resolved at 20\,GHz, but not in the 
lower-frequency SUMSS image. 

\noindent
{\bf J125438-123255} (NGC\,4783 = PKS\,1251-12, 3C\,278) \\
This is an extended AT20G source associated with an FR-1 radio galaxy (Morganti et al.\ 1993). 
The optical ID is one of a close pair of elliptical galaxies, NGC\,4782 and 4783.  
Baum et al.\ (1988) identify NGC\,4782 as the optical ID, but both the AT20G 
catalogue position and the core position measured by Ricci et al.\ (2006) are 
closer to NGC\,4783, so we tentatively identify this galaxy as the optical counterpart 
of the AT20G source. 

\noindent
{\bf J130100-322628} (ESO\,443-G24 = PKS\,1258-321) \\
This object has extended emission at 1.4\,GHz in the NVSS image, and is classified as a low-power 
FR-1 radio galaxy by Marshall et al.\ (2005). 

\noindent
{\bf J130527-492804} (NGC\,4945) \\
This is a nearby, well-studied spiral galaxy with an active nucleus. 

\noindent
{\bf J130841-242259} (IRAS\,13059-2407) \\
This edge-on disk galaxy (tentatively classified as an Sc spiral in 
the NASA Extragalactic Database) has the lowest infrared K-band 
luminosity of the 202 galaxies in our AT20G-6dFGS sample. 
It is also classified as a `radio-excess IRAS galaxy' by 
Drake et al.\ (2003).  Allison et al.\ (2012; 2013) recently detected 
associated 21\,cm HI absorption against the central radio source. 

\noindent
{\bf J131124-442240} (PKS\,1308-441) \\ 
The 843\,MHz SUMSS image shows low-frequency emission extending over 
more than 15\, arcmin on the sky.  Tingay (1997) notes that this is 
a giant radio galaxy with a morphology intermediate 
between FR-1 and FR-2.  We tentatively classify it as FR-1 on the 
basis of the SUMSS image. 

\noindent
{\bf J131931-123925} (NGC\,5077) \\
This elliptical galaxy has an extended disk of ionised gas along its minor axis which 
has been studied in detail by Bertola et al.\ (1991), who note that this galaxy also 
contains a substantial amount of neutral hydrogen (see also Serra \& Oosterloo 2010). 

\noindent
{\bf J131949-272437} (NGC\,5078) \\
This is a nearby spiral galaxy with a prominent dust lane. 

\noindent
{\bf J132112-434216} (NGC\,5090) \\
The catalogued AT20G source is the core of an FR-1 galaxy with very extended 
low-frequency radio emission (Morganti et al.\ 1993). NGC\,5090 has a close 
(probably interacting) spiral companion, NGC\,5091 (Smith \& Bicknell 1986).   

\noindent
{\bf J132527-430108} (NGC\,5128, Centaurus A) \\
The AT20G source is the core of the well-studied nearby radio galaxy NGC\,5128. 
The radio luminosity used in this paper was calculated using the total 20\,GHz flux 
density of 28.350\,Jy listed in the AT20G catalogue (Murphy et al.\ 2010). 
Israel et al.\ (2008) measured a higher flux density of 112$\pm$13\,Jy at 23\,GHz 
using WMAP images which include the emission from the outer radio lobes. 

\noindent
{\bf J133639-335756} (IC\,4296) \\ 
The catalogued AT20G source is the core of a well-studied FR-1 radio galaxy 
which has very extended emission at low frequencies (Killeen et al.\ 1986).  
The listed 1.4\,GHz flux density is the sum of four NVSS components, and 
the 843\,MHz flux density is from Jones \& McAdam (1992). 

\noindent
{\bf J134624-375816} (ESO\,325-G16, PKS\,1343-377) \\
This AT20G source has complex, highly extended low-frequency emission, 
and appears to be the core of a head-tail radio galaxy 
in the cluster Abell\,3570.  

\noindent
{\bf J141949-192825} (PKS\,1417-19) 
The 20\,GHz emission is centred on the 6dFGS galaxy, while the lower-frequency 
NVSS source has a centroid significantly offset from the optical position. 

\noindent
{\bf J145509-365508} (PKS\,1452-367) \\
The galaxy has extended low-frequency emission on scales of 3--4\,arcmin 
in the SUMSS and NVSS images, and we tentatively classify it as an FR-1 
radio galaxy based on the low-frequency morphology. 

\noindent
{\bf J145924-16413} (NGC\,5793) \\
This AT20G source is the core of a spiral galaxy.  Gardner et al.\ (1992) detected 
a VLBI core less than 0.03 arcsec in size, with HI seen in absorption against the 
core. Hagiwara et al.\ (1997) detected an H$_2$O megamaser in NGC\,5793. 
The HI absorption system has been studied in detail with the VLBA by 
Pihlstr\"om, et al.\ (2000). 

\noindent
{\bf J151741-242220} (AP Lib) \\ 
This well-studied BL Lac object is the most luminous AT20G source in the 
local ($z<0.1$) universe, and is known to be variable at optical as well 
as radio wavelengths (Carini et al.\ 1991). 

\noindent
{\bf J164416-771548} (PKS\,1637-77) \\
The cataloged AT20G source is the core of an FR-2 radio galaxy with low-frequency 
radio emission extending over at least 4-5 arcmin in the SUMSS image. 
There is no 6dFGS spectrum, but Tadhunter et al.\ (1993) note that the optical 
spectrum shows strong emission lines.  

\noindent
{\bf J172019-005851} and {\bf J172034-005824} (PKS\,1717-00, 3C253) \\
Two hotspots of this FR-2 radio galaxy are detected as separate AT20G sources 
3.6\,arcmin apart. The galaxy core was not detected in the AT20G survey. 

\noindent
{\bf J172341-650036} (NGC\,6328) \\
This well-studied galaxy is a strong, compact radio source at 20\,GHz, and has 
been identified by Tingay (1997) as one of the closest GPS radio galaxies. 
The galaxy shows faint spiral structure in the optical and has been detected 
in HI by Veron-Cetty et al.\ (1995), who suggest that this object is the result 
of a recent merger involving a gas-rich spiral galaxy. 

\noindent
{\bf J173722-563630} and {\bf J173742-563246} (PKS\,1733-56) \\
FR-2 radio galaxy with the two hotspots detected as separate AT20G 
sources 4.7 arcmin apart. The galaxy core was not detected in the AT20G 
survey. 

\noindent
{\bf J180207-471930} (MRC\,1758-473) \\
This galaxy has extended radio emission in the 843\,MHz 
SUMSS image, and we tentatively classify it as an FR-1 radio galaxy. 

\noindent
{\bf J180712-701234} (PKS\,1801-702) \\
This source is a compact double at 20\,GHz, with the two components separated by 
about 20\,arcsec.  It appears to be unresolved in the lower-frequency SUMSS image.  

\noindent
{\bf J180957-455241}  (PKS\,1806-458) \\ 
Kollgaard et al.\ (1995) found that this source had a GHz-peaked spectrum with a maximum around 4\,GHz, 
and remarked that it was much brighter than the vast majority of other GHz-peaked 
sources and should be studied further.  
More recently Massardi et al.\ (2011b) made multi-frequency observations at several epochs, 
which showed the radio spectrum peaking above 10\,GHz.  The high radio luminosity of this 
source, together with the presence of broad emission lines in the optical spectrum, suggest that 
(as discussed in\S6.2.1 above) this is probably a blazar seen in a flaring state rather than 
a genuine GPS radio galaxy. 

\noindent
{\bf J181857-550815} (PMN\,J1818-5508) \\
This 20\,GHz source is the core of a radio galaxy which is a resolved double in 
the 843\,MHz SUMSS image. We tentatively classify this as an FR-1 radio galaxy.  

\noindent
{\bf J181934-634548} (PKS 1814-63) \\
This is a very strong compact source which is a member of the Morganti 
et al.\ (1993) 2\,Jy sample and has classified as a CSS radio source by 
Tzioumis et al.\ (2002).  The galaxy has been studied in detail by Morganti 
et al.\ (2011), who note that it is a rare example of a powerful radio 
AGN hosted by a disk galaxy. In the optical, there is a very bright 
foreground star whose light contaminates the 6dFGS spectrum. 

\noindent
{\bf J184314-483622}  (PKS\,1839-48) \\
The AT20G source is the core of an FR-1 radio galaxy in the 2\,Jy 
sample of Morganti et al.\ (1993). 

\noindent
{\bf J191457-255202} (PMN\,J1914-2552) \\
The NVSS image shows diffuse emission extending up to 5\,arcmin 
to the west of the galaxy core, suggesting that this may be a 
head-tail radio source in a cluster. 

\noindent
{\bf J192817-293145} (ESO\,460-G04, MRC\,1925-296) \\
The NVSS image shows extended low-frequency emission, with a morphology 
suggesting that this is a wide-angle tail (WAT) radio galaxy in a cluster. 

\noindent
{\bf J195817-550923} (PKS\,1954-55) \\
This source is a resolved double in the AT20G image at 20\,GHz, with the 
components separated by about 1 arcmin.  A detailed ATCA image made by 
Morganti et al.\ (1999) at 8\,GHz shows a core together with complex, 
extended jet structures which are not well imaged by the limited uv coverage 
of the ATCA snapshot observations. This is one of the most highly-polarized sources 
in the AT20G catalogue, with a fractional linear polarisation of 43.5\% at 20\,GHz 
(Murphy et al.\ 2010). 

\noindent
{\bf J200954-482246} (NGC\,6868) \\
This is a well-studied dust-lane elliptical galaxy in a group. A detailed X-ray study has 
been carried out by Machacek et al.\ (2010). 

\noindent
{\bf J203444-354857} (ESO\,400-IG40) \\
The AT20G source is associated with the northern member of 
a close pair of galaxies in the cluster Abell\,3695. 

\noindent
{\bf J204552-510627} (ESO\,234-G68) \\
The SUMSS image shows very extended low-frequency radio emission, from which 
we tentatively classify this object as an FR-1 radio galaxy. 

\noindent
{\bf J205202-570407} (IC\,5063) \\
This is a well-known dust-lane S0 galaxy recently studied in detail by Morganti et al.\ (2013). 

\noindent
{\bf J205306-162007} (IC\,1335, PKS\,2050-16) \\
The NVSS image shows extended low-frequency radio emission with a 
total angular total extent of at least 10\,arcmin.  We tentatively classify this 
object as an FR-1 radio galaxy based on the NVSS morphology. 

\noindent
{\bf J205603-195646}  (PKS\,2053-20) \\
This source is a compact double, about 0.5\,arcmin in extent, at 20\,GHz. 
The two components appear to be hotspots, and no core is detected. 
The catalogued AT20G position corresponds to the southern hotspot. 
The source is barely resolved in the lower-frequency NVSS image.  

\noindent
{\bf J205754-662919} (ESO\,106-IG15) \\
The SUMSS image shows low-frequency radio emission extending over 
at least 5\,arcmin on the sky, from which we tentatively classify 
this object as an FR-1 radio galaxy.

\noindent
{\bf J212222-560014} (PMN\,J2122-5600) \\
The AT20G source is offset by more than 40\,arcsec from the 
centroid of the extended low-frequency emission seen in the SUMSS 
image.  This offset, together with the morphology of the low-frequency 
emission, suggests this is probably a head-tail radio source 
in a cluster. As noted in \S3.4.1 of this paper, Hancock 
et al.\ (2010) find that the compact AT20G source has a radio 
spectrum peaking above 40\,GHz, suggesting that this may be 
a rare example of a recently-restarted radio galaxy. 

\noindent
{\bf J213133-383703} (NGC\,7075) \\
This galaxy has extended low-frequency emission in the NVSS and 
SUMSS images, and is tentatively classified as an FR-1 radio 
galaxy on the basis of the low-frequency morphology. 

\noindent
{\bf J213741-143241} (PKS\,2135-13)  \\ 
The catalogued AT20G source is a hotspot of the well-studied FR-2 radio galaxy 
PKS\,2135-13 (see Table\,\ref{tab_mult}).  
The core is also detected at 20\,GHz, and the 6dFGS spectrum shows strong, 
broad optical emission lines. 

\noindent
{\bf J214824-571351} (PMN\,J2148-5714) \\
This source has extended, asymmetric radio emission at 843\,MHz, 
and may be a head-tail radio source in a cluster. 

\noindent
{\bf J215129-552013} (PKS\,2148-555) \\
The catalogued AT20G source is the core of an extended FR-1 radio 
galaxy, with low-frequency emission extending over at least 12\,arcmin 
in the 843\,MHz SUMSS image. The AT20G 20\,GHz image shows extended emission 
to the south-west of the core, possibly originating in a jet. Raimann et al.\ 
(2005) note that the host galaxy has an absorption-line (Aa) optical spectrum. 

\noindent
{\bf J215706-694123} (ESO\,075-G041, PKS\,2153-69)  \\
This is a very bright (37\,Jy) double source in SUMSS, and is classified 
as an FR-2 radio galaxy by Morganti et al.\ (1993). As noted by Ricci et al. 
(2004), both the core and a fainter southern hotspot are seen in the 
AT20G 20\,GHz image. 

\noindent
{\bf J220113-374654} (PKS\,2158-30) \\ 
The catalogued AT20G source is a hotspot of the FR-2 radio galaxy PKS\,2158-30 
(see Table\,\ref{tab_mult}).  The core is not detected in the AT20G image. 

\noindent
{\bf J220916-471000} (NGC\,7213) \\
This is a bright, nearby early-type spiral galaxy which has recently been 
studied in detail at both radio and X-ray wavelengths by Bell et al.\ (2011). 

\noindent
{\bf J224559-173724} (PKS\,2243-179) \\ 
Although not flagged as extended in the AT20G catalogue, this source appears extended or 
double in the 20\,GHz image.  

\noindent
{\bf J225710-362744} (IC\,1459) \\ 
This is a nearby and well-studied giant elliptical galaxy which has been identified 
by Tingay et al.\ (2003) as one of the closest GPS radio galaxies. Oosterloo et al.\ (2007) 
note that IC\,1459 lies in a gas-rich galaxy group and has HI tidal tails extending over 
at least 200\,kpc on the sky.  Franx \& Illingworth (1988) discovered a fast counter-rotating 
stellar core in IC\,1459, and Cappellari et al.\ (2002) have studied the kinematics of this 
galaxy in detail.  

\noindent
{\bf J231905-420648} (PKS\,2316-423) \\
This galaxy is a bright X-ray source (Crawford \& Fabian 1994), and was 
identified as a radio-selected BL Lac object by Roberts et al. 
(1998).  The SUMSS image shows extended low-frequency emission 
with a complex structure. 

\noindent
{\bf J231915-533159}  (PKS\,2316-538) \\
This galaxy has extended low-frequency emission in the 
SUMSS image, and is tentatively classified as an FR-1 radio 
galaxy on the basis of the low-frequency morphology. 

\noindent
{\bf J234205-160840}  (PKS\,2339-164) \\  
The NVSS image shows complex, extended emission at low-frequency, 
and the NVSS morphology suggests that this is a wide-angle tail 
radio galaxy in a cluster. 

\noindent
{\bf J235156-010909} (PKS\,2349-01) \\
The AT20G source is a resolved double roughly 15-20\,arcsec in extent. 
The low-frequency emission is barely resolved in NVSS, but the 
higher-resolution FIRST image at 1.4 GHz shows a compact double 
with similar morphology to the AT20G image. 

\end{document}


\setcounter{table}{2}
\clearpage\begin{table*}
\vspace*{-2cm}
\begin{minipage}{175mm}
\caption{6dFGS galaxies detected as AT20G radio sources. The meanings of the columns are defined in \S2.5 of the paper.   } 
\begin{tabular}{lllrrrrrrrrrrrrcll}
\hline
(1) & \multicolumn{2}{c}{(2)} & (3) & 
(4) & (5) & (6) & 
(7) & (8) & (9) & 
(10) & \multicolumn{1}{c}{(11)} & (12) & (13) & (14) & (15) & (16) & (17) \\ 
\multicolumn{1}{c}{AT20G name} & \multicolumn{2}{c}{AT20G position} & \multicolumn{1}{c}{S$_{20}$} & $\pm$
& \multicolumn{1}{c}{S$_{8}$} & $\pm$ &\multicolumn{1}{c}{S$_{5}$} & $\pm$ & \multicolumn{1}{c}{NVSS} & 
 \multicolumn{1}{c}{SUMSS} & \multicolumn{1}{c}{6dFGS name} & Offset & \multicolumn{1}{c}{K$_{\rm tot}$} & 
\multicolumn{1}{c}{Redshift} & Q & Class & \multicolumn{1}{l}{Notes} \\
 & \multicolumn{2}{c}{(J2000)} & \multicolumn{2}{c}{(mJy)} & \multicolumn{2}{c}{(mJy)} & \multicolumn{2}{c}{(mJy)} & 
\multicolumn{1}{c}{(mJy)} & \multicolumn{1}{c}{(mJy)} & \multicolumn{2}{r}{(arcsec)} & (mag) &  & & &  \\  
\hline
\multicolumn{15}{l}{(a)6dFGS galaxies identified with AT20G point sources} \\
J000922-513011 & 00 09 22.06 & -51 30 11.9 &  49 &  3 &  59 &  3 &  56 &  3 &    ... &   45.1 & g0009220-513012 & 0.5 & 12.30 & 0.1170 & 4 & Aa  &   \\
J000935-321637 & 00 09 35.67 & -32 16 37.3 & 221 & 10 & 275 & 14 & 313 & 16 &  521.6 &  558.8 & g0009355-321637 & 1.5 &  9.55 & 0.0255 & 4 & Aae & IC\,1531 \\ 
J001215-384618 & 00 12 15.13 & -38 46 18.8 &  80 &  4 & 105 &  5 & 103 &  5 &   65.4 &   52.2 & g0012152-384619 & 0.5 & 11.82 & 0.0755 & 4 & Aae &   \\ 
J001605-234352 & 00 16 05.81 & -23 43 52.1 &  68 &  4 &  -  &  - &  -  &  - &  275.8 &    ... & g0016057-234352 & 0.9 & 12.22 & 0.0640 & 4 & Ae  & ESO\,473-G07  \\
J002301-565005 & 00 23 01.58 & -56 50 05.2 &  99 &  5 &  99 &  5 & 106 &  5 &    ... &  208.4 & g0023015-565005 & 0.6 & 10.08 & 0.0342 & 4 & Aae &  \\
J002736-540907 & 00 27 36.69 & -54 09 07.7 &  43 &  2 &  -  &  - &  -  &  - &    ... &   58.3 & g0027366-540908 & 0.8 & 12.46 & 0.1235 & 4 & Aa  & \\
J002901-011341 & 00 29 01.01 & -01 13 41.7 & 207 &  9 &  -  &  - &  -  &  - &  284.8 &    ... & g0029010-011342 & 0.6 & 11.06 & 0.0830 & ..& ... &  \\ 
J003033-581914 & 00 30 33.78 & -58 19 14.7 &  51 &  3 &  53 &  3 &  29 &  2 &    ... &   35.6 & g0030338-581915 & 0.4 & 12.97 & 0.1740 & 4 & Aa  &   \\
J003908-222001 & 00 39 08.17 & -22 20 01.3 &  49 &  3 &  -  &  - &  -  &  - &  113.5 &    ... & g0039082-222001 & 0.6 & 11.99 & 0.0644 & ..& ..  &  \\
J003939-130059 & 00 39 39.94 & -13 00 59.7 &  68 &  3 &  -  &  - &  -  &  - &   47.3 &    ... & g0039399-130060 & 0.8 & 12.03 & 0.1075 & 4 & Aae &   \\
J004048-204339 & 00 40 48.16 & -20 43 39.1 & 103 &  7 & 215 & 11 & 333 & 17 &  641.2 &    ... & g0040482-204339 & 0.5 & 12.54 & 0.0918 & 4 & Ae  &   \\ 
J005452-043704 & 00 54 52.67 & -04 37 04.5 &  49 &  3 &  -  &  - &  -  &  - &  177.0 &    ... & g0054526-043703 & 1.4 & 11.59 & 0.0689 & 3 & Aa  &  \\ 
J005620-093631 & 00 56 20.14 & -09 36 31.1 &  78 &  4 &  -  &  - &  -  &  - &  199.2 &    ... & g0056201-093630 & 1.3 & 11.95 & 0.1031 & ..& ..  & \\ 
J005734-012328 & 00 57 34.88 & -01 23 28.0 &  55 &  2 &  -  &  - &  -  &  - & 5365.4 &    ... & g0057349-012328 & 0.5 & 10.55 & 0.0450 & J & Aa  &  \\ 
J010249-795604 & 01 02 49.45 & -79 56 04.9 & 113 &  6 & 110 &  6 & 233 & 11 &    ... &  837.9 & g0102490-795604 & 1.6 & 11.67 & 0.0570 & ..& ..  &  \\ 
J010633-463836 & 01 06 33.06 & -46 38 36.5 &  47 &  3 &  49 &  3 &  52 &  3 &    ... &   85.2 & g0106330-463837 & 0.5 & 10.99 & 0.0307 & ..& ..  & ESO\,243-G29 \\ 
J011132-730209 & 01 11 32.25 & -73 02 09.9 &  74 &  4 &  72 &  4 &  72 &  4 &    ... &   86.4 & g0111325-730210 & 1.2 & 11.12 & 0.0666 & 4 & Aa  &  \\ 
J012600-012041 & 01 26 00.61 & -01 20 41.9 & 147 &  7 &  -  &  - &  -  &  - & 3730.7 &    ... & g0126006-012043 & 0.7 &  8.49 & 0.0184 & J & Aa  & NGC\,547 \\ 
J012820-564939 & 01 28 20.38 & -56 49 39.7 & 189 & 10 & 141 &  7 & 130 &  7 &    ... &   82.6 & g0128205-564940 & 0.9 & 12.24 & 0.0666 & 4 & Ae  &   \\
J013241-080405 & 01 32 41.06 & -08 04 05.1 & 131 &  5 &  -  &  - &  -  &  - &  306.8 &    ... & g0132411-080405 & 1.3 & 12.15 & 0.1485 & ..& ..  & \\
J014648-520233 & 01 46 48.63 & -52 02 33.5 & 226 & 11 & 212 & 11 & 225 & 12 &    ... &  349.5 & g0146486-520234 & 0.4 & 11.79 & 0.0981 & 4 & Aa  &  \\ 
J015422-023454 & 01 54 22.79 & -02 34 54.6 & 118 &  5 &  -  &  - &  -  &  - &  193.6 &    ... & g0154228-023454 & 0.5 & 11.79 & 0.0823 & 4 & Aae &  \\
J021316-294326 & 02 13 16.42 & -29 43 26.5 &  84 &  6 & 167 &  9 & 175 &  9 &  114.6 &    ... & g0213164-294326 & 0.5 & 10.93 & 0.0351 & 4 & Aae & ESO\,415-G07 \\ 
J023109-575505 & 02 31 09.32 & -57 55 05.8 & 104 &  5 & 120 &  6 & 140 &  7 &    ... &  254.4 & g0231093-575506 & 0.4 & 10.18 & 0.0320 & ..& ..  & ESO\,115-G15 \\  
J023137-204021 & 02 31 37.01 & -20 40 21.4 & 160 & 11 & 212 & 11 & 170 &  9 &  169.1 &    ... & g0231370-204022 & 0.6 & 12.03 & 0.0898 & 4 & Aa  &   \\
J023241-112020 & 02 32 41.77 & -11 20 20.1 &  69 &  4 &  -  &  - &  -  &  - &  242.8 &    ... & g0232419-112020 & 2.7 & 12.86 & 0.2087 & 3 & Aa  &   \\
J023749-823252 & 02 37 49.54 & -82 32 52.8 &  50 &  3 &  41 &  2 &  74 &  4 &    ... &  457.3 & g0237503-823256 & 3.9 & 12.48 & 0.0754 & 4 & Aae &  \\
J024104-081520 & 02 41 04.77 & -08 15 20.5 &1199 & 59 &  -  &  - &  -  &  - &  912.5 &    ... & g0241048-081521 & 0.6 &  7.45 & 0.0053 & 4 & Ae  &  NGC\,1052 \\ 
J024326-561242 & 02 43 26.57 & -56 12 42.3 & 114 &  6 & 224 & 11 & 312 & 16 &    ... &  181.9 & g0243265-561242 & 0.3 & 12.73 & 0.0637 & 4 & Ae  &   \\
J024524-110717 & 02 45 24.96 & -11 07 17.1 & 167 &  8 &  -  &  - &  -  &  - &  141.2 &    ... & g0245250-110717 & 0.2 & 11.89 & 0.0989 & 4 & Aae &  \\
J024554-445939 & 02 45 54.07 & -44 59 39.5 & 581 & 29 & 591 & 30 & 688 & 34 &    ... & 1972.0 & g0245541-445939 & 0.4 & 12.61 & 0.2829 & 4 & AeB & \\
J025544-141227 & 02 55 44.13 & -14 12 27.9 &  60 &  3 &  -  &  - &  -  &  - &   94.7 &    ... & g0255442-141228 & 1.2 &  9.21 & 0.0290 & ..& ..  & IC\,270 \\
J025926-394037 & 02 59 26.51 & -39 40 37.7 &  71 &  4 & 149 &  8 & 304 & 15 &  984.8 & 1552.0 & g0259267-394038 & 2.2 & 11.53 & 0.0662 & 4 & Aae &  \\ 
J025955-123635 & 02 59 55.24 & -12 36 35.1 &  92 &  5 &  -  &  - &  -  &  - &  518.1 &    ... & g0259553-123635 & 0.8 & 12.05 & 0.0867 & 4 & Aae &  \\
J031357-395403 & 03 13 57.83 & -39 54 03.4 &  78 &  4 & 129 &  7 & 236 & 12 &  597.1 &  917.6 & g0313579-395404 & 1.0 & 12.16 & 0.0803 & 4 & Aa  &  \\
J031552-190644 & 03 15 52.16 & -19 06 44.5 & 108 &  7 & 121 &  6 &  99 &  5 &  100.1 &    ... & g0315521-190644 & 0.7 & 12.65 & 0.0671 & 4 & Ae? &\\ 
J031757-441416 & 03 17 57.66 & -44 14 16.8 & 258 & 13 & 449 & 23 & 611 & 31 &    ... & 1915.0 & g0317577-441417 & 0.6 & 10.47 & 0.0759 & ..& ..  &  ESO 248-G06  \\
J033114-524148 & 03 31 14.99 & -52 41 48.2 &  55 &  3 &  59 &  3 &  56 &  3 &    ... &   51.0 & g0331150-524148 & 0.1 & 12.10 & 0.0665 & ..& ..  &  \\ 
J033913-173600 & 03 39 13.73 & -17 36 00.9 &  97 &  6 & 120 &  6 & 130 &  7 &  170.0 &    ... & g0339137-173601 & 0.5 & 11.57 & 0.0656 & 4 & Aa  &  \\
J034630-342246 & 03 46 30.56 & -34 22 46.1 & 102 &  4 &  -  &  - &  -  &  - & 1724.5 & 1789.4 & g0346306-342246 & 0.3 & 12.05 & 0.0535 & ..& ..  &  \\ 
\hline
\end{tabular}
\end{minipage}
\end{table*}

\setcounter{table}{2}
\clearpage\begin{table*}
\vspace*{-2cm}
\begin{minipage}{175mm}
\caption{6dFGS galaxies detected as AT20G radio sources. } 
\begin{tabular}{lllrrrrrrrrrrrrcll}
\hline
(1) & \multicolumn{2}{c}{(2)} & (3) & 
(4) & (5) & (6) & 
(7) & (8) & (9) & 
(10) & (11) & (12) & (13) & (14) & (15) & (16) & (17) \\ 
\multicolumn{1}{c}{AT20G name} & \multicolumn{2}{c}{AT20G position} & \multicolumn{1}{c}{S$_{20}$} & $\pm$
& \multicolumn{1}{c}{S$_{8}$} & $\pm$ &\multicolumn{1}{c}{S$_{5}$} & $\pm$ & \multicolumn{1}{c}{NVSS} & 
 \multicolumn{1}{c}{SUMSS} & \multicolumn{1}{c}{6dFGS name} & Offset & \multicolumn{1}{c}{K$_{\rm tot}$} & 
\multicolumn{1}{c}{Redshift} & Q & Class & \multicolumn{1}{c}{Notes} \\
 & \multicolumn{2}{c}{(J2000)} & \multicolumn{2}{c}{(mJy)} & \multicolumn{2}{c}{(mJy)} & \multicolumn{2}{c}{(mJy)} & 
\multicolumn{1}{c}{(mJy)} & \multicolumn{1}{c}{(mJy)} & & (arcsec) & (mag) &  & &  \\  
\hline
J035145-274311 & 03 51 45.09 & -27 43 11.4 & 122 &  8 &  -  &  - &  -  &  - & 5340.2 &    ..  & g0351358-274435 &$>$60& 12.54 & 0.0657 & 4 & Ae  & Hotspot of PKS\,0349-27 \\
J035410-265013 & 03 54 10.07 & -26 50 13.7 &  98 &  6 & 114 &  6 & 104 &  5 &   86.5 &    ... & g0354102-265014 & 1.4 & 11.78 & 0.0650 & 4 & Aa  &  \\
J040106-160640 & 04 01 06.75 & -16 06 40.0 & 403 & 26 & 452 & 24 & 236 & 12 &   88.2 &    ... & g0401066-160639 & 1.9 & 10.73 & 0.0317 & 4 & Aae &  \\
J042203-562127 & 04 22 03.84 & -56 21 27.3 &  95 &  5 & 105 &  5 &  85 &  4 &    ... &   25.5 & g0422038-562128 & 0.7 & 11.97 & 0.0380 & 4 & Aa  &  \\
J043022-613201 & 04 30 22.00 & -61 32 01.0 & 148 &  7 & 177 &  9 & 156 &  9 &    ... & 2790.  & g0430220-613201 & 0.5 & 11.07 & 0.0555 & 4 & Aa  &  \\ 
J043754-425853 & 04 37 54.73 & -42 58 53.9 & 119 &  4 & 119 &  6 & 122 &  6 &    ... &  173.2 & g0437547-425854 & 0.2 & 10.75 & 0.0475 & 4 & Aae &  \\
J051103-255450 & 05 11 03.77 & -25 54 50.9 & 123 &  8 & 113 &  6 &  92 &  5 &   43.4 &    ... & g0511038-255451 & 0.5 & 12.22 & 0.0929 & 4 & Aa  &  \\
J052200-285608 & 05 22 00.78 & -28 56 08.4 &  43 &  3 &  61 &  5 & 211 & 11 &  574.1 &    ... & g0522008-285610 & 1.1 & 12.43 & 0.0670 & 3 & Ae  &  \\ 
J052223-072513 & 05 22 23.21 & -07 25 13.7 &  91 &  4 &  -  &  - &  -  &  - &   54.7 &    ... & g0522232-072513 & 0.6 & 12.68 & 0.1645 & 4 & AeB & \\
J052851-104114 & 05 28 51.23 & -10 41 14.0 &  82 &  4 &  -  &  - &  -  &  - &   43.4 &    ... & g0528512-104114 & 0.3 & 12.12 & 0.1130 & 4 & Aa  & \\
J053533-120222 & 05 35 33.28 & -12 02 22.6 & 212 & 10 &  -  &  - &  -  &  - & 1356.7 &    ... & g0535333-120223 & 0.5 & 12.27 & 0.1570 & ..& ..  & \\ 
J054828-331331 & 05 48 28.51 & -33 13 31.5 &  53 &  3 &  38 &  3 &  32 &  2 &    6.0 &    ... & g0548285-331331 & 0.4 & 12.87 & 0.0407 & 4 & Aa  &  \\ 
J055712-372836 & 05 57 12.45 & -37 28 36.3 &  82 &  4 & 120 &  6 & 165 &  8 &  301.8 &  457.5 & g0557126-372837 & 1.2 & 10.42 & 0.0448 & 4 & Ae  & ESO\,364-G18 \\
J060554-351808 & 06 05 54.10 & -35 18 08.3 & 117 &  5 & 283 & 14 & 531 & 27 & 1337.0 & 1799.0 & g0605540-351808 & 0.9 & 12.46 & 0.1412 & 4 & Aa  & \\
J060555-392905 & 06 05 55.98 & -39 29 05.0 &  79 &  4 & 101 &  5 & 110 &  6 &  108.8 &   84.3 & g0605560-392906 & 0.7 & 11.48 & 0.0454 & 3 & Aa  & \\
J061324-421824 & 06 13 24.09 & -42 18 24.1 &  95 &  4 & 156 &  8 & 243 & 12 &    ... &  650.2 & g0613241-421824 & 0.2 & 12.76 & 0.1097 & 4 & Aa  & \\
J062706-352916 & 06 27 06.73 & -35 29 16.0 & 688 & 29 & 962 & 48 &1301 & 65 & 4632.9 & 4592.0 & g0627067-352915 & 0.7 & 10.51 & 0.0549 & 4 & Aa  & \\
J065359-415144 & 06 53 59.97 & -41 51 44.8 &  56 &  3 &  72 &  4 & 102 &  5 &    ... &  727.5 & g0653599-415145 & 0.7 & 11.84 & 0.0908 & 4 & Aa  & \\
J070240-284149 & 07 02 40.37 & -28 41 49.9 & 194 & 13 &  -  &  - &  -  &  - &  239.4 &    ... & g0702404-284150 & 0.4 &  7.89 & 0.0072 & ..& ..  & NGC\,2325  \\
J070459-490459 & 07 04 59.20 & -49 04 59.5 &  84 &  4 &  94 &  6 &  92 &  5 &    ... &   41.2 & g0704593-490460 & 0.6 &  9.84 & 0.0419 & ..& ..  & ESO\,207-G19 \\ 
J070914-360121 & 07 09 14.07 & -36 01 21.6 &  94 & 5  &  -  &  - &  -  &  - & 1865.4 & 3420.  & g0709141-360122 & 1.8 & 12.60 & 0.1108 & ..& ..  & \\ 
J081611-703944 & 08 16 11.64 & -70 39 44.4 &  62 &  3 &  -  &  - &  -  &  - &    ... & 1940.1 & g0816118-703945 & 1.0 & 10.86 & 0.0332 & 4 & Aa  & ESO\,060-IG02 \\ 
J084452-100059 & 08 44 52.73 & -10 00 59.0 &  46 &  4 &  -  &  - &  -  &  - &  311.9 &    ... & g0844527-100059 & 0.8 & 11.18 & 0.0429 & 4 & Aa  &  \\
J090802-095937 & 09 08 02.23 & -09 59 37.5 &  60 &  4 &  -  &  - &  -  &  - &  182.3 &    ... & g0908022-095938 & 0.7 & 11.39 & 0.0535 & 4 & Aa &  \\
J090825-093332 & 09 08 25.37 & -09 33 32.5 &  46 &  2 &  -  &  - &  -  &  - &  264.6 &    ... & g0908253-093334 & 1.6 & 12.58 & 0.1590 & 2 & .. & \\ 
J091300-210320 & 09 13 00.22 & -21 03 20.6 & 135 &  9 & 170 &  9 & 198 & 10 &  328.4 &    ... & g0913002-210321 & 0.3 & 12.65 & 0.1980 & 3 & Aa & \\
J092338-213544 & 09 23 38.95 & -21 35 44.9 & 328 & 22 & 390 & 20 & 307 & 20 &  267.7 &    ... & g0923389-213547 & 2.5 & 12.00 & 0.0520 & ..& .. &  \\
J093626-123649 & 09 36 26.47 & -12 36 49.0 &  50 &  3 &  -  &  - &  -  &  - &   69.8 &    ... & g0936265-123647 & 1.8 & 11.82 & 0.1133 & 4 & Aa & \\  
J093955-241516 & 09 39 55.72 & -24 15 16.8 &  40 &  2 &  39 &  2 &  32 &  2 &   30.2 &    ... & g0939558-241517 & 0.5 & 11.48 & 0.0658 & 4 & Aae & \\
J094409-015116 & 09 44 09.86 & -01 51 16.3 &  78 &  5 &  -  &  - &  -  &  - &   88.5 &    ... & g0944099-015115 & 1.3 & 12.39 & 0.0418 & ..& .. &  \\
J095513-212303 & 09 55 13.59 & -21 23 03.3 &  93 &  6 &  81 &  4 &  61 &  3 &   58.3 &    ... & g0955136-212303 & 0.5 & 12.31 & 0.1087 & 4 & Ae & \\
J095824-265538 & 09 58 24.45 & -26 55 38.7 & 142 &  9 & 152 &  8 & 187 &  9 &  313.1 &    ... & g0958246-265536 & 3.6 &  7.88 & 0.0083 & ..& .. & NGC\,3078 \\
J100040-313952 & 10 00 40.82 & -31 39 52.2 & 266 & 12 & 404 & 20 & 488 & 24 &  530.2 &  724.2 & g1000408-313952 & 0.7 &  8.08 & 0.0086 & 4 & Aae & NGC\,3100\\
J100833-042839 & 10 08 33.63 & -04 28 39.8 &  44 &  3 &  -  &  - &  -  &  - &  250.1 &    ... & g1008337-042840 & 0.8 & 11.23 & 0.0509 & ..& .. &  \\ 
J103859-422353 & 10 38 59.73 & -42 23 53.4 &  85 &  4 & 191 & 10 & 339 & 17 &   ...  & 1157.0 & g1038597-422355 & 1.4 & 12.16 & 0.1137 & 4 & Aa &  \\
J110957-373220 & 11 09 57.70 & -37 32 20.3 &  52 &  4 &  23 &  2 &  77 &  4 &  582.8 &  567.4 & g1109577-373221 & 0.8 &  7.20 & 0.0101 & 4 & Aa &  NGC\,3557   \\ 
J112119-001316 & 11 21 19.47 & -00 13 16.0 & 108 &  5 &  -  &  - &  -  &  - &  648.9 &    ... & g1121194-001316 & 0.9 & 12.12 & 0.0993 & ..& .. &  \\ 
J113305-040046 & 11 33 05.21 & -04 00 46.4 & 108 &  6 &  -  &  - &  -  &  - &  979.1 &    ... & g1133051-040048 & 2.0 & 10.47 & 0.0520 & 4 & Aa &  \\
J114503-325824 & 11 45 03.50 & -32 58 24.1 &  75 &  4 &  91 &  5 &  69 &  4 &   31.5 &   39.7 & g1145035-325824 & 0.7 & 10.47 & 0.0384 & 4 & Aae &  \\
J114539-105350 & 11 45 39.02 & -10 53 50.4 &  73 &  4 &  -  &  - &  -  &  - &  102.4 &    ... & g1145389-105352 & 2.5 & 11.50 & 0.0774 & 4 & Aa &  \\
\hline
\end{tabular}
\end{minipage}
\end{table*}

\setcounter{table}{2}
\clearpage\begin{table*}
\vspace*{-2cm}
\begin{minipage}{175mm}
\caption{6dFGS galaxies detected as AT20G radio sources. } 
\begin{tabular}{lllrrrrrrrrrrrrcll}
\hline
(1) & \multicolumn{2}{c}{(2)} & (3) & 
(4) & (5) & (6) & 
(7) & (8) & (9) & 
(10) & (11) & (12) & (13) & (14) & (15) & (16) & (17) \\ 
\multicolumn{1}{c}{AT20G name} & \multicolumn{2}{c}{AT20G position} & \multicolumn{1}{c}{S$_{20}$} & $\pm$
& \multicolumn{1}{c}{S$_{8}$} & $\pm$ &\multicolumn{1}{c}{S$_{5}$} & $\pm$ & \multicolumn{1}{c}{NVSS} & 
 \multicolumn{1}{c}{SUMSS} & \multicolumn{1}{c}{6dFGS name} & Offset & \multicolumn{1}{c}{K$_{\rm tot}$} & 
\multicolumn{1}{c}{Redshift} & Q & Class & \multicolumn{1}{c}{Notes} \\
 & \multicolumn{2}{c}{(J2000)} & \multicolumn{2}{c}{(mJy)} & \multicolumn{2}{c}{(mJy)} & \multicolumn{2}{c}{(mJy)} & 
\multicolumn{1}{c}{(mJy)} & \multicolumn{1}{c}{(mJy)} & & (arcsec) & (mag) &  & & \\  
\hline
J121044-435437 & 12 10 44.67 & -43 54 37.4 & 129 &  6 & 168 &  9 & 136 &  7 &    ... &  122.7 & g1210448-435437 & 1.5 & 12.29 & 0.0693 & 4 & Aae & \\
J122343-423532 & 12 23 43.40 & -42 35 32.2 & 330 & 17 & 507 & 19 & 999 & 41 &    ... & 3105.0 & g1223434-423532 & 0.8 & 12.34 & 0.1706 & 4 & Ae  &  \\ 
J123148-321314 & 12 31 48.73 & -32 13 14.1 & 106 &  6 & 184 &  9 & 176 &  9 &  165.1 &  137.5 & g1231487-321314 & 0.3 & 12.59 & 0.0884 & 4 & Aae & \\
J123449-243232 & 12 34 49.79 & -24 32 32.1 &  73 &  4 & 104 &  5 &  58 &  4 &   29.0 &    ... & g1234497-243233 & 1.4 & 12.63 & 0.1812 & 3 & Aa? & \\
J123959-113721 & 12 39 59.32 & -11 37 21.4 &  81 &  4 &  -  &  - &  -  &  - &   93.4 &    ... & g1239594-113723 & 2.2 &  4.96 & 0.0036 & ..& ..  & NGC\,4594 \\ 
J125457-442456 & 12 54 57.67 & -44 24 56.9 & 344 & 15 & 356 & 18 & 330 & 17 &    ... &  368.1 & g1254575-442457 & 2.1 & 10.69 & 0.0411 & 4 & Aae & \\  
J125615-114635 & 12 56 15.96 & -11 46 35.7 & 123 &  6 &  -  &  - &  -  &  - &   54.4 &    ... & g1256160-114637 & 1.4 & 11.88 & 0.0579 & 4 & Aa  & \\
J125711-172434 & 12 57 11.59 & -17 24 34.0 &  83 &  5 & 132 &  7 & 164 &  8 &   98.4 &    ... & g1257116-172434 & 0.2 &  9.65 & 0.0470 & 4 & Aa  & \\
J130031-441442 & 13 00 31.06 & -44 14 42.6 &  78 &  3 & 104 &  5 &  68 &  4 &    ... &   26.9 & g1300310-441442 & 1.2 & 10.45 & 0.0322 & ..& ..  & \\ 
J130058-231215 & 13 00 58.52 & -23 12 15.6 & 188 & 10 &  -  &  - &  -  &  - &  777.0 &    ... & g1300589-231210 & 7.9 & 12.73 & 0.1294 & 4 & Aa  & \\
J130100-322628 & 13 01 00.80 & -32 26 28.0 & 176 &  9 & 378 & 19 & 574 & 29 & 1387.0 & 1574.0 & g1301008-322629 & 0.9 &  8.51 & 0.0170 & ..& ..  & ESO\,443-G24 \\ 
J130527-492804 & 13 05 27.47 & -49 28 04.7 & 726 & 36 & 786 & 39 &1566 & 78 &    ... & 5549.0 & g1305273-492805 & 1.7 &  4.48 & 0.0019 & ..& ..  & NGC\,4945 \\ 
J130715-760245 & 13 07 15.46 & -76 02 45.6 &  71 &  1 & 118 &  6 & 114 &  6 &    ... &  139.1 & g1307156-760245 & 0.6 & 12.50 & 0.1833 & 4 & AeB & \\
J130841-242259 & 13 08 41.92 & -24 22 59.9 &  57 &  4 &  -  &  - &  -  &  - &  474.5 &    ... & g1308420-242258 & 2.3 & 11.78 & 0.0142 & 4 & Ae  & \\ 
J131124-442240 & 13 11 24.04 & -44 22 40.5 &  44 &  2 &  50 &  4 &  80 &  5 &    ... &  208.3 & g1311238-442240 & 2.4 & 10.22 & 0.0506 & 4 & Aae & \\ 
J131931-123925 & 13 19 31.42 & -12 39 25.5 & 149 &  8 &  -  &  - &  -  &  - &  156.7 &    ... & g1319317-123925 & 3.3 &  8.22 & 0.0084 & ..& ..  & NGC\,5077 \\ 
J131949-272437 & 13 19 49.86 & -27 24 37.5 &  54 &  4 &  35 &  5 &  86 &  5 &  262.3 &    ... & g1319500-272436 & 2.8 &  7.12 & 0.0071 & 4 & Aae & NGC\,5078 \\ 
J132112-434216 & 13 21 12.81 & -43 42 16.7 & 705 & 35 & 496 & 18 & 447 & 27 &    ... & 1325.8 & g1321129-434217 & 0.6 &  7.57 & 0.0112 & J & Aae & NGC\,5090 \\ 
J132340-410123 & 13 23 40.83 & -41 01 23.4 & 156 &  8 & 244 & 12 & 326 & 16 &    ... &  208.3 & g1323408-410124 & 0.5 & 12.89 & 0.0941 & 3 & Aae & \\ 
J132920-264022 & 13 29 20.70 & -26 40 22.2 & 101 &  6 & 118 &  6 & 118 &  6 &  102.4 &    ... & g1329208-264022 & 2.0 & 11.50 & 0.0502 & 4 & Aa  & \\
J133608-082952 & 13 36 08.30 & -08 29 52.3 & 591 & 29 &  -  &  - &  -  &  - &  388.8 &    ... & g1336083-082952 & 0.8 &  9.17 & 0.0228 & 4 & Aae & NGC\,5232 \\ 
J133639-335756 & 13 36 39.01 & -33 57 56.3 & 323 & 14 & 298 & 22 & 262 & 34 & 2334.2 &14900.  & g1336390-335757 & 1.0 &  7.50 & 0.0125 & J & Aae & IC\,4296 \\ 
J134624-375816 & 13 46 24.13 & -37 58 16.0 &  71 &  4 &  55 &  3 &  68 &  5 &  697.1 & 1225.7 & g1346240-375816 & 1.7 & 10.70 & 0.0376 & 4 & Aa  & ESO\,325-G16 \\ 
J135036-163449 & 13 50 36.28 & -16 34 49.4 & 186 & 12 & 220 & 12 & 173 &  9 &  109.2 &    ... & g1350362-163450 & 2.0 & 12.38 & 0.0877 & 4 & Ae  & \\ 
J135607-172433 & 13 56 07.02 & -17 24 33.1 & 239 & 16 & 271 & 14 & 255 & 13 &  180.1 &    ... & g1356070-172432 & 1.7 & 11.80 & 0.0747 & 4 & Aae & \\ 
J140729-270105 & 14 07 29.76 & -27 01 05.4 & 202 & 13 & 415 & 22 & 552 & 28 &  645.5 &    ... & g1407298-270104 & 1.1 &  9.57 & 0.0218 & ..& ..  & IC\,4374 \\
J140912-231550 & 14 09 12.03 & -23 15 50.1 & 137 &  9 & 356 & 19 & 521 & 26 &  599.4 &    ... & g1409120-231550 & 0.9 & 12.68 & 0.0864 & 4 & Aae & \\ 
J145509-365508 & 14 55 09.60 & -36 55 08.6 & 198 & 10 & 205 & 11 & 241 & 12 & 1207.5 & 1869.0 & g1455096-365508 & 0.9 & 11.33 & 0.0946 & 4 & Aae & \\ 
J145924-164136 & 14 59 24.78 & -16 41 36.5 &  83 &  5 &  -  &  - &  -  &  - & 1200.5 &    ... & g1459248-164137 & 0.1 &  9.37 & 0.0117 & ..& ..  & NGC\,5793 \\ 
J145928-181045 & 14 59 28.74 & -18 10 45.2 & 118 &  7 &  -  &  - &  -  &  - &  105.4 &    ... & g1459288-181045 & 0.2 & 12.59 & 0.2344 & 3 & Ae  & \\
J151741-242220 & 15 17 41.76 & -24 22 20.2 &3449 &226 &3647 &190 &3173 &158 & 2041.9 &    ... & g1517418-242219 & 1.2 & 10.77 & 0.0490 & 4 & Aae & AP Lib \\ 
J152433-301221 & 15 24 33.40 & -30 12 21.8 & 255 & 13 & 351 & 18 & 323 & 16 &  174.3 &  187.0 & g1524334-301221 & 0.4 & 10.91 & 0.0195 & 4 & Aae & \\
J155821-141000 & 15 58 21.83 & -14 10 00.0 & 172 &  9 &  -  &  - &  -  &  - &  460.6 &    ... & g1558220-140959 & 2.1 & 11.12 & 0.0974 & 4 & Ae  & \\
J164416-771548 & 16 44 16.03 & -77 15 48.5 & 399 & 20 & 238 & 13 & 245 & 14 &    ... & 1165.9 & g1644161-771549 & 0.5 & 11.20 & 0.0430 & J & Aae & \\ 
J165710-735544 & 16 57 10.08 & -73 55 44.5 &  42 &  1 & 137 &  7 & 128 &  7 &    ... &   99.3 & g1657105-735542 & 2.6 & 12.35 & 0.0712 & 4 & Aa  & \\
J170241-774156 & 17 02 41.29 & -77 41 56.3 &  50 &  1 & 172 & 15 & 306 & 27 &    ... & 2912.0 & g1702410-774157 & 1.2 & 12.51 & 0.0947 & 4 & Aa  & \\
J171522-652018 & 17 15 22.19 & -65 20 18.6 &  53 &  3 &  86 &  4 &  97 &  5 &    ... &  190.5 & g1715223-652018 & 0.8 & 10.77 & 0.0492 & 4 & Aae & \\
J172341-650036 & 17 23 41.10 & -65 00 36.3 &2872 &143 &4603 &230 &5605 &280 &    ... & 3699.0 & g1723410-650037 & 0.8 &  9.18 & 0.0142 &.. & ..  & NGC\,6328 \\ 
J180207-471930 & 18 02 07.68 & -47 19 30.3 & 114 &  6 & 119 &  6 & 123 &  7 &    ... &  681.1 & g1802076-471932 & 2.0 & 12.57 & 0.1227 & 3 & Aa  & \\ 
J180228-523645 & 18 02 28.50 & -52 36 45.9 & 100 &  5 & 196 & 10 & 257 & 13 &    ... &  686.9 & g1802285-523646 & 0.7 & 11.98 & 0.1306 & 4 & Ae  & \\
\hline
\end{tabular}
\end{minipage}
\end{table*}

\setcounter{table}{2}
\clearpage\begin{table*}
\vspace*{-2cm}
\begin{minipage}{175mm}
\caption{6dFGS galaxies detected as AT20G radio sources. } 
\begin{tabular}{lllrrrrrrrrrrrrcll}
\hline
(1) & \multicolumn{2}{c}{(2)} & (3) & 
(4) & (5) & (6) & 
(7) & (8) & (9) & 
(10) & (11) & (12) & (13) & (14) & (15) & (16) & (17) \\ 
\multicolumn{1}{c}{AT20G name} & \multicolumn{2}{c}{AT20G position} & \multicolumn{1}{c}{S$_{20}$} & $\pm$
& \multicolumn{1}{c}{S$_{8}$} & $\pm$ &\multicolumn{1}{c}{S$_{5}$} & $\pm$ & \multicolumn{1}{c}{NVSS} & 
 \multicolumn{1}{c}{SUMSS} & \multicolumn{1}{c}{6dFGS name} & Offset & \multicolumn{1}{c}{K$_{\rm tot}$} & 
\multicolumn{1}{c}{Redshift} & Q & Class & \multicolumn{1}{c}{Notes} \\
 & \multicolumn{2}{c}{(J2000)} & \multicolumn{2}{c}{(mJy)} & \multicolumn{2}{c}{(mJy)} & \multicolumn{2}{c}{(mJy)} & 
\multicolumn{1}{c}{(mJy)} & \multicolumn{1}{c}{(mJy)} & & (arcsec) & (mag) &  & & \\  
\hline
J180957-455241 & 18 09 57.79 & -45 52 41.2 &1087 & 54 &  -  &  - &  -  &  - &    ... & 1530.0 & g1809579-455241 & 0.6 & 12.01 & 0.0697 & 4 & AeB & \\
J181857-550815 & 18 18 57.99 & -55 08 15.0 &  75 &  4 &  54 &  4 &  42 &  5 &    ... &  185.6 & g1818580-550815 & 0.5 & 11.41 & 0.0726 & 4 & Aa  & \\ 
J181934-634548 & 18 19 34.99 & -63 45 48.2 & 1704& 85 & 3567&178 &5562 &277 &    ... &19886.0 & g1819351-634548 & 0.2 & 12.10 & 0.0647 & J & Ae  & PKS\,1814-63 \\
J182125-763459 & 18 21 25.15 & -76 34 59.6 & 172 &  9 & 246 & 12 & 331 & 17 &    ... &  704.5 & g1821247-763500 & 1.7 &  9.25 & 0.0184 & 4 & Aae & NGC\,6557 \\
J182507-631453 & 18 25 07.31 & -63 14 53.7 &  88 &  4 & 117 &  6 & 131 &  7 &    ... &  141.6 & g1825072-631454 & 0.9 &  9.22 & 0.0144 & 4 & Aae & NGC\,6614 \\
J184038-770929 & 18 40 38.59 & -77 09 29.1 & 224 & 11 & 229 & 12 & 408 & 20 &    ... & 1152.0 & g1840386-770929 & 0.6 &  9.41 & 0.0182 & 4 & Aa  & ESO\,045-G11 \\
J192043-383107 & 19 20 43.13 & -38 31 07.1 &  63 &  2 &  -  &  - &  -  &  - &  239.7 &  247.8 & g1920430-383106 & 1.5 & 11.23 & 0.0453 & 4 & Aae & \\
J192234-163253 & 19 22 34.88 & -16 32 53.3 & 110 &  7 &  -  &  - &  -  &  - &  482.9 &    ... & g1922348-163254 & 2.0 & 12.29 & 0.1266 & 4 & AeB & \\
J192817-293145 & 19 28 17.05 & -29 31 45.0 &  78 &  4 & 168 &  9 & 252 & 14 & 2108.5 &    ... & g1928170-293144 & 0.8 &  9.56 & 0.0244 & 4 & Aa  & ESO\,460-G04 \\ 
J194524-552049 & 19 45 24.22 & -55 20 49.2 & 779 & 39 &1009 & 51 &1059 & 53 &    ... &  596.3 & g1945242-552049 & 0.4 &  9.00 & 0.0152 & 4 & Aae & NGC\,6812 \\ 
J200954-482246 & 20 09 54.04 & -48 22 46.2 & 113 &  6 & 135 &  7 & 143 &  7 &    ... &  138.8 & g2009541-482246 & 0.5 &  7.32 & 0.0095 & 4 & Aae & NGC\,6868 \\ 
J204008-711457 & 20 40 08.46 & -71 14 57.1 &  53 &  3 &  65 &  3 &  88 &  4 &    ... &  494.2 & g2040083-711460 & 2.8 & 12.38 & 0.1618 & 4 & Aa  & \\ 
J204552-510627 & 20 45 52.29 & -51 06 27.7 &  54 &  3 &  63 &  3 &  65 &  3 &    ... &  620.9 & g2045523-510627 & 0.9 & 10.19 & 0.0485 & 4 & Aa  & ESO\,234-G068 \\ 
J205202-570407 & 20 52 02.32 & -57 04 07.9 & 130 &  6 & 230 & 12 & 453 & 23 &    ... & 1975.0 & g2052023-570408 & 0.5 &  8.75 & 0.0118 & 4 & Ae  & IC\,5063 \\ 
J205306-162007 & 20 53 06.00 & -16 20 07.4 &  56 &  3 &  -  &  - &  -  &  - &  116.5 &    ... & g2053061-162008 & 2.1 & 10.73 & 0.0427 & 4 & Aa  & IC\,1335 \\
J205401-424238 & 20 54 01.79 & -42 42 38.7 &  86 &  5 & 125 &  6 & 154 &  8 &    ... &  160.2 & g2054018-424240 & 0.8 & 11.34 & 0.0429 & 4 & Aae & \\
J205754-662919 & 20 57 54.01 & -66 29 19.6 &  49 &  3 &  42 &  2 &  51 &  3 &    ... &  282.7 & g2057540-662920 & 0.2 & 11.85 & 0.0754 & 4 & Aa  & ESO\,106-IG15 \\ 
J205837-575636 & 20 58 37.39 & -57 56 36.5 &  97 &  3 & 203 &  1 & 333 & 10 &    ... &  846.2 & g2058376-575636 & 1.6 & 11.40 & 0.0524 & 4 & Aa  & \\
J205846-144304 & 20 58 46.69 & -14 43 04.7 &  57 &  3 &  -  &  - &  -  &  - &  146.6 &    ... & g2058468-144305 & 1.1 & 12.13 & 0.0778 & 4 & Aa  & \\
J210353-093341 & 21 03 53.28 & -09 33 41.2 & 103 &  5 &  -  &  - &  -  &  - &  145.7 &    ... & g2103533-093342 & 0.2 & 12.20 & 0.0882 & 4 & Aa  & \\
J210602-712218 & 21 06 02.92 & -71 22 17.9 & 246 & 13 & 447 & 22 & 619 & 31 &    ... & 1206.0 & g2106029-712218 & 0.2 & 11.88 & 0.0745 & 4 & Aa  & \\
J212222-560014 & 21 22 22.81 & -56 00 14.6 &  58 &  3 &  34 &  2 &  28 &  2 &    ... &  100.9 & g2122229-560014 & 1.0 & 11.56 & 0.0518 & 4 & Aae &  \\ 
J213133-383703 & 21 31 33.09 & -38 37 03.9 &  46 &  2 &  47 &  2 &  66 &  3 &  979.2 & 1145.8 & g2131330-383705 & 1.5 &  9.56 & 0.0182 & 4 & Aa  & NGC\,7075 \\ 
J214824-571351 & 21 48 24.09 & -57 13 51.5 &  48 &  3 &  51 &  3 &  67 &  4 &    ... &  447.1 & g2148243-571352 & 1.3 & 12.09 & 0.0806 & ..& ..  & \\ 
J220156-332103 & 22 01 56.41 & -33 21 03.3 &  71 &  2 & 139 &  6 & 224 & 11 & 1291.4 & 2111.0 & g2201564-332103 & 0.4 & 12.31 & 0.1537 & 4 & Aa  & \\
J220253-563543 & 22 02 53.31 & -56 35 43.0 &  69 &  4 &  81 &  4 &  84 &  4 &    ... &   58.4 & g2202533-563542 & 1.1 & 12.16 & 0.0489 & ..& ..  &  \\ 
J220538-053531 & 22 05 38.59 & -05 35 31.9 &  67 &  4 &  -  &  - &  -  &  - &  187.2 &    ... & g2205386-053533 & 1.1 & 11.46 & 0.0576 & ..& ..  &  \\
J220916-471000 & 22 09 16.25 & -47 10 00.3 & 123 &  6 & 161 &  8 & 136 &  7 &    ... &  119.6 & g2209162-471000 & 0.3 &  7.04 & 0.0060 & ..& ..  & NGC\,7213  \\
J221220-251829 & 22 12 20.77 & -25 18 29.0 &  71 &  5 &  -  &  - &  -  &  - &  304.4 &    ... & g2212207-251829 & 1.0 & 10.81 & 0.0626 & 4 & Aa  & \\
J223931-360912 & 22 39 31.26 & -36 09 12.5 &  64 &  2 & 138 &  7 & 182 &  9 &  661.0 &  842.5 & g2239311-360912 & 1.4 & 11.32 & 0.0575 & 4 & Aa  & \\
J224559-173724 & 22 45 59.13 & -17 37 24.2 &  43 &  3 &  -  &  - &  -  &  - &  395.8 &    ... & g2245590-173732 & 8.4 & 11.07 & 0.0684 & ..& ..  &  \\
J225710-362744 & 22 57 10.49 & -36 27 44.5 & 554 & 28 & 851 & 43 &1082 & 54 & 1279.7 &  966.2 & g2257106-362744 & 1.3 &  6.80 & 0.0057 & 4 & Aae & IC\,1459\\
J230136-591320 & 23 01 36.31 & -59 13 20.5 & 155 &  8 & 143 &  7 & 137 &  7 &    ... &   61.9 & g2301363-591321 & 0.7 & 12.26 & 0.1501 & 4 & Ae  & \\
J231546-230744 & 23 15 46.98 & -23 07 44.2 &  73 &  4 &  77 &  4 &  54 &  3 &   13.7 &    ... & g2315469-230744 & 0.5 & 12.17 & 0.0703 & 4 & Aae & \\
J231905-420648 & 23 19 05.92 & -42 06 48.9 & 150 &  6 &  -  &  - &  -  &  - &   ...  &  911.6 & g2319059-420649 & 0.6 & 10.49 & 0.0543 & 4 & Aae & \\ 
J231915-533159 & 23 19 15.92 & -53 31 59.1 &  75 &  3 & 113 & 10 & 177 & 10 &   ...  &  501.8 & g2319158-533159 & 1.4 & 12.27 & 0.0958 & 4 & Aa  & \\ 
J232519-120727 & 23 25 19.74 & -12 07 27.1 &  84 &  4 &  -  &  - &  -  &  - & 1874.6 &    ... & g2325197-120727 & 0.4 & 12.10 & 0.0822 & 4 & Ae  & \\
J233355-234340 & 23 33 55.28 & -23 43 40.8 & 957 & 48 & 772 & 39 & 656 & 33 &  782.1 &    ... & g2333552-234341 & 0.9 & 11.97 & 0.0477 & ..& ..  &  \\
J234129-291915 & 23 41 29.72 & -29 19 15.3 & 120 &  8 & 182 & 10 & 153 &  8 &  239.6 &    ... & g2341298-291915 & 1.1 & 10.37 & 0.0517 & ..& ..  &   \\ 
J234205-160840 & 23 42 05.82 & -16 08 40.4 &  43 &  3 &  -  &  - &  -  &  - &  416.7 &    ... & g2342059-160841 & 0.7 & 11.74 & 0.0649 & 4 & Aa  & \\ 
\hline
\end{tabular}
\end{minipage}
\end{table*}

\setcounter{table}{2}
\clearpage\begin{table*}
\vspace*{-2cm}
\begin{minipage}{175mm}
\caption{6dFGS galaxies detected as AT20G radio sources. } 
\begin{tabular}{lllrrrrrrrrrrrrcll}
\hline
(1) & \multicolumn{2}{c}{(2)} & (3) & 
(4) & (5) & (6) & 
(7) & (8) & (9) & 
(10) & (11) & (12) & (13) & (14) & (15) & (16) & (17) \\ 
\multicolumn{1}{c}{AT20G name} & \multicolumn{2}{c}{AT20G position} & \multicolumn{1}{c}{S$_{20}$} & $\pm$
& \multicolumn{1}{c}{S$_{8}$} & $\pm$ &\multicolumn{1}{c}{S$_{5}$} & $\pm$ & \multicolumn{1}{c}{NVSS} & 
 \multicolumn{1}{c}{SUMSS} & \multicolumn{1}{c}{6dFGS name} & Offset & \multicolumn{1}{c}{K$_{\rm tot}$} & 
\multicolumn{1}{c}{Redshift} & Q & Class & \multicolumn{1}{c}{Notes} \\
 & \multicolumn{2}{c}{(J2000)} & \multicolumn{2}{c}{(mJy)} & \multicolumn{2}{c}{(mJy)} & \multicolumn{2}{c}{(mJy)} & 
\multicolumn{1}{c}{(mJy)} & \multicolumn{1}{c}{(mJy)} & & (arcsec) & (mag) &  & &  \\  
\hline
\multicolumn{15}{l}{(b) 6dFGS galaxies identified with extended or multiple AT20G sources} \\
J000311-544516 & 00 03 11.04 & -54 45 16.8 &  95 &  3 & 313 &  3 & 552 &  9 &    ... & 1549.0 & g0003112-544458 &18.9 & 10.22 & 0.0327 & 4 & Aa  & Resolved double \\
J000413-525458 & 00 04 13.97 & -52 54 58.7 &  65 &  3 &  98 &  4 & 192 &  6 &    ... &  934.2 & g0004140-525459 & 0.4 & 10.06 & 0.0328 & ..& ... & \\ 
J003704-010907 & 00 37 04.24 & -01 09 07.2 & 404 & 17 &  -  &  - &  -  &  - & 4067.1 &    ... & g0037041-010908 & 2.5 & 11.91 & 0.0734 & 4 & Aa  & \\ 
PKS\,0043-42$*$  & 00 46 17.8  & -42 07 52.  & 438 & 20 &  -  &  - &  -  & -  &  ... &11260. & g0046178-420752 & ..  & 12.66 & 0.116  & J  & Aae & Two AT20G sources \\ 
J004733-251717 & 00 47 33.08 & -25 17 17.7 & 608 & 29 &1411 & 30 &2190 & 60 & 6831.0 &    ... & g0047331-251719 & 1.4 &  3.77 & 0.0008 & J & SF  & NGC\,253 \\ 
J013357-362935 & 01 33 57.90 & -36 29 35.3 & 440 &  - &  -  &  - &  -  &  - & 8400.  &10300.  & g0133577-362936 & 2.0 &  9.58 & 0.0305 & 4 & Ae  & NGC\,612 \\ 
J021645-474908 & 02 16 45.07 & -47 49 08.9 &  93 &  3 &  -  &  - &  -  &  - &    ... & 3900.  & g0216451-474909 & 0.4 & 10.54 & 0.0643 & 4 & Aa  & ESO\,198-G01 \\ 
J024240-000046 & 02 42 40.72 & -00 00 46.7 & 474 & 17 &  -  &  - &  -  &  - & 4610.  &    ... & g0242407-000048 & 1.2 &  5.79 & 0.0038 & J & Ae  & NGC\,1068 \\ 
J035257-683117 & 03 52 57.46 & -68 31 17.4 &  68 &  3 & 114 &  4 & 165 &  5 &    ... &  409.4 & g0352575-683117 & 0.6 & 11.70 & 0.0870 & ..& ..  & \\ 
J042908-534940 & 04 29 08.24 & -53 49 40.2 & 145 &  4 & 244 &  6 & 782 & 10 &    ... & 8410.  & g0429082-534940 & 0.3 &  9.76 & 0.0380 & J & Aa  & IC\,2082 \\ 
J045523-203409 & 04 55 23.42 & -20 34 09.8 & 331 & 10 &  -  &  - &  -  & -  & 4563.9 &    ... & g0455237-203416 & 7.5 & 10.06 & 0.0354 & J & Aa  & NGC\,1692 \\ 
J050453-101451 & 05 04 53.01 & -10 14 51.5 & 123 &  5 &  -  &  - &  -  &  - & 1482.4 &    ... & g0504531-101453 & 1.3 & 11.20 & 0.0403 & 4 & Ae  & Arp\,187 \\
Pictor A$*$    & 05 19 49.70 & -45 46 43.7 &6320 &110 &  -  &  - &  -  &  - &    ... &85700.  & g0519497-454644 & ..  & 11.95 & 0.0351 & J & AeB & Three AT20G sources \\ 
J054754-195805 & 05 47 54.59 & -19 58 05.9 &  42 &  3 &  93 &  6 & 163 & 13 &  863.1 &    ... & g0547546-195805 & 1.1 & 11.96 & 0.0551 & 4 & Aa  &  \\
J055049-314428 & 05 50 49.12 & -31 44 28.9 & 133 &  3 & 277 &  6 & 448 & 11 & 1034.3 & 1540.0 & g0550498-314427 & 9.5 & 11.04 & 0.0400 & ..& ..  & ESO\,424-G27  	   \\
J062143-524132 & 06 21 43.41 & -52 41 32.3 & 266 &  9 & 442 & 11 & 928 & 13 &    ... & 4840.3 & g0621433-524133 & 1.4 &  9.80 & 0.0511 & J & Aae &  \\ 
J062620-534151 & 06 26 20.58 & -53 41 51.4 & 253 &  4 &  -  &  - &  -  &  - &    ... &11600.  & g0626205-534136 &15.7 &  9.66 & 0.0551 & J & Aa  & ESO\,161-IG07 \\ 
J062648-543214 & 06 26 48.91 & -54 32 14.0 & 106 &  3 & 336 &  5 & 773 & 13 &    ... & 5000.  & g0626496-543234 &21.1 & 10.78 & 0.0517 & ..& ..  & Resolved triple \\ 
PKS\,0634-20$*$& 06 36 31.24 & -20 28 57.6 & 238 &  9 &1551 & 20 &2993 & 40 & 9625.0 &    ... & g0636323-203453 & ..  & 11.70 & 0.0551 & 4 & Ae  & Two AT20G sources  \\
J065153-602158 & 06 51 53.67 & -60 21 58.4 &  44 &  2 &  36 &  2 & 102 &  5 &    ... & 1849.2 & g0651549-602217 &20.7 & 12.26 & 0.1339 & 3 & Aa  & \\ 
J080537-005819 & 08 05 37.69 & -00 58 19.0 &  81 &  3 &  -  &  - &  -  &  - & 1329.1 &    ... & g0805378-005818 & 2.0 & 10.70 & 0.0902 & 4 & Aa  & \\ 
J080852-102832 & 08 08 52.49 & -10 28 31.9 & 131 &  5 &  -  &  - &  -  &  - & 4256.1 &    ... & g0808536-102740 &55.2 & 12.33 & 0.1090 & 4 & Ae  & Hotspot of PKS\,0806-10  \\ 
J091805-120532 & 09 18 05.82 & -12 05 32.5 &1056 & 52 &  -  &  - &  -  &  - &40849.9 &    ... & g0918057-120544 &11.6 & 10.90 & 0.0548 & J & Aae & Resolved double, Hydra A \\
J094110-120450 & 09 41 10.74 & -12 04 50.6 &  44 &  2 &  -  &  - &  -  &  - &  728.7 &    ... & g0941102-120536 &46.7 & 12.24 & 0.1500 & 3 & Aa  & Resolved triple \\ 
J105533-283134 & 10 55 33.39 & -28 31 34.7 & 181 &  8 & 712 &  4 & 996 & 17 & 1792.1 &    ... & g1055334-283134 & 0.7 & 11.29 & 0.0611 & 4 & Aae &  \\
J124849-411840 & 12 48 49.37 & -41 18 40.3 & 278 & 10 & 819 & 19 &1421 & 45 &    ... & 5674.0 & g1248493-411840 & 1.5 &  7.14 & 0.0102 & J & Aae &  NGC\,4696 \\  
J125438-123255 & 12 54 38.55 & -12 32 55.7 &  83 &  2 &  -  &  - &  -  &  - & 8200.  &    ... &  g1254361-123353 &67.5 &  7.75 & 0.015 & J & Aae & Resolved double, NGC\,4783 \\ 
J132527-430104 & 13 25 27.60 & -43 01 04.9 &28350&  - &  -  &  - &  -  &  - &    ... &342000. & g1325277-430108 & 3.2 &  3.94 & 0.0018 & J & Aae & Centaurus\,A \\ 
J141949-192825 & 14 19 49.80 & -19 28 25.3 & 117 &  8 &  -  &  - &  -  &  - & 1973.1 &    ... & g1419497-192826 & 1.2 & 12.35 & 0.1200 & ..& ..  & \\   
PKS\,1717-00$^*$ & 17 20 28.15 & -00 58 46.8 & 435 & 13 &  -  &  - &  -  &  - &56980. &   ... & g1720282-005847 & ..  & 10.82 & 0.0304 & J & Aae & Two AT20G sources \\ 
PKS\,1733-56$^*$ & 17 37 22.24 & -56 36 30.0 & 696 & 27 &  -  & -  &  - &  -&    ... & 12560. & g1737358-563403 & ..  & 12.67 & 0.0985 & 4 & Ae  & Two AT20G sources \\ 
J180712-701234 & 18 07 12.55 & -70 12 34.5 &  62 &  3 &  59 &  3 &  98 &  4 &    ... & 1278.0 & g1807149-701238 &12.3 & 10.56 & 0.0402 & 4 & Aae & Resolved double \\ 
J184314-483622 & 18 43 14.77 & -48 36 22.8 & 305 &  9 & 590 & 10 &1157 &  9 &    ... &  5880. & g1843146-483623 & 1.4 & 11.84 & 0.1108 & 4 & Aa  &  \\ 
J191457-255202 & 19 14 57.79 & -25 52 02.8 & 147 &  9 & 275 & 12 & 322 & 11 &  612.2 &    ... & g1914577-255201 & 2.1 & 10.88 & 0.0631 & 4 & Aa  & \\ 
J195817-550923 & 19 58 17.08 & -55 09 23.3 & 581 & 12 &1132 & 12 &2063 & 35 &    ... & 9700.  & g1958185-550930 &13.6 & 11.81 & 0.0581 & J & Aae & Resolved double  \\ 
J203444-354857 & 20 34 44.68 & -35 48 57.6 & 104 &  3 &  -  &  - &  -  &  - & 1717.1 & 2609.0 & g2034447-354902 & 4.3 & 11.79 & 0.0888 & 4 & Aa  & ESO\,400-IG40 \\
J205603-195646 & 20 56 03.52 & -19 56 46.8 & 159 &  3 &  -  &  - &  -  &  - & 2726.2 &    ... & g2056043-195635 &15.7 & 12.69 & 0.1566 & 4 & Aa? & Resolved double  \\
J213741-143241 & 21 37 41.17 & -14 32 41.9 & 256 &  7 &  -  &  - &  -  &  - & 3864.0 &    ... & g2137452-143256 &59.9 & 12.61 & 0.1999 & 4 & AeB & Hotspot of PKS\,2135-147 \\ 
J215129-552013 & 21 51 29.86 & -55 20 13.1 &  88 &  2 &  93 &  3 & 154 &  3 &    ... & 2450.  & g2151299-552013 & 0.7 & 10.43 & 0.0388 & ..& ..  & \\ 
J215706-694123 & 21 57 06.10 & -69 41 23.3 &3400 &210 &  -  &  - &  -  &  - &    ... &37300.  & g2157060-694124 & 0.9 & 10.05 & 0.0285 & J & AeB & ESO\,075-G41 \\  
J220113-374654 & 22 01 13.79 & -37 46 54.3 & 174 &  5 & 460 & 10 & 536 & 12 & 1530.  & 2140.  & g2201171-374624 & 49.5& 10.86 & 0.0334 & 4 & Ae  & Hotspot of PKS\,2158-380 \\
J235156-010909 & 23 51 56.29 & -01 09 09.3 & 141 &  7 &  -  &  - &  -  &  - & 1608.4 &    ... & g2351561-010913 & 4.5 & 11.87 & 0.1737 & 4 & AeB &  \\
\hline
\end{tabular}
\end{minipage}
\end{table*}